\documentclass{aa}
\usepackage{times}
\usepackage{epsfig}
\begin{document}

   \thesaurus{04  
              {03.20.1 
               10.19.1 
               10.15.1 
               10.11.1 
               10.19.3 }}

   \title{The distribution of nearby stars in phase space mapped by Hipparcos\thanks{Based on data from the Hipparcos astrometry satellite}}

   \subtitle{III. Clustering and streaming among A-F type stars}

   \author{E. Chereul\inst{1}, M.~Cr\'ez\'e\inst{1,2} \and O. Bienaym\'e\inst{1}}

    \institute{Centre de Donn\'ees de Strasbourg, Observatoire Astronomique de Strasbourg,
11, rue de l'Universit\'e, F-67000 Strasbourg, France
    \and  I.U.P de Vannes, Tohannic, rue Yves Mainguy,  F-56 000 Vannes, France}

    \offprints{chereul@cdsxb6.u-strasbg.fr}

    \date{Received / accepted}
   \authorrunning{E. Chereul et al}
   \titlerunning{The distribution of nearby stars in phase space mapped by Hipparcos. III}
   \maketitle
  \begin{abstract}
This paper presents the detailed results obtained in the search of density-velocity inhomogeneities 
in a volume limited and absolute magnitude limited sample of A-F type dwarfs within 125 parsecs of the Sun.
A 3-D wavelet analysis is used to extract inhomogeneities, both in the density and velocity distributions. 
Thus, a real picture of the phase space is produced. Not only are some clusters and streams detected, 
but the fraction of clumped stars can be measured. By estimating individual stellar ages one can relate 
the streams and clusters to the state of the interstellar medium (ISM) at star formation time and provide a 
quantitative view of cluster evaporation and stream mixing.\\
Having established this picture without assumption we come back to previously known observational facts 
regarding clusters and associations, moving groups or so-called {\it superclusters}.\\
In the 3-D position space,  well known open clusters (Hyades, Coma Berenices and Ursa Major), 
associations (parts of the Scorpio-Centaurus association) as well as the Hyades evaporation track are 
retrieved. Three new probably loose clusters are identified (Bootes, Pegasus 1 and 2). The sample is relatively 
well mixed in the position space since less than 7 per cent of the stars belong to structures with 
coherent kinematics, most likely gravitationally bound.\\
The wavelet analysis exhibits strong velocity structuring at typical scales of velocity dispersion 
$\overline{\sigma}_{stream}$ $\sim$ 6.3, 3.8 and 2.4 $km\cdot s^{-1}$. 
The majority of large scale velocity structures ($\overline{\sigma}_{stream}$ $\sim$ 6.3 $km\cdot s^{-1}$) are Eggen's 
{\em superclusters} (Pleiades SCl, Hyades SCl and Sirius SCl) with the whole Centaurus association. 
A new {\em supercluster}-like structure is found with a mean velocity between the Sun and Sirius SCl velocities. 
These structures are all characterized by a large age range which reflects the overall sample age distribution. 
Moreover, a few old streams of $\sim$ 2 Gyr are also extracted at this scale with high U components. 
We show that all these large velocity dispersion structures represent 63$\%$ of the sample. 
This percentage drops to 46$\%$ if we remove the velocity background created by a smooth velocity ellipsoid in 
each structure. Smaller scales ($\overline{\sigma}_{stream}\sim$ 3.8 and 2.4 $km\cdot s^{-1}$) reveal that 
{\em superclusters} are always substructured by 2 or more streams which generally exhibit a coherent age distribution. 
At these scales, background stars are negligible and percentages of stars in streams are 38$\%$ and 18$\%$ 
respectively.\\

   \end{abstract}
%
\section{Introduction}
The context of this study as well as implications of the results for the solar circle kinematics understanding 
are in Chereul et al (1998) (hereafter Paper II). This third paper aims to provide to the reader of Paper II the 
details of the phase space inhomogeneities brought to the fore by means of wavelet analysis and technical 
details on several features of the analysis.\\
Wavelet analysis is extensively used to extract deviations from smooth homogeneity, both in 
density (clustering) and velocity distribution (streaming). A 3-D wavelet analysis tool is first 
developed and calibrated to recognize physical inhomogeneities among random fluctuations. It 
is applied separately in the position space (Section \ref{sec:density}) and in the velocity space 
(Section \ref{sec:velocity}). Once significant features are identified (whether in density or in velocity), 
feature members can be identified and their behaviour can be traced in the complementary space 
(velocity or density). Thus, a real picture of the phase space is produced. Not only are some clusters 
and streams detected, but the fraction of stars involved in clumpiness can be measured. 
Then, estimated stellar ages help connecting streams and clusters to the state of the ISM at star formation 
time and providing a quantitative view of cluster melting and stream mixing at work.\\
Only once this picture has been established on a prior-less basis we come back to previously 
known observational facts such as clusters and associations, moving groups or so-called {\it superclusters}.
A sine qua non condition for this analysis to make sense is the completeness of data within well 
defined limits. In so far as positions, proper motions, distances, magnitudes and colours are concerned, 
the Hipparcos mission did care. Things are unfortunately not so simple concerning 
radial velocities and ages. More than half radial velocities are missing and the situation regarding 
Str\"omgren photometry data for age estimation is not much better. So we developed palliative 
methods which are calibrated and tested on available true data to circumvent the completeness 
failure. The palliative for radial velocities is based on an original combination of the classical 
convergent point method (Section \ref{sec:pcvg}) with the wavelet analysis. The palliative for 
ages (Section \ref{sec:agemvbv}) is an empirical relationship between age and an absolute-magnitude/colour 
index. It has only statistical significance in a very limited range of the HR diagram.\\
\\
The sample was pre-selected inside the Hipparcos Input Catalogue (\cite{Hic92}) among the ``Survey stars''. 
The limiting magnitude is $m_{v} \leq 7.9 + 1.1\cdot\sin\mid b\mid$ for spectral types 
earlier than G5 (\cite{Tur86}). 
Spectral types from A0 to G0 with luminosity classes V and VI were kept. Within this 
pre-selection the final choice was based on Hipparcos \cite{Hip97} magnitude ($m_{v} \leq 8.0$), 
colours ($-0.1\leq B-V \leq 0.6$), and parallaxes ($\pi \geq 8 mas$). The sample studied 
(see sample named h125 in Paper I, \cite{Creze98}) is a slice in absolute magnitude of this selection 
containing 2977 A-F type dwarf stars with absolute magnitudes brighter than 2.5. It is complete within 
125 pc from the Sun.\\
\\
Details concerning the wavelet analysis implementation through the ``\`a trou'' algorithm and the thresholding 
procedure are given in Section \ref{sec:implement}. The palliative age determination  
method is fully described in Section \ref{sec:agemvbv}. The results of the density analysis in position space (clustering) 
are given in Section \ref{sec:density}. The convergent point method is explained in Section \ref{sec:pcvg} and 
the procedures estimating spurious members and field stars among detected streams are given in 
Section \ref{sec:noisestim} and \ref{sec:fieldstar} respectively. The results of the velocity space analysis (streaming) 
are in Section \ref{sec:phenom} including a detailed review of observational facts and a comparison with evidences 
collected by previous investigators. Conclusion in Section \ref{sec:conclu} summarizes the different results.
%
\section{Implementation of the wavelet analysis}
\label{sec:implement}
Main reasons for the choice of this algorithm developed by  \cite{Holschneider89} as well as theoretical way to compute 
wavelet coefficients are provided in Paper II. Here we focus on the practical computation of the wavelet coefficients 
(Section \ref{sec:atrou}) and the thresholding procedure adopted (Sections \ref{sec:extract} and \ref{sec:calib}). 
\subsection{The ``\`a trou'' algorithm}
\label{sec:atrou}
 \begin{figure}[ht]
  \begin{center}
    \leavevmode
  \vspace{1.5cm}
  \centerline{\epsfig{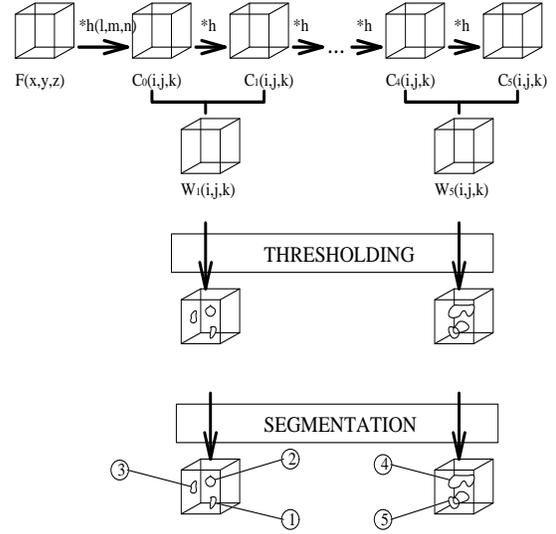}}
  \end{center}
  \vspace{-1.5cm}
  \caption{\em Implementation of the wavelet analysis}
  \label{fig:algo}
\end{figure}
The adopted {\em  mother wavelet} $\Psi(x)$ is defined as the difference at two different scales of a same 
{\em scaling function } $\Phi(x)$ (or smoothing function):
\begin{equation}
\Psi(x)=\Phi(x)-\frac{1}{2}\Phi(\frac{x}{2})\\
\label{eq:relation}
\end{equation} 
This particularity allows to use the ``\`a trou'' algorithm implementation of the wavelet transform. This algorithm 
avoid the direct computation of the scalar product between $\Psi(x)$ and the signal $F(x)$ to obtain wavelet 
coefficients. It uses an iterative schema based both on the relation existing between the {\em mother wavelet} 
and the {\em scaling function} (Equation \ref{eq:relation}) and the following relation
\begin{equation}
\frac{1}{2^{s+1}}\Phi(\frac{x}{2^{s+1}})=\sum_{l=-2}^{2}h(l)\Phi(\frac{x}{2^{s}}-l)\\
\end{equation} 
where $h(l)$=\{$\frac{1}{16},\frac{1}{4},\frac{3}{8},\frac{1}{4},\frac{1}{16}$\} is a one-dimensional discrete 
low-pass filter. $h(l)$ is applied in the algorithm to compute iteratively the different approximations 
(or smoothing) $C_{s}$ of the signal at each scale $s$. For a 3D signal like our density or velocity distributions, $h(l)$ 
is applied separately in each dimension to obtain the signal approximation at scale $s$ and pixel (i,j,k):
\begin{eqnarray}
C_{s}(i,j,k)=\sum_{l=-2}^{2}\sum_{m=-2}^{2}\sum_{n=-2}^{2} h(l)h(m)h(n)\nonumber \\
\, \cdot C_{s-1}(i+2^{s-1}l,j+2^{s-1}m,k+2^{s-1}n)
\end{eqnarray}
The distance between two bins  increases by a factor 2 from scale $s-1$ to scale $s$.
The result at each scale $s$ ($C_{s-1}(i,j,k)$-$C_{s}(i,j,k)$) is the signal difference which contains the 
information between these two scales and is the discrete set of 128$^{3}$ wavelet coefficients $W_{s}(i,j,k) $
associated with the wavelet transform by $\Psi(x,y,z)$ (Figure \ref{fig:algo}):
\begin{equation}
W_{s}(i,j,k)=C_{s-1}(i,j,k)-C_{s}(i,j,k)\\
\end{equation}
\subsection{Extracting significant coefficients}
\label{sec:extract}
In order to remove non-zero wavelet coefficients generated by noise fluctuations in the 3D distributions, a 
local thresholding is applied at each scale in the space of wavelet coefficients. Thresholds are set at each scale 
and each position by estimating the noise level generated at the same scale by a uniform random signal built 
with the same gross-characteristics as the observed one at the position considered.
\begin{itemize}
\item  Local hard-thresholding procedure\\
The noise $N_{s}(i,j,k)$ is estimated at the position of each pixel $(i,j,k)$ from the observed distribution 
in a filter area corresponding to scale $s$. The signal estimation $S_{s}(i,j,k)$, in this area, is:\\
\begin{equation}
S_{s}(i,j,k)=\sum_{l=i-2^{s}}^{i+2^{s}} \sum_{m=j-2^{s}}^{j+2^{s}} \sum_{n=k-2^{s}}^{k+2^{s}} F(l,m,n) \\
\end{equation}
the noise estimation is:\\
\begin{equation}
N_{s}(i,j,k)=\sqrt{\frac{S_{s}(i,j,k)}{(2^{s+1})^{3}}} \\
\end{equation}
At scale $s$, the number of pixels included in the convolution filter volume is (2$^{s+1}$)$^{3}$.\\
The local threshold $\lambda_{s}^{+}(i,j,k)$ is inferred from the local noise $N_{s}(i,j,k)$ through 
calibration curves established by numerical simulations (see Section \ref{sec:calib}) which link it to 
the noise $N_{s}(i,j,k)$. 
Then the hard-thresholding on wavelet coefficients follows:\\
\[  \mbox{$W_{s}(i,j,k) $}=  \left\{  \begin{array}{ll}
        \mbox{$W_{s}(i,j,k) $} & \mbox{if $W_{s}(i,j,k)$ $\geq$ $\lambda_{s}^{+}(i,j,k)$} \\
        0 & \mbox{otherwise}\\
        \end{array}
        \right. \]
\item  Segmentation procedure\\
A segmentation procedure is applied on thresholded coefficients to localize significant over-densities.
This segmentation algorithm detects all maxima in a moving box of filter size volume. Neighbouring 
pixels are aggregated around each maximum until the intensity of wavelet coefficient stop decreasing. 
This is the classical method of valley lines. All aggregated coefficients around a maximum are considered 
to sign one structure. All structures are labelled and defined pixel by pixel. This procedure has the 
advantage of splitting coalescent structures.
\end{itemize}
\subsection{Thresholding calibration}
\label{sec:calib}
\begin{itemize}
\item  Choice of significant probability\\
Following Lega et al (1996), thresholds $\lambda$ are defined with respect to the probability $P$ that 
the coefficient $W_{s}(i,j,k)$ is larger than $\lambda_{s}^{+}$ (or smaller than $\lambda_{s}^{-}$) at a 
given scale $s$ in the distribution of positive coefficients (or negative coefficients).\\
The probability is
\begin{equation}
\hspace{-0.1cm}P(W_{s}(i,j,k)\geq \lambda_{s}^{+})= \frac{\int_{W_{s}=\lambda_{s}^{+}}^{+\infty} f(W_{s}) dW_{s}}
{ \int_{W_{s}=0}^{+\infty} f(W_{s}) dW_{s}}=\epsilon\\
\end{equation}
where $f$ is the frequency of positive wavelet coefficients at scale $s$.
$P$ was set to achieve the best compromise between false detections and losses at each scale
with a set of uniform background simulations containing gaussian structures of different sizes and signal
to noise ratio. This optimum was found between 5$\cdot$10$^{-4}$ and 10$^{-4}$ in all experiments and 
in all scales  under reasonable signal to noise ratio. Then $\epsilon = 10^{-4}$ was fixed for the rest of the
calibration.\\
\item  Calibration curves\\
A calibration curve is obtained for each scale with an other set of different Poisson noise simulations.
Noises $N_{s}(i,j,k)$ are estimated at each pixel $(i,j,k)$ from the observed distribution in the same way as 
described in Section \ref{sec:extract}.
The mean $\overline{N_{s}}$ of Poisson noise distribution obtained is linked with threshold 
$\lambda_{s}^{+}$ (or $\lambda_{s}^{-}$) defined from $P$ giving the calibration curve for scale $s$.\\
\end{itemize}
%
%
\section{Palliative age determination method}
 \label{sec:agemvbv}
We succeed in estimating individual ages from Str\"omgren photometry (\cite{Figue91},
Masana, 1994, Asiain et al, 1997) for a third of the sample (see Paper II). 
To rule out any bias coming from an analysis of this sub-sample alone, we propose an empirical palliative 
age estimation method based on the Hipparcos absolute magnitude and colour 
for the rest of the sample. On a first step, we use existing ages to draw a plot of ages versus 
$(Mv, (B-V))$. A primary age parameter is assigned as the mean age associated to a 
given range of $(Mv, (B-V))$ (Section \ref{sec:roughage}). Then Str\"omgren age data are used a second time 
to assign a probability distribution of  palliative ages as a function of the primary age parameter 
(Section \ref{sec:corage}).
 \subsection{Primary age parameter}
\label{sec:roughage}
\begin{figure}
  \vspace*{-0.7cm}
  \begin{center}
    \centerline{\epsfig{file=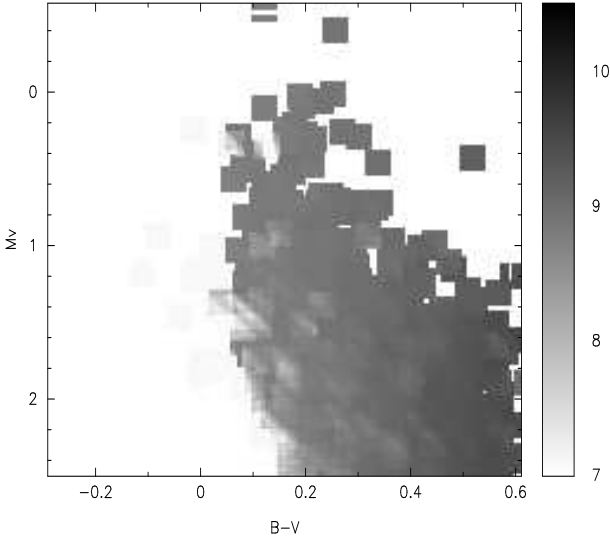,width=9.cm,height=10.cm,angle=-90.}}
  \end{center}
  \caption{\em Primary age parameter versus $(M_{v},(B-V))$ relationship calibrated with the Str\"omgren sub-sample. Grey levels represent the log(age) value.} 
  \label{fig:agemvbv_cal}
 \end{figure}

\begin{figure}
  \begin{center}
  \centerline{\epsfig{file=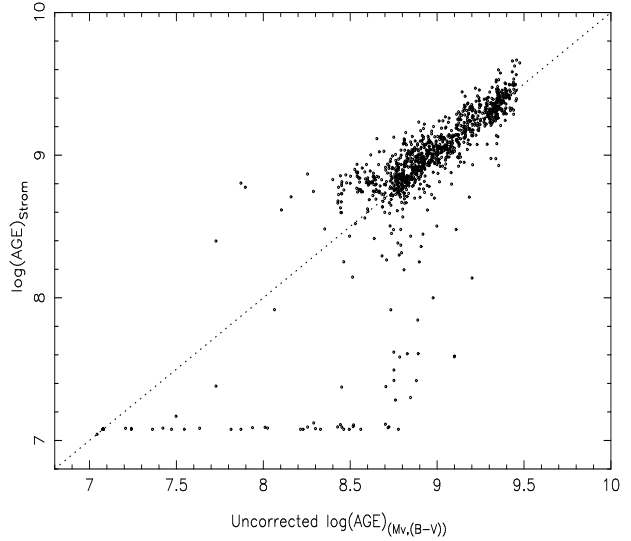,width=8.cm,height=9.cm,angle=-90.}}
  \end{center}
  \caption{\em Primary age parameter versus Str\"omgren age.}
   \label{fig:unextra_strom}
  \end{figure}

Str\"omgren ages are available for 1077 A-F type stars with spectral types later than A3 ($B-V$ $\geq$ 0.08) because 
the metallicity cannot be determined for A0-A3 spectral types. With this sub-sample a relation between the 
absolute magnitude $M_{v}$, the colour indice $B-V$ and the age has been calibrated on a grid to extrapolate a 
less accurate age for the rest of our sample (1900 stars). Since the relation cannot be calibrated with Str\"omgren 
ages for spectral types between A0 and A3, we add 30 known very young stars ($\sim$$10^{7}$ yr) belonging to the 
Centaurus-Crux association (see Section \ref{sec:density}) which have $B-V$ $\leq$ 0.08. Consequences on the 
infered age distribution are investigated in Section \ref{sec:corage}. Available Str\"omgren ages are averaged on a 
200x200 grid, ranging from [-0.5,2.5] in absolute magnitude and [-0.1,0.6] in $B-V$. The result is smoothed and 
extrapolated to empty cells, where feasible, using a 10x10 moving window. This process produces a primary age 
parameter  (Figure \ref{fig:agemvbv_cal}). In Figure \ref{fig:unextra_strom}, Str\"omgren age indicators are plotted 
against this primary age parameter. Obviously, there is not a one to one relation between the primary age parameter 
and the Str\"omgren age. Very young Str\"omgren ages are the most affected by the degeneracy of this relation; 
while the bulk of primary age parameters older than $log(age)= 8.7$ are in relatively good agreement with 
Str\"omgren ages. Only 14 among the 1900 stars have $(M_{v},(B-V))$ out of the calibrated grid and do not have 
primary age parameter. On the basis of Figure \ref{fig:unextra_strom}, we can assign a probability distribution 
for the Str\"omgren age to each primary age parameter.
 \subsection{Palliative ages}
\label{sec:corage}
  \begin{figure}
   \begin{center}
   \epsfig{file=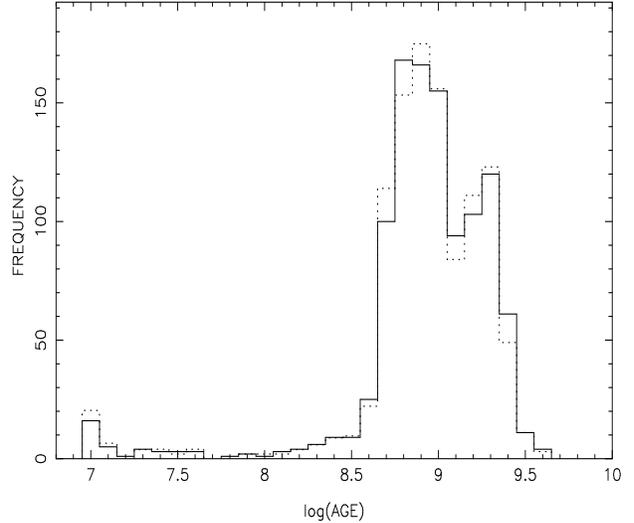,width=8.cm,height=9.cm,angle=-90.}
   \end{center}
   \caption{\em Str\"omgren age distribution obtained with 1077 stars (full line) and palliative age distribution for the same sub-sample (dotted line).}
   \label{fig:agedist1}
 \end{figure}
  \begin{figure}
   \vspace{-1cm}
   \begin{center}
   \epsfig{file=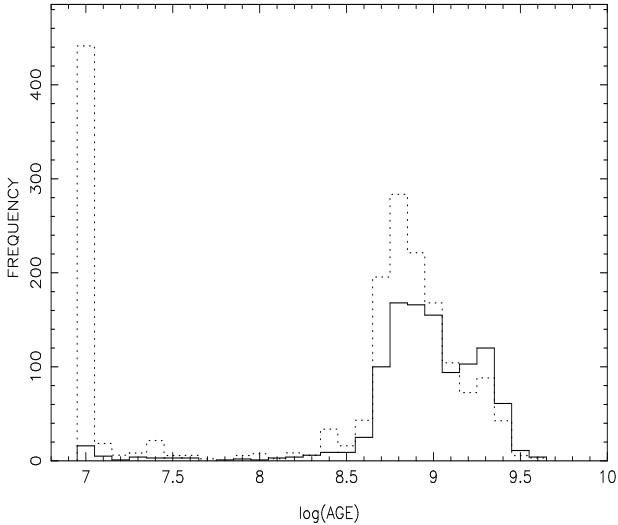,width=8.cm,height=9.cm,angle=-90.}
   \end{center}
   \caption{\em Str\"omgren age distribution obtained with 1077 stars (full line) and palliative age distribution for the remainder, i.e stars without Str\"omgren photometry (dotted line).}
   \label{fig:agedist2}
 \end{figure}
Given a primary age parameter $i$, Figure \ref{fig:unextra_strom} can be read as giving a discrete 
probability distribution $P(j/i)$ of Str\"omgren ages $j$ under $i$. $P(j/i)$ is given by: 
\begin{equation}
 P(j/i)=\frac{n(i/j)}{q(i)}\\
 \end{equation}
where $n(i,j)$ is the number of stars in cell $(i,j)$ and 
 \begin{equation}
 q(i)=\sum_{j}n(i,j)\\
 \end{equation}
These probabilities are discrete on a grid with an adaptive bin size to keep at least 3 stars per bin. 
This process produces a palliative age distribution which is free from possible biases affecting the 
sub-sample of stars with Str\"omgren photometry for stars with $B-V$ $\geq$ 0.08.\\
The small deviations from the distribution of original Str\"omgren ages (Figure \ref{fig:agedist1}) are due to finite 
bin steps used in the discretisation process. The palliative age distribution of the sub-sample without Str\"omgren 
photometry (Figure \ref{fig:agedist2}) shows a great difference for very young ages with respect to the Str\"omgren 
age distribution. The great peak at $\sim$$10^{7}$ yr partly results from the poor calibration of the relation between 
age and ($M_{v}$,$(B-V)$) for $B-V$ $\leq$ 0.08. If the calibration of the relation is realized without the additional 
Centaurus-Crux stars, we would obtain a frequency, at $\sim$$10^{7}$ yr, of $\sim$ 170 stars instead of $\sim$ 440 stars. 
It means that there are at least 170 stars of few $10^{7}$ year old and that the difference of $\sim$ 270 stars is 
composed of A0-A3 type stars which may be older. How older are they? Using theoretical isochrones 
from Bertelli et al (1994), a relation between the total lifetime of a star at the turnoff point and the color 
indice $(B-V)$ can be infered. We can show that the total lifetimes of A0-A3 type stars are between 4 - 6 $10^{8}$ yr, 
if the metallicity of the stars are between Z=0.008 and Z=0.02. Taking the half lifetime of the star as the most 
probable age, we obtained, for these 270 stars, an age distribution in the range 2 - 3 $10^{8}$ yr ($log(age)=8.3-8.5$). 
This distribution of these half lifetimes remains very well separated from the older stars in the sample.\\
To summarize, the great peak at $10^{7}$ yr contains at least 170 stars of $\sim$$10^{7}$ yr and 270 stars for which 
ages may spread up to 3 $10^{8}$ yr. These stars belong to the younger part of the sample.\\
\\
The palliative ages are statistical ages. Hence, they sometimes produce artifacts or young 
ghost peaks in some stream age distributions. Such dummy young peaks will always appear as the 
weak counterpart of a heavy peak around $log(age)=8.7$. Nevertheless palliative ages permit 
to shed light on the age content of the phase space structures when Str\"omgren data are sparse.\\
%
%
\section{Clustering}
\label{sec:density} 
\subsection{Searching for clusters}
\label{sec:implementdens}
(X,Y,Z) distributions range from -125 pc to +125 pc and are binned in a Sun centered orthonormal 
frame, X-axis towards the galactic center, Y-axis in the rotation direction and Z-axis towards the 
north galactic pole. The discrete wavelet analysis is performed on five scales: 9.7, 13.6, 21.5, 37.1 
and 68.3 pc. These values correspond to the size of the dilated filter $h$ at each scale.\\
Stars belonging to each volume defined by significant coefficients are collected. Due to the over-sampling 
of the signal by the ``\`a trou'' algorithm, some structures are detected  on several scales. A cross-correlation 
has been done between all scales to keep the structure at the largest scale provided that there is no 
sub-structure at a lower scale (higher resolution).  We control a posteriori the efficiency of the detection 
by calculating for each over-density the probability to find an identical concentration from the total 
volume of the sample in a uniform random population (see probability $P_{2}$ in Table \ref{tab:table1}). 
The method also succeeds in finding over-densities with a high probability to be generated by a poissonian 
process. An iterative 2.5 sigma clipping procedure is done on tangential velocity distributions for each group 
to remove field stars and select structures with coherent kinematics (Table \ref{tab:table1} and 
Figures \ref{fig:cencrux}, \ref{fig:coma}, \ref{fig:new1}, \ref{fig:new21} and \ref{fig:new22}).\\
 \begin{table*}[!ht]
  \vspace{-0.1cm}
   \caption{\em Main characteristics of detected spatial structures after iterative 2.5sigma clipping procedure on tangential velocities: galactic coordinates {\bf (l,b)}, distance {\bf D}, mean tangential velocities {\bf  ($\overline{T_{l}}$,$\overline{T_{b}}$)}, tangential velocity dispersions {\bf ($\sigma_{T_{l}}$,$\sigma_{T_{b}}$)}, number of selected stars {\bf N$_{sel}$}. Other parameters are given: scales of detection {\bf S}, volume {\bf v} occupied by the total over-density,  number of expected stars {\bf N$_{e}$} in this volume  according to the mean density, number of observed stars {\bf N$_{o}$} before selection on tangential velocities, probability {\bf P$_{1}$} to find such a concentration from a poissonian distribution in this v volume and probability {\bf P$_{2}$} to find this volume v with this concentration within the 125 pc radius sphere. A value of .001 indicates a probability P$\leq$0.001.}
   \label{tab:table1}
     \leavevmode
     \footnotesize 
    \vspace{-0.5cm}
    \begin{center}
     \begin{tabular}[h]{lrrrccccrrrrrcc}
       \hline \\[-5pt]
       Structure &l&b&D&$\overline{T_{l}}$&$\sigma_{T_{l}}$&$\overline{T_{b}}$&$\sigma_{T_{b}}$&N$_{sel}$&S &v&N$_{e}$&N$_{o}$&P$_{1}$&P$_{2}$\\[+5pt]
        &(deg)&(deg)&(pc)& \multicolumn{4}{c}{($km\cdot s^{-1}$)}&&&(pc$^{3}$)&&&& \\[+5pt]
       \hline \\[-5pt]
        \ 1  &227.65  & -6.14 & 117 & \multicolumn{4}{c}{Possible double system}&2&2&831 & 0.302& 2&.037&.999\\
        \ 2  & 58.52 &+35.89  &54 &-4.9  & 0.6 & 10.1 & 6.8 & 2 &2 & 20& 0.007& 2& .001& .999\\
        \ 3  & 83.96 & -42.34& 81 & \multicolumn{4}{c}{Possible double system}&2&2& 5&0.002 & 2&.001&.933\\
        \ 4  & 40.71& -0.60 &121  & 6.0 & 3.3 & -5.5 &5.4 &4 &2 & 696& 0.253& 4&.001&.807\\
        \ 5  & 231.58 & -20.34 & 101 & \multicolumn{4}{c}{Possible double system}&2&2 & 77& 0.02& 2&.001&.999\\
        \ 6~U Ma &128.44  & +59.97 & 25 & -12.5 & 1.7 & 1.5 &  2.8& 6& 2-3& 1191& 0.43& 6&.001&.042\\
        \ 7~Cen-Crux & 303.91 & +2.95 & 105 &  -17.9&  3.9&  -7.4&  2.1& 33& 2-4& 52229& 19.0& 36&.001&.050\\
        \ 8  &  228.98&  +28.84&  110&  -7.6&  11.4&  -17.0&  8.3& 8&4 & 16093& 5.8& 8&.236&.999\\
        \ 9 &  325.40&  -34.41&  111&  -20.9&  13.9&  -8.1&  8.5& 6& 4& 11718& 4.2& 6&.257&.999\\
        10 Cen-Lup   & 335.27 & +17.30 &117  & -17.7 &  10.5&  -5.2&  10.5& 9& 4& 33028& 12.0& 12&.540&.999\\
        11 Coma&  214.50&  +83.6&  88&  2.9&  1.8&  -6.4&  2.3& 16& 2-4& 10968& 3.9& 21&.001&.001\\
        12 Hyades&  180.26&  -21.38&  47&  19.4&  2.8&  14.2&  2.1& 22& 2-4& 63986& 23.3& 45&.001&.005\\
        13  &  34.39& +33.89 &  108&  -9.5&  7.6&  -0.5&  14.3& 7& 4& 18316& 6.6& 7&.499&.999\\
        14 {\bf Bootes}&  97.64&  +56.68&  105&  5.0&  1.6&  6.8& 1.3& 5&2-4& 12574& 5.9& 13&.001&.455\\
        15 {\bf Pegasus 1}&  60.77&  -32.57&97&  0.6&  3.5&  -12.0& 3.9 &10 &4-5 & 71687& 26.0& 35&.054&.998\\
        \hspace*{0.3cm} {\bf Pegasus 2}&  54.20&  -31.33& 89 & 14.2 & 2.8 &-21.2  & 5.2& 8 &4-5& ''& ''& ''&''&''\\
        16  & 80.44 & +32.61 & 109 & 4.7 &  16.3&  0.3&  3.7& 8& 4& 17651& 6.4& 8&.316&.999\\
        17 Hyades' tail&  163.56& -8.34& 67 &  20.8&  8.5& -2.6 & 4.4 &39 & 5& 246202& 89.5& 90&.496&.999\\
       \hline \\
       Total &  & & &  &  & & &{\bf 189}& &&&299&&\\
        Percentage &  & & &  &  & & &{\bf 7$\%$}&&&&10$\%$&&\\
      \hline \\
      \end{tabular}
      \end{center}
      \end{table*}
\subsection{Detailed cluster characteristics}
\label{sec:clustdet}
The volumes selected by the segmentation procedure content 10 per cent of the stars. 
After the 2.5 sigma clipping procedure on the tangential velocities, only 7 per cent are still in clusters or groups. 
Most of them are well known: Hyades, 
Coma Berenices, Ursa Major open clusters (hereafter OCl) and the Scorpio-Centaurus association. 
The following cases are the most interesting either because of newly discovered features associated 
to well known structures (Scorpio-Centaurus association) or simply because they 
were undetected up to now (Bootes and Pegasus 1 and 2).\\
\begin{enumerate}
\item The {\bf Scorpio-Centaurus association} shows three precisely limited over-densities with an elongated 
shape crossing the galactic disc (see scale 4 on Paper II, Figure 5): lower Cen-taurus-Crux, 
upper Centaurus-Lupus which both have identical proper motions and a third unknown clump 
at $(l=325.4^{o}$, $b=-34.4^{o})$ with slightly different proper motions along $l$ axis (see cluster 9 
in Table \ref{tab:table1}). We do not find previous mention of this extension of the Scorpio-Centaurus. 
 \begin{figure}
  \begin{center}
  \epsfig{file=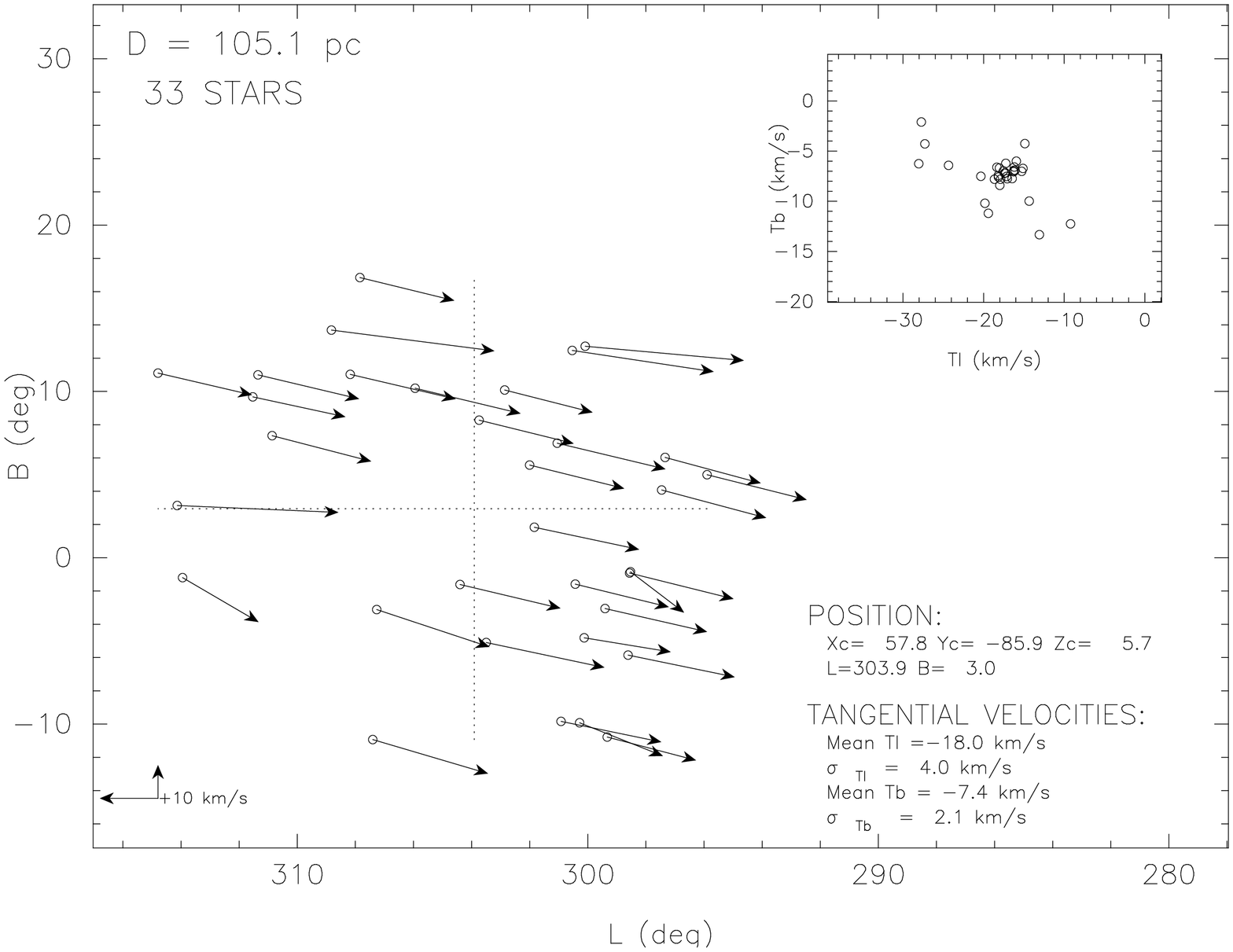,width=7.cm,height=8.cm,angle=-90.}
  \epsfig{file=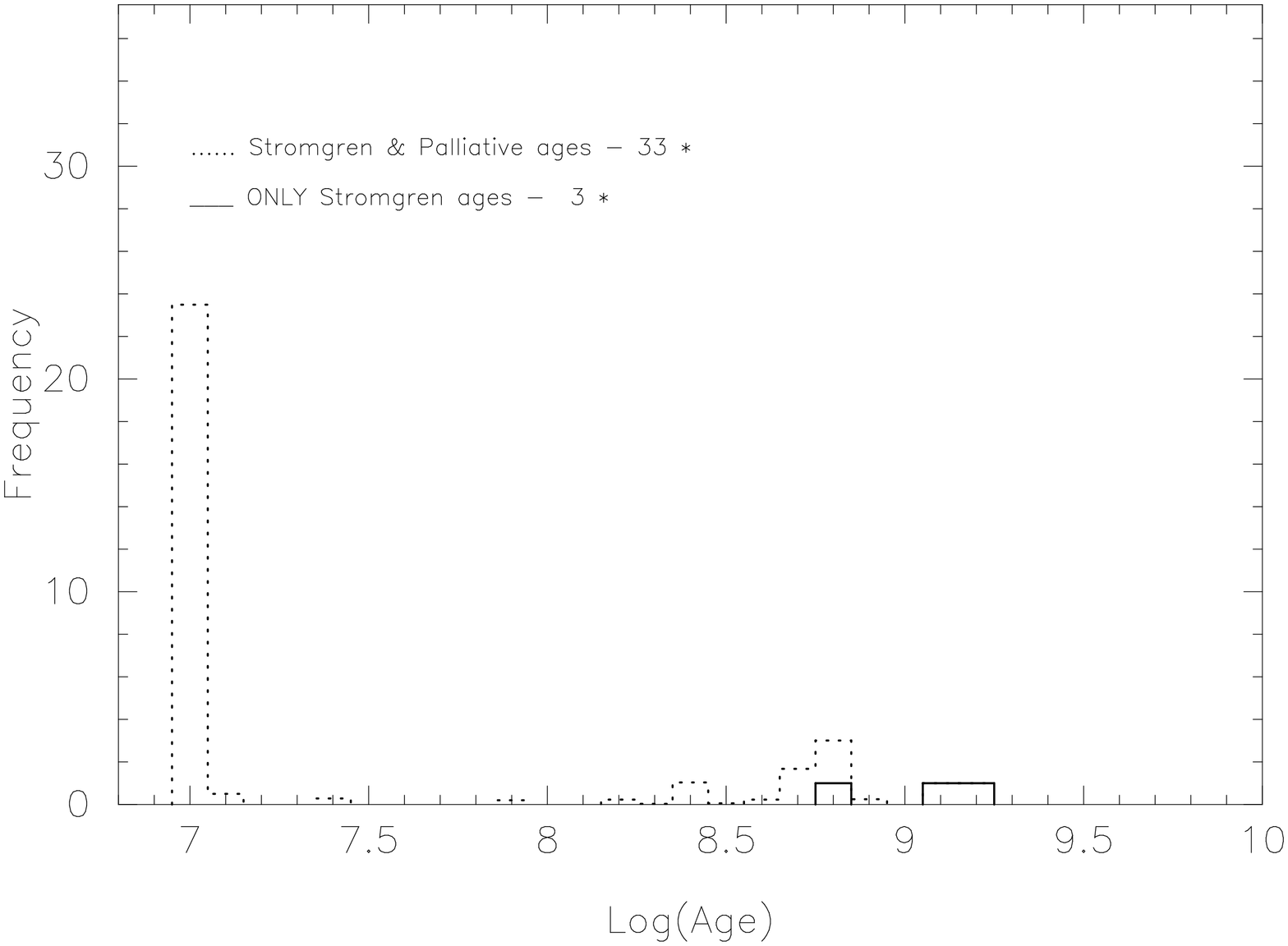,width=7.cm,height=8.cm,angle=-90.}
  \end{center}
  \caption{\em Lower Centaurus-Crux after selection on tangential velocities {\bf (top)} and its age distribution {\bf (bottom)}.}
  \label{fig:cencrux}
 \end{figure}
Unfortunately very few of these stars have Str\"omgren photometry. In the case of Centaurus-Crux 
(see Figures \ref{fig:cencrux}), among 33 stars, 3 stars having a Str\"omgren age are rather old. 
One of these stars is flagged as a double system in the Hipparcos catalogue, which could alter the 
Str\"omgren photometry. 
The two others probably do not belong to the association but are not rejected by the selection on 
tangential velocities because of proper motions close to the association ones. However this association 
is known to be an O-B association, which means that it is younger than a few tens of million years.\\
The palliative age distributions show that all the three clumps are composed by the youngest stars of 
the sample. The analysis of the velocity field (see Section \ref{sec:velfield}) reveals that these density 
inhomogeneities are embedded in a stream-like structure moving through all the 125 pc sphere.\\
\\
Two new structures are found both probably on their way towards disruption. In the following 
these loose clusters are designated after the constellation they are found in.\\
 \begin{figure}
  \begin{center}
  \epsfig{file=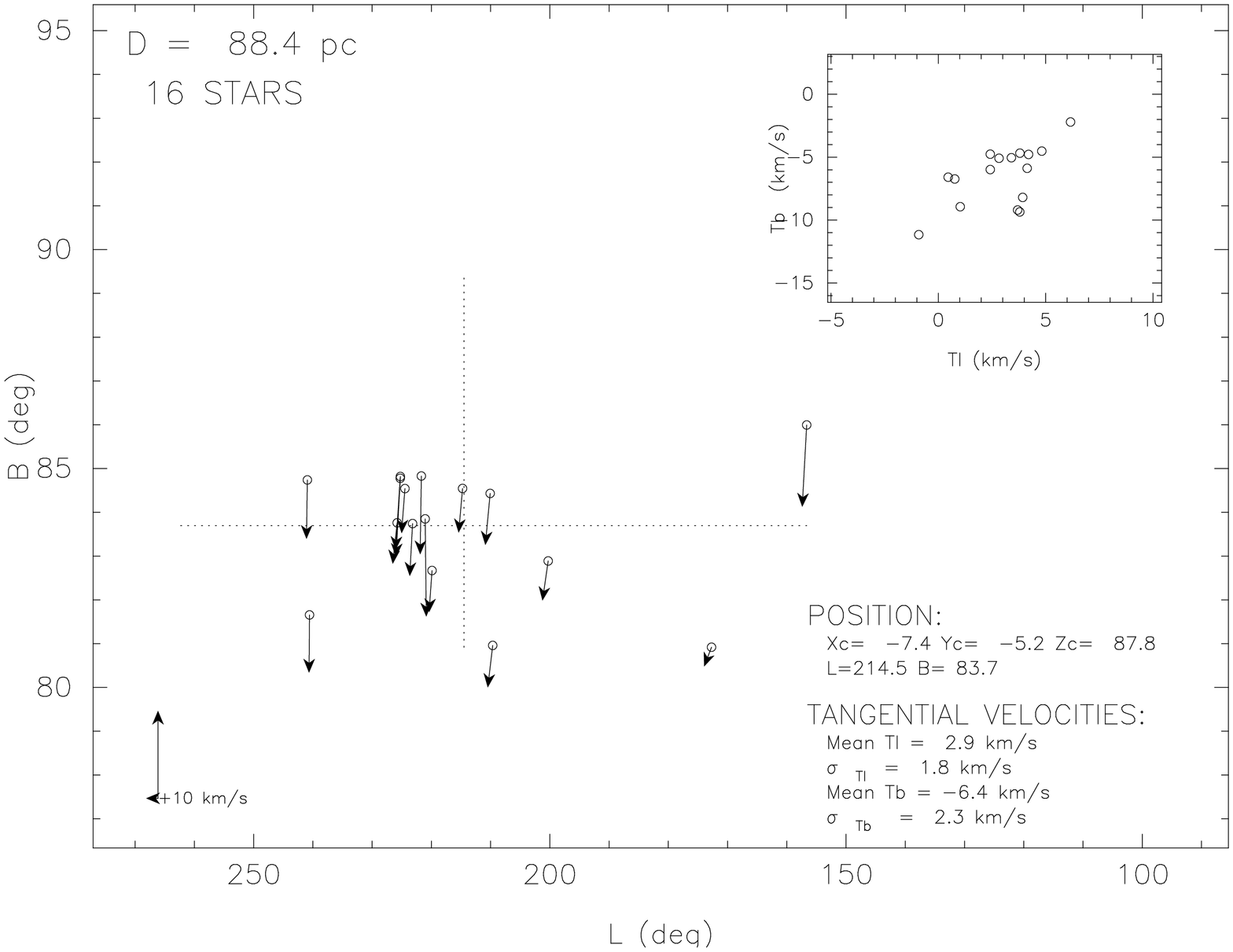,width=7.cm,height=8.cm,angle=-90.}
  \epsfig{file=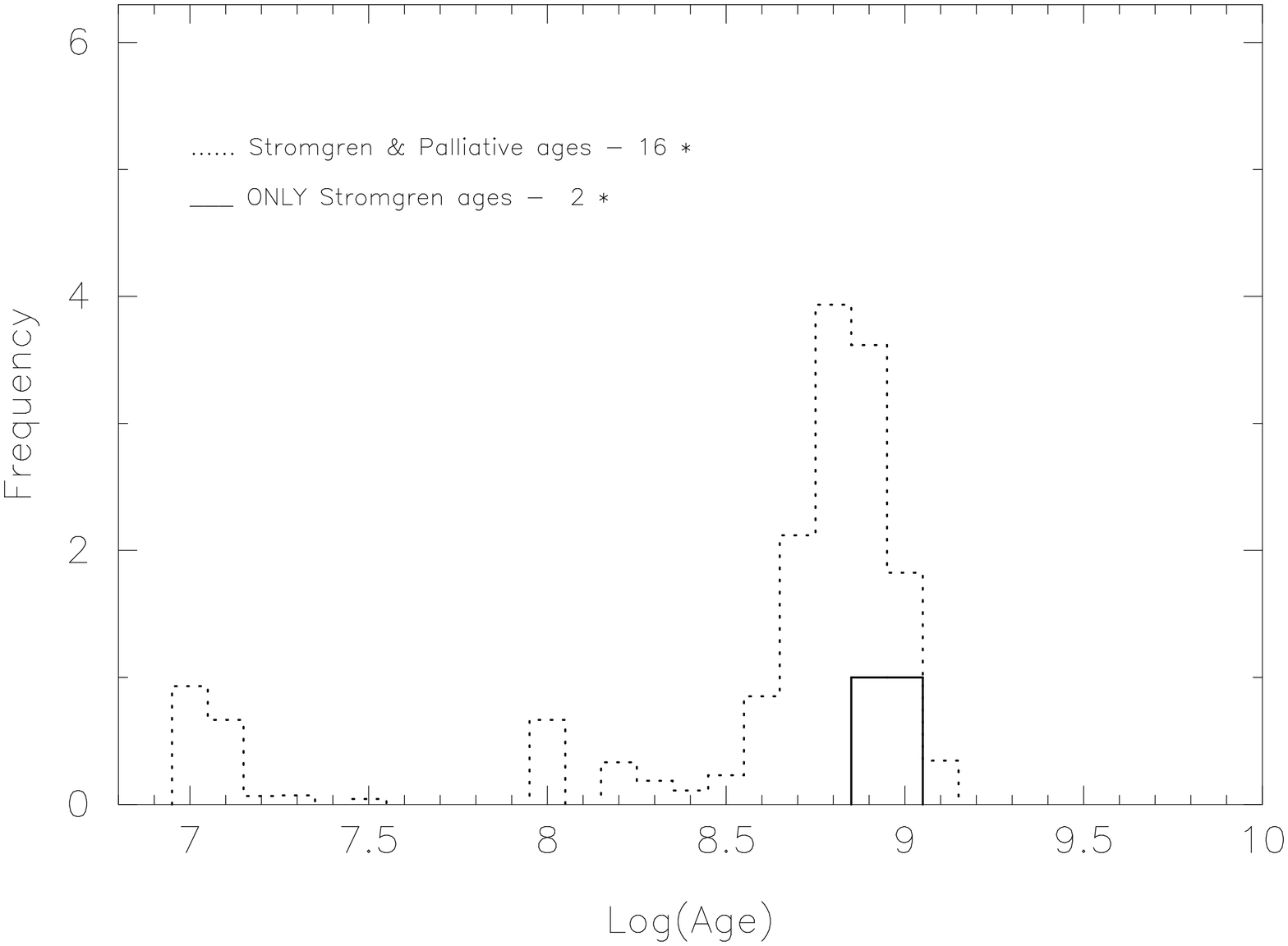,width=7.cm,height=8.cm,angle=-90.}
  \end{center}
  \caption{\em Coma Berenices open cluster after selection on tangential velocities ({\bf top}) and its age distribution ({\bf bottom}).}
  \label{fig:coma}
 \end{figure}
 \begin{figure}
  \begin{center}
  \epsfig{file=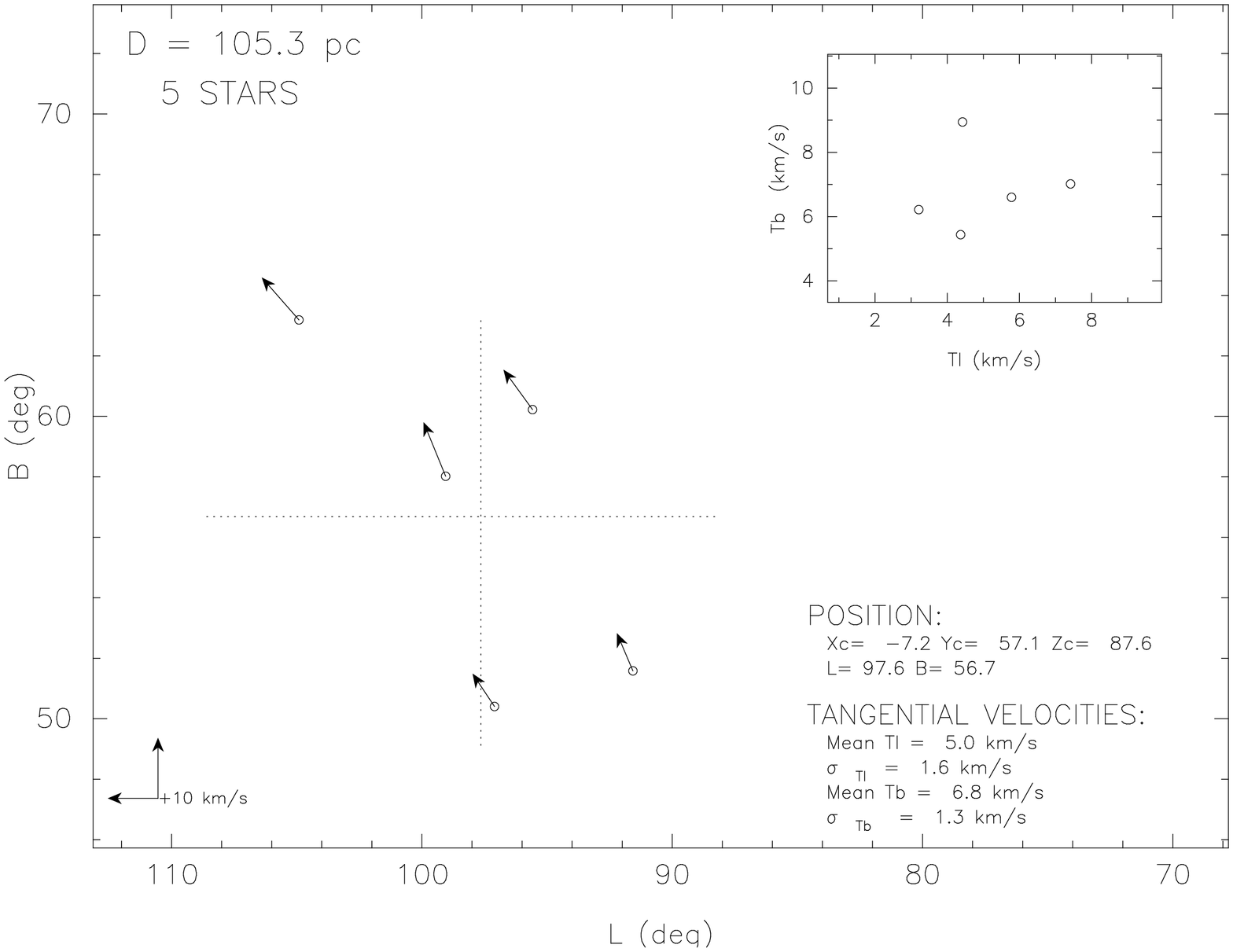,width=7.cm,height=8.cm,angle=-90.}
  \epsfig{file=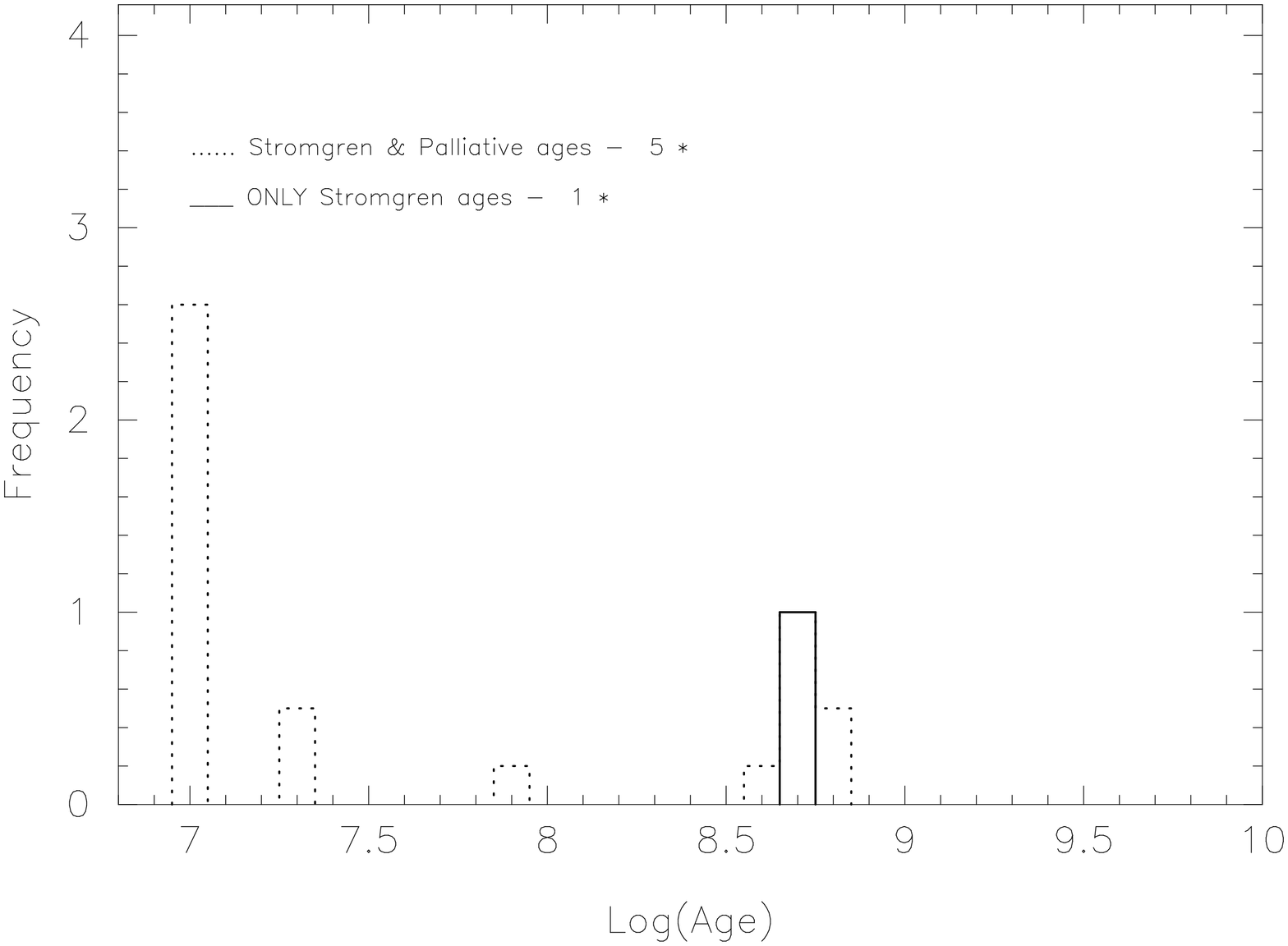,width=7.cm,height=8.cm,angle=-90.}
  \end{center}
  \vspace{-0.3cm}
  \caption{\em Bootes loose cluster after selection on tangential velocities ({\bf top}) and its age distribution ({\bf bottom}).}
  \label{fig:new1}
 \end{figure}
\item {\bf Bootes} (Figure \ref{fig:new1}) is composed of only 5 stars at a mean distance of 105 pc and 
centred on $(l=97.6^{o},b=+56.6^{o})$. This small clump is interesting for several reasons. It is located at 
the same x and z coordinates as Coma Ber. OCl (Figure \ref{fig:coma}) but 62 pc ahead along y-axis with 
exactly the same space velocity components (U,V,W)=(-3.1, -7.8, -0.8) $km\cdot s^{-1}$. These coincidences 
suggest a common origin for these two structures. But their respective age distributions (Figure \ref{fig:coma} 
and  \ref{fig:new1}) show that Bootes is probably younger (peak at $10^{7}$ yr for the palliative age distribution) 
than Coma Ber. OCl which is $\sim 6\cdot 10^{8}$ yr old. Moreover, the smaller tangential velocity dispersions of 
Bootes tends to confirm this hypothesis. A more detailed investigation among other spectral types is necessary to 
better define Bootes in velocity space as well as in age content.\\
 \begin{figure}
  \begin{center}
  \epsfig{file=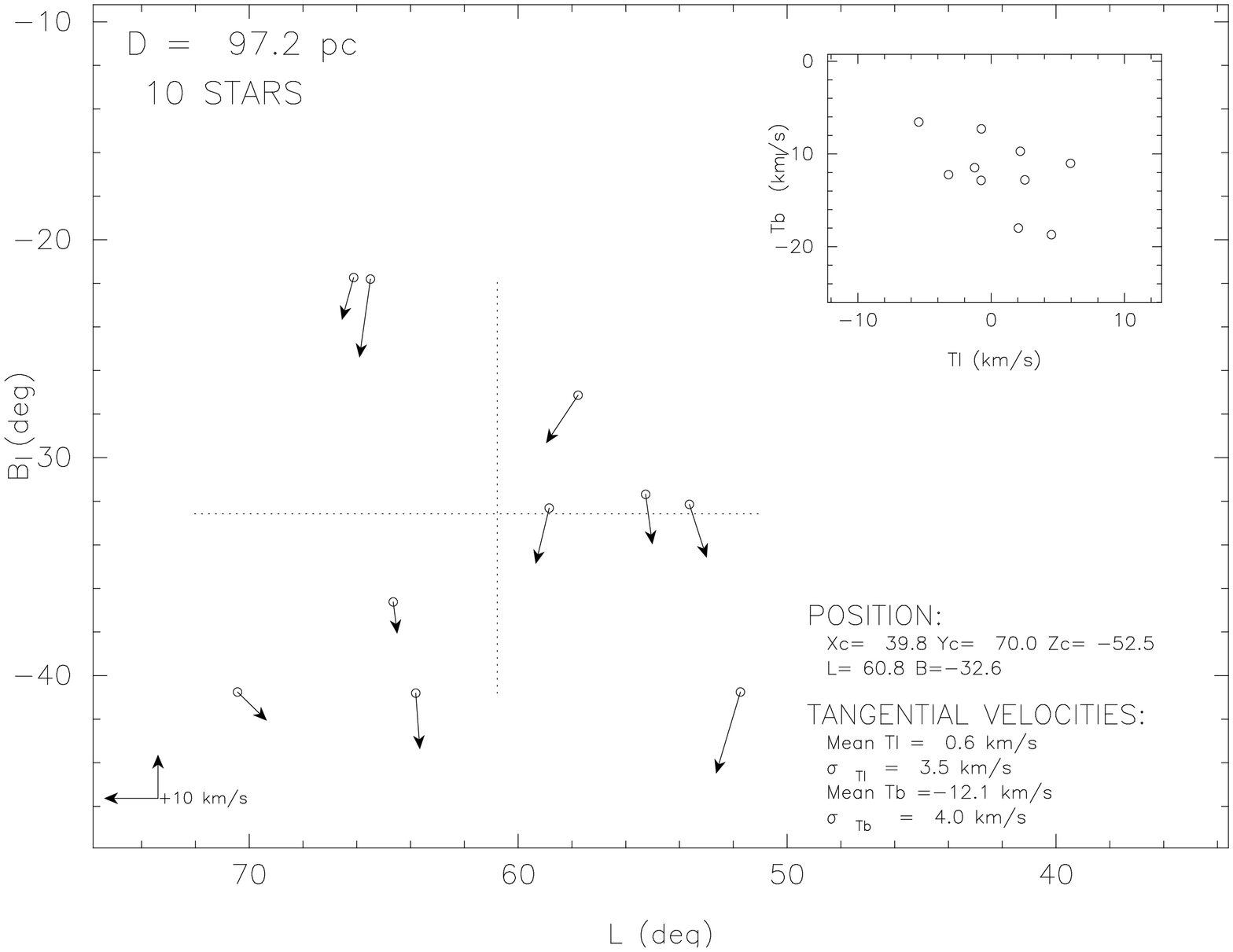,width=7.cm,height=8.cm,angle=-90.}
  \epsfig{file=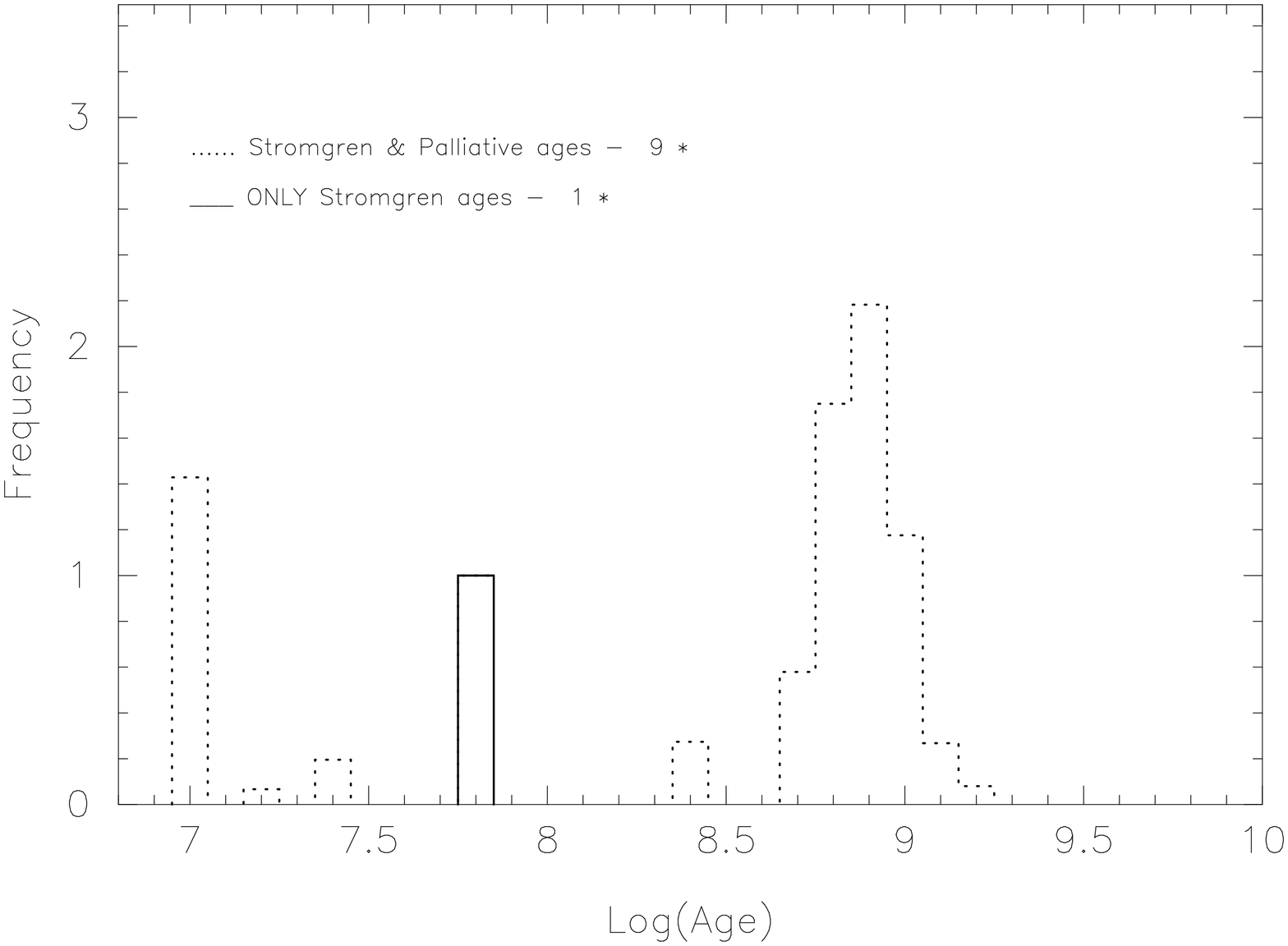,width=7.cm,height=8.cm,angle=-90.}
  \end{center}
  \vspace{-0.3cm}
  \caption{\em Pegasus 1 after selection on tangential velocities ({\bf top}) and its age distribution ({\bf bottom}).}
  \label{fig:new21}
 \end{figure}
 \begin{figure}
  \begin{center}
  \epsfig{file=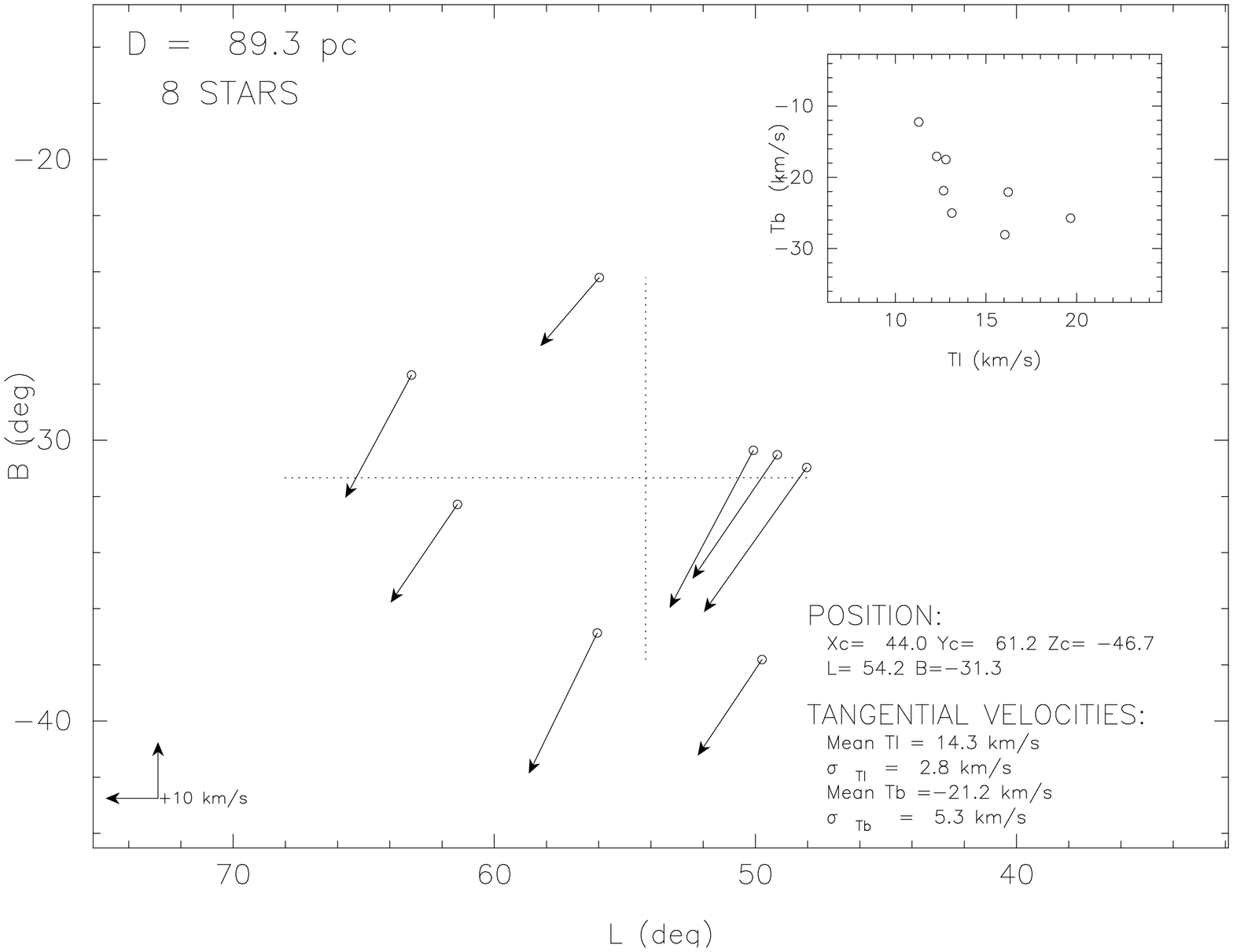,width=7.cm,height=8.cm,angle=-90.}
  \epsfig{file=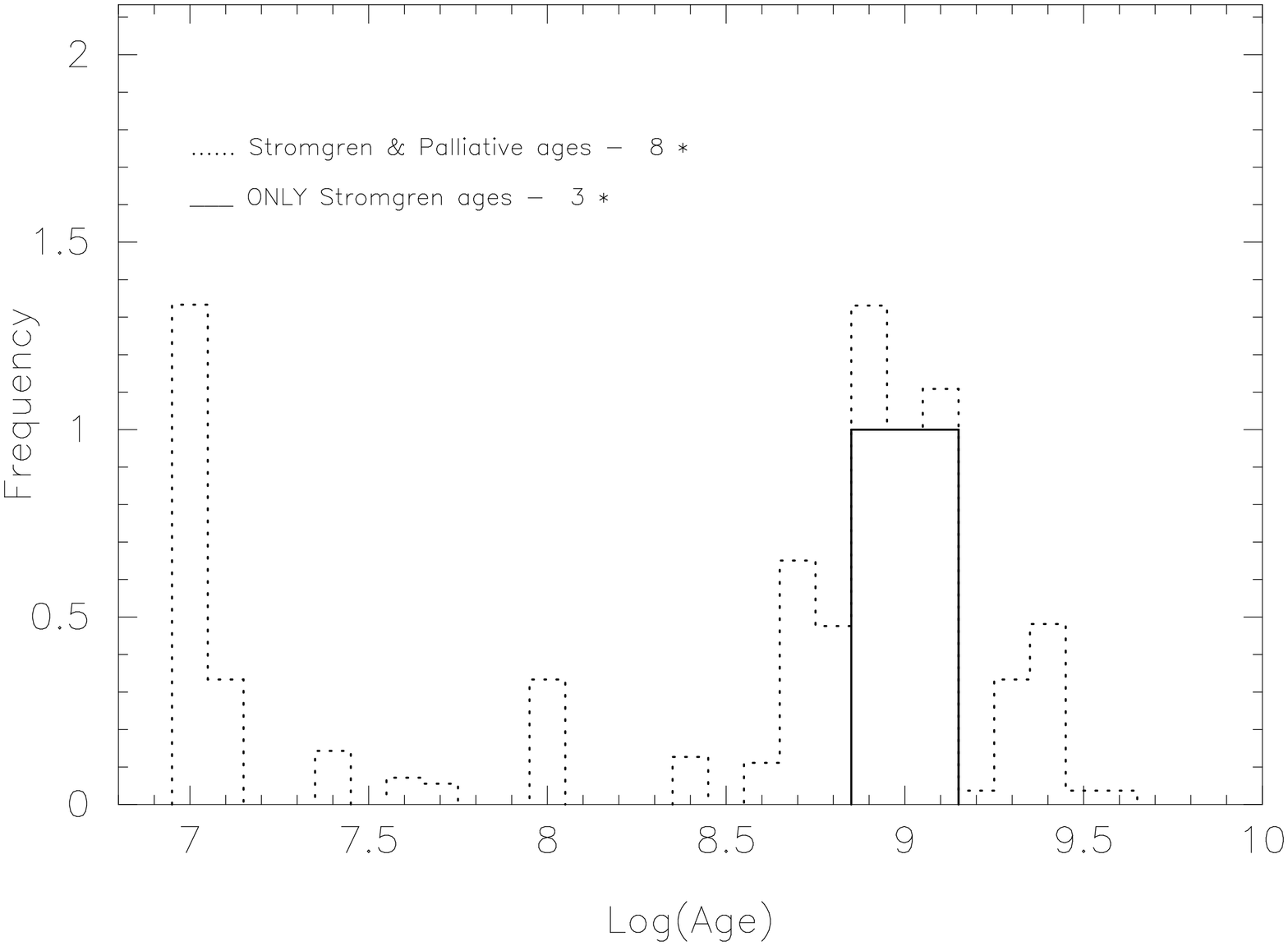,width=7.cm,height=8.cm,angle=-90.}
  \end{center}
  \vspace{-0.3cm}
  \caption{\em Pegasus 2 after selection on tangential velocities ({\bf top}) and its age distribution ({\bf bottom}).}
  \label{fig:new22}
 \end{figure}
\item {\bf Pegasus} appears to be composed of two different velocity components (Figure \ref{fig:new21}, 
\ref{fig:new22}). This feature do not permit to extract coherent velocity structures by means of the sigma clipping procedure. A hand selection was realized on the basis of the whole tangential velocity distribution. Pegasus 1 with 10 stars and Pegasus 2 with 8 stars are obtained with mean distances of respectively 97 and 89 pc 
and centred on $(l=60.7^{o},-32.5^{o})$ and $(54.2^{o},-31.3^{o})$. Their age distributions 
are not very precise because of the lack of Str\"omgren ages. However, it seems that Pegasus 1 has an age peak at 
$8\cdot 10^{8}$ yr. It has no very young ages but 3 Str\"omgren ages are between $8\cdot 10^{8}$ 
and $1.2\cdot 10^{9}$ yr which would mean an age older than Pegasus 1. This complex structure gives an 
instantaneous snapshot of the interactions affecting groups, clusters or associations: merging-like process 
can take part in dissolving spatial inhomogeneities.\\
\end{enumerate}
%
%
\section{Streaming} 
\label{sec:velocity} 
The sub-sample with observed radial velocities is incomplete and contains biases since the stars are 
not observed at random. To keep the benefits of the sample completeness, a statistical convergent point 
method is developed to analyze all the stars (Section \ref{sec:pcvg}). However, this method creates spurious 
members among detected streams with wavelet analysis, Section \ref{sec:noisestim} describe the procedure 
to handle their proportion. Moreover the fraction of field stars detected as stream members is also evaluated through 
the procedure in Section \ref{sec:fieldstar}.
 \subsection{Recovering 3-D velocities from proper motions: a convergent point method}
\label{sec:pcvg}
U, V and W velocities are reconstructed from Hipparcos tangential velocities by a convergent point method for stars
 which belong to streams. All pairs of stars are considered; each pair gives a possible convergent point (assuming
that both stars move exactly parallel) and the radial velocity is inferred for each component. Then fully
reconstructed velocities of the two stars, $V_{1}$ and $V_{2}$, are considered only if $\mid V_{1}-V_{2} \mid$
does not exceed a fixed selection criterion. This pre-selection of reconstructed velocities eliminates most of
false reconstructions. In the following, the criterion is fixed to 0.5 $km\cdot s^{-1}$.
\begin{figure}[!h]
  \begin{center}
   \leavevmode
    \epsfig{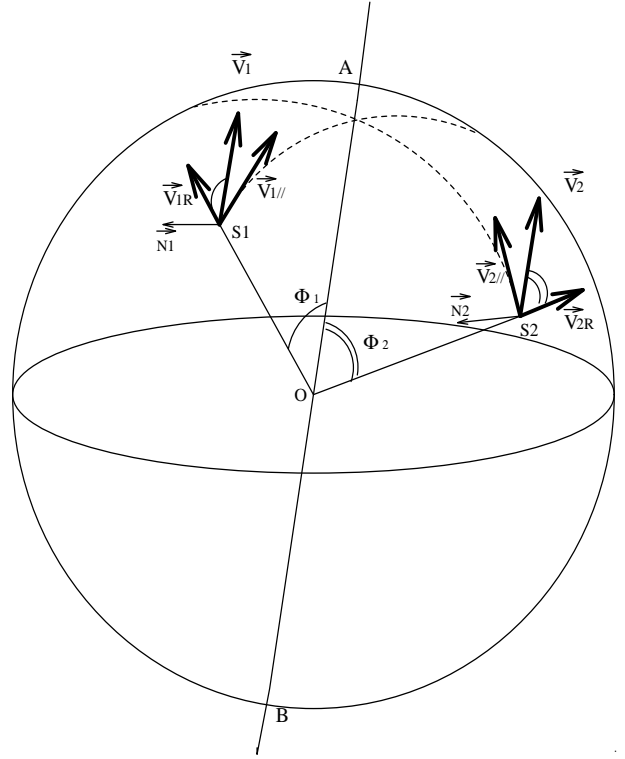}
  \caption{\em Implementation of the convergent point method. $\vec{V_{1\parallel}}$, $\vec{V_{2\parallel}}$ are the tangential velocities measured by Hipparcos for two stars. $\vec{V_{1R}}$, $\vec{V_{2R}}$, are deduced radial velocities assuming that stars $S_{1}$ and $S_{2}$ belong to the same stream.}
\label{fig:pcvg}
  \end{center}
\end{figure}
The convergent point algorithm (see Figure \ref{fig:pcvg}) follows the steps:
\begin{enumerate}
\item keep a pair of stars $(S_{1},S{2})$
\item define vectors $\vec{N_{1}}$ and $\vec{N_{2}}$ which are perpendicular respectively to the plane 
($\vec{OS_{1}}$,$\vec{V_{1\parallel}}$) and ($\vec{OS_{2}}$,$\vec{V_{2\parallel}}$)
\item obtain $\vec{N_{1}} \times \vec{N_{2}} = \vec{OA}$ which is the direction of the hypothetical convergent 
point A
\item test the coherence of this direction with both tangential velocities: if sgn($\vec{OA} \cdot \vec{V_{1\parallel}}$)
$\neq$ sgn($\vec{OA} \cdot \vec{V_{2\parallel}}$), it is not a convergent point. Go to 1
\item calculate angles between star directions and the convergent point:
$\Phi_{1}$=($\widehat{\vec{OS_{1}},\vec{OA}}$) and $\Phi_{2}$=($\widehat{\vec{OS_{2}},\vec{OA}}$)
\item infer moduli and signs of radial velocities:
\begin{eqnarray}
V_{R}=\frac{V_{\parallel}}{\tan\Phi}&if&\Phi < \frac{\pi}{2} \nonumber\\
or & &\nonumber \\
V_{R}=\frac{V_{\parallel}}{\tan(\Phi-\pi)}&if&\Phi > \frac{\pi}{2} \ and \ \Phi \neq \pi
\end{eqnarray}
\item calculate space velocities $V_{1}$ and $V_{2}$ which vectors are strictly parallels by construction
\begin{equation}
V = \sqrt{V_{\parallel}^{2}+V_{R}^{2}}
\end{equation}
\item test agreement between V$_{1}$ and V$_{2}$ within tolerance $\eta$\\
$\mid V_{1} - V_{2} \mid \leq \eta$  with fixed $\eta$ = 0.5 $km\cdot s^{-1}$.\\
\end{enumerate}
Following this process, a large number of may-be velocities are calculated. Several definitions of each star velocity 
are obtained. All are distributed along a line in the 3D velocity space, part of them being spurious. Real streams 
produce over-density clumps formed by line intersections. The wavelet analysis detects such clumps in the 
(U, V, W)  distributions. The detection sensitivity is tested numerically by simulating a variety of stream amplitudes and 
velocity dispersions over a velocity ellipsoid background. Simulations show that our wavelet analysis implementation is
able to discriminate streams formed by at least 16 stars non-spatially localized and moving together with
a typical velocity dispersion of 2 $km\cdot s^{-1}$ in a velocity background matching the sample's one.
The scale at which the stream velocity is detected is a measure of the stream velocity dispersion.
A more accurate  knowledge of this parameter is obtained after the full identification of the members
(see Section \ref{sec:phenom}).\\
Velocities of open cluster stars are poorly reconstructed by this method because their members are spatially close. 
For such stars, even a small internal velocity dispersion results a poor determination of the convergent point.  
For this reason, we have removed stars belonging to the 6 main identified space concentrations (Hyades OCl, 
Coma Berenices OCl, Ursa Major OCl and Bootes 1, Pegasus 1, Pegasus 2 groups) found in the previous spatial analysis. 
Eventually, the reconstruction of the velocity field is performed with 2910 stars.\\
Reconstructed (U,V,W) distributions are given in an orthonormal frame centred in the Sun velocity in the range 
[-50,50] $km\cdot s^{-1}$ on each component. The wavelet analysis is performed on five scales: 3.2, 5.5, 8.6, 14.9 
and 27.3 $km\cdot s^{-1}$. In the following, the analysis focuses
on the first three scales revealing the stream-like structures because larger ones reach the typical size of the velocity 
ellipsoid. Once the segmentation procedure is achieved, stars belonging to velocity clumps are identify. 
The set of velocity definitions of a star may cross two velocity clumps. In this case, the star is associated with the 
clump in which it appears most frequently.
\subsection{Cleaning stream statistics}
\label{sec:cleanstat}
\subsubsection{Estimating the fraction of spurious members in velocity clumps}
\label{sec:noisestim}
Despite the pre-selection of reconstructed velocities ($\mid$V$_{1}$-V$_{2}$$\mid \leq \eta$) some spurious
velocities are still present in the field. For this reason, real velocity clumps do include some proportion of spurious 
members generated by the convergent point method. Estimating the proportion of spurious members in each 
velocity clump is essential and is done by comparing the mean of reconstructed radial velocities of each star with 
its observed radial velocity whenever this data is available (1362 among 2910 stars). For each star in a stream, 
the following procedure is adopted:
\begin{enumerate}
\item Only radial velocities reconstructed with other suspected members of the same stream are considered.
\item Mean $\overline{V}_{R rec.}$ and dispersion $\sigma_{V_{R rec.}}$ of the reconstructed radial velocity
distribution are calculated (Figure \ref{fig:distvr}).
\begin{figure}
 \begin{center}
    \epsfig{file=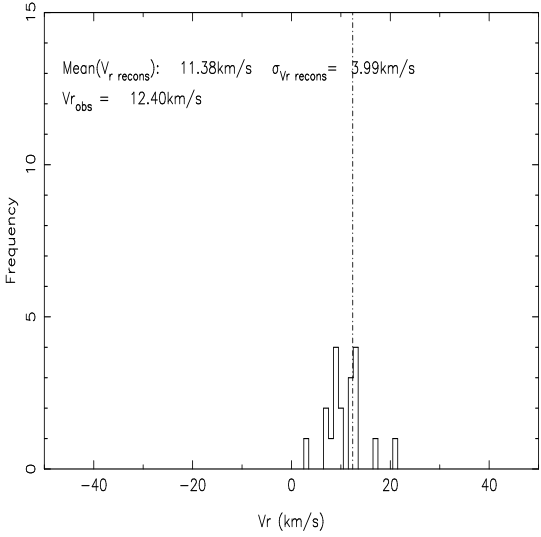,height=8.cm,width=8.cm,angle=-90.}
  \caption{\em Example of V$_{R_{rec.}}$ distribution for one star suspected to be member of a stream and its observed radial velocity (dot dashed line)}
  \label{fig:distvr}
  \end{center}
\end{figure}
\item The star is confirmed to belong to the stream when the normalized residual
\begin{equation}
R_{N}=\frac{\mid V_{R obs.}-\overline{V}_{R rec.}\mid}{\sqrt{\sigma_{V_{R rec.}}^{2}+\sigma_{V_{R obs.}}^{2}}} \nonumber
\end{equation}
doesn't exceed a value $\kappa$. Residuals take into account errors $\sigma_{V_{R obs.}}$ on the observed radial velocities 
given in the Hipparcos Input Catalogue. These errors are classified into 4 main values: 0.5, 1.2, 2.5, 5 $
km\cdot s^{-1}$.\\
The threshold $\kappa$ is fixed empirically on the basis of the normalized residual histogram of all suspected stream 
members with observed radial velocities (Figure \ref{fig:distkappa}) at scale 2.  It is set to $\mid\kappa\mid$=
 3 in order to keep the bulk of the central peak. The results of this paper are robust to any reasonable change of
 $\mid\kappa\mid$ between 2.5 and 4.\\
\begin{figure}
 \begin{center}
    \epsfig{file=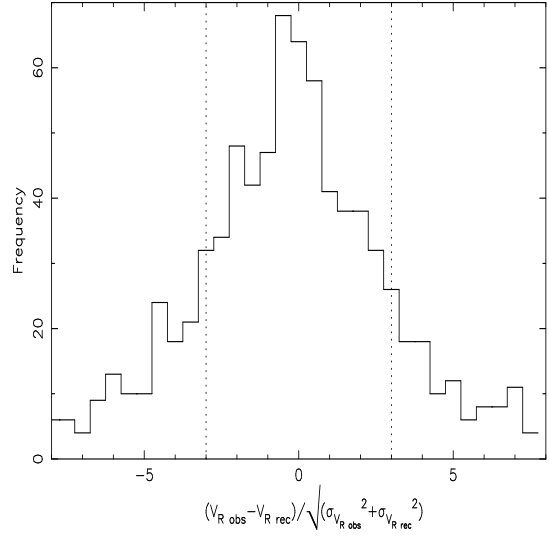,height=8.cm,width=8.cm,angle=-90.}
  \caption{\em Distribution of weighted radial velocity deviates (normalized residual) for all suspected stream members with observed radial velocity at scale 2. Dashed line delimits the area of high probability of false detection.}
  \label{fig:distkappa}
  \end{center}
\end{figure}
\end{enumerate}
Noise estimations in each velocity clump for the three first scales (provided that they have at least 3 stars
with observed $V_{R}$), following the procedure quoted above, are shown in Figure \ref{fig:noise}. There are
two regimes in these noise estimations: streams with more than 50 initial suspected members ($N_{init}$) which
have a contamination by spurious members around 30$\%$; streams with less than 50 initial suspected
members which may have a contamination up to 85$\%$. In this extreme case, should we say that these
streams are false detection? Not necessarily, because our denoising method is too drastic towards small
streams. Indeed, in those, each star has very few reconstructions of its radial velocity with the other
members of the same stream. The dispersion of the reconstructed $V_{R}$ distribution is necessarily small,
implying an important normalized residual. Then, the star is often rejected.
If we refer to our previous simulations there is a high probability that under 16 initial members, streams are
false detection.\\
\begin{figure}
 \begin{center}
    \epsfig{file=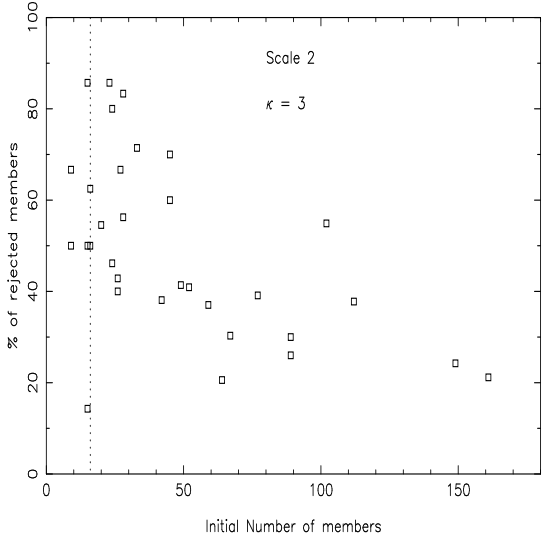,height=8.cm,width=8.cm,angle=-90.}
  \caption{\em Proportion of rejected members for all the detected velocity clumps at scale 2 by the selection on observed radial velocity function of the initial number of members. Dashed line indicates the minimum number of stream members requires to detect streams in simulations.}
  \label{fig:noise}
  \end{center}
\end{figure}
 At the end of this selection we obtain the number of confirmed members among stars with observed
$V_{R}$ in each stream for the three first scales (column $N_{sel}$ in Tables \ref{tab:table3}, \ref{tab:table4} and 
\ref{tab:table5}) and their sum (line $Total 2$ in column $N_{sel}$).
\subsubsection{Estimating the proportion of field stars in velocity clumps}
\label{sec:fieldstar}
The fraction of a smooth distribution filling the velocity ellipsoid of our complete sample,
expected inside the velocity volume spanned by the 6 {\em superclusters} described bellow,
range between 2$\%$ and 4$\%$ depending on the position of the structure with respect to
the distribution centroid. Adding up these contributions, 19.2$\%$ of field stars should be
expected to fill the total volume occupied by {\em superclusters} with pure random coincidence.
This is about 20-30$\%$ of the stars detected as {\em supercluster} members at scale 3.
However, streams with smaller velocity dispersions (scales 1 and 2) are not significantly
affected by this background. Proportions of field stars for the largest structures found at scale 3
are given in \ref{tab:table3} at column $\%_{field}$ while it is neglected for the remaining streams.
\subsection{Stream phenomenology}
\label{sec:phenom}
Tables \ref{tab:table3}, \ref{tab:table4} and \ref{tab:table5} give mean velocities, velocity 
dispersions and numbers of stars remaining after correction of spurious members (procedure \ref{sec:noisestim}) 
and field stars (procedure \ref{sec:fieldstar}) for streams at respectively scale 3, 2 and 1. 
Each stream has $N_{V_{r}}$ observed radial velocity members. Out of the $N_{V_{r}}$ stars with 
radial velocities among suspected stream members, only $N_{sel}$ get confirmed by procedure \ref{sec:noisestim}. 
So the ratio $N_{sel}$/$N_{V_{r}}$  is an estimate of the confirmed /suspected 
ratio in each stream. Applying this ratio to $N_{init}$ (total stream member candidates) we get the 
expected number of real stream members in each stream, and the total number of stream members in 
the sample ({\em Total 3}). The percentage of stars in streams in the total sample follows ({\em Total 4}). 
The correction for the uniform background contribution is negligible at scales 1 and 2; 
it is significant at scale 3 where the fraction of stars in streams drops from 63.0$\%$ to 46.4$\%$. 
In the case of large velocity dispersion structures at scale 3 proportions of field stars is also given in 
column {\em $\%_{field}$} and the percentage of remaining stream stars is done in column {\em $N_{init}$} 
line {\em Total 5}.\\
\subsubsection{Particulars on superclusters}
\label{sec:velfield}
Streams appearing at scale 3 ($\overline{\sigma}_{stream} \sim$ 6.3 $km\cdot s^{-1}$) correspond 
to the so-called Eggen {\em superclusters}. Four already known such structures are found: the Pleiades, 
Hyades and Sirius {\em superclusters} (hereafter SCl) and the whole Centaurus association. Moreover, 
evidence is given for one additional structure not detected yet. The reason why this {\em supercluster} 
remained undetected is probably the small velocity offset with respect to the Sun's. At smaller scales 
($\overline{\sigma}_{stream} \sim$ 3.8 and 2.4 $km\cdot s^{-1}$) {\em superclusters} split into distinct 
streams of smaller velocity dispersions.\\
The analysis of the age distribution inside each stream is performed on three different data sets: 
\begin{itemize}
\item the whole sample (ages are either Str\"omgren or palliative),
\item the sample restricted to stars with Str\"omgren photometry data (without selection on radial velocity),
\item the sample restricted to stars with {\em observed } (as opposed to {\em  reconstructed}) radial velocity data 
(ages are either  Str\"omgren or palliative).
\end{itemize}
The selection on photometric ages gives a more accurate description of the stream age content while the 
last sample permits to obtain a reliable kinematic description since stream members are selected through 
the \ref{sec:noisestim} procedure. All mean velocities and velocity dispersions of the streams are calculated 
with the radial velocity data set. Combining results from these selected data sets generally brings 
unambiguous conclusions. \\
\begin{enumerate}
\item {\bf Pleiades SCl} \\
The analysis of the Pleiades SCl is realized in Paper II where it is found to be composed of two main streams 
of few $10^{7}$ and $10^{9}$ yr.\\
\item {\bf Hyades SCl} and {\bf NGC 1901 stream}\\
\begin{figure*}
  \begin{center}
    \epsfig{file=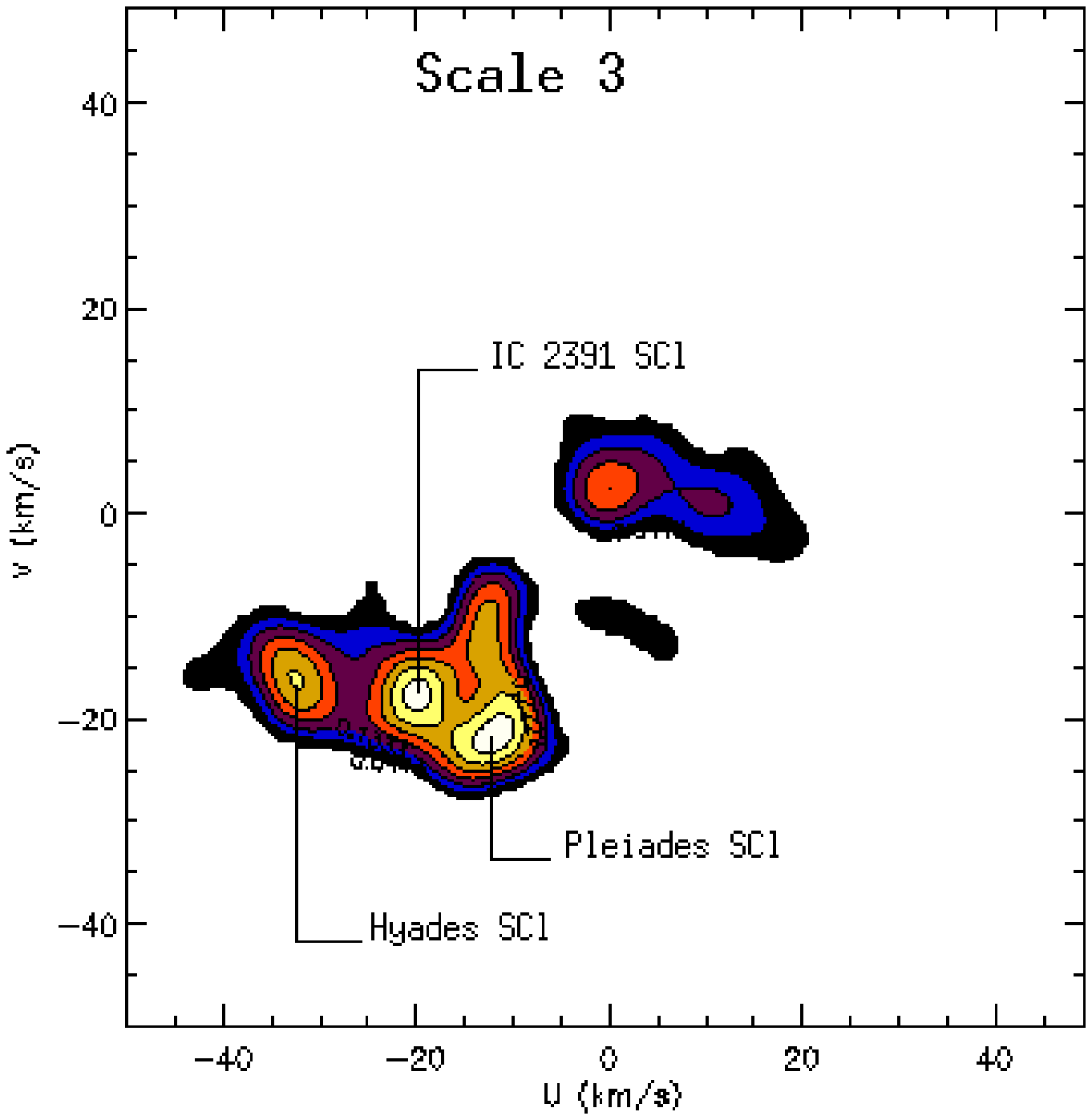,height=7cm,width=7cm,angle=0.}
  \epsfig{file=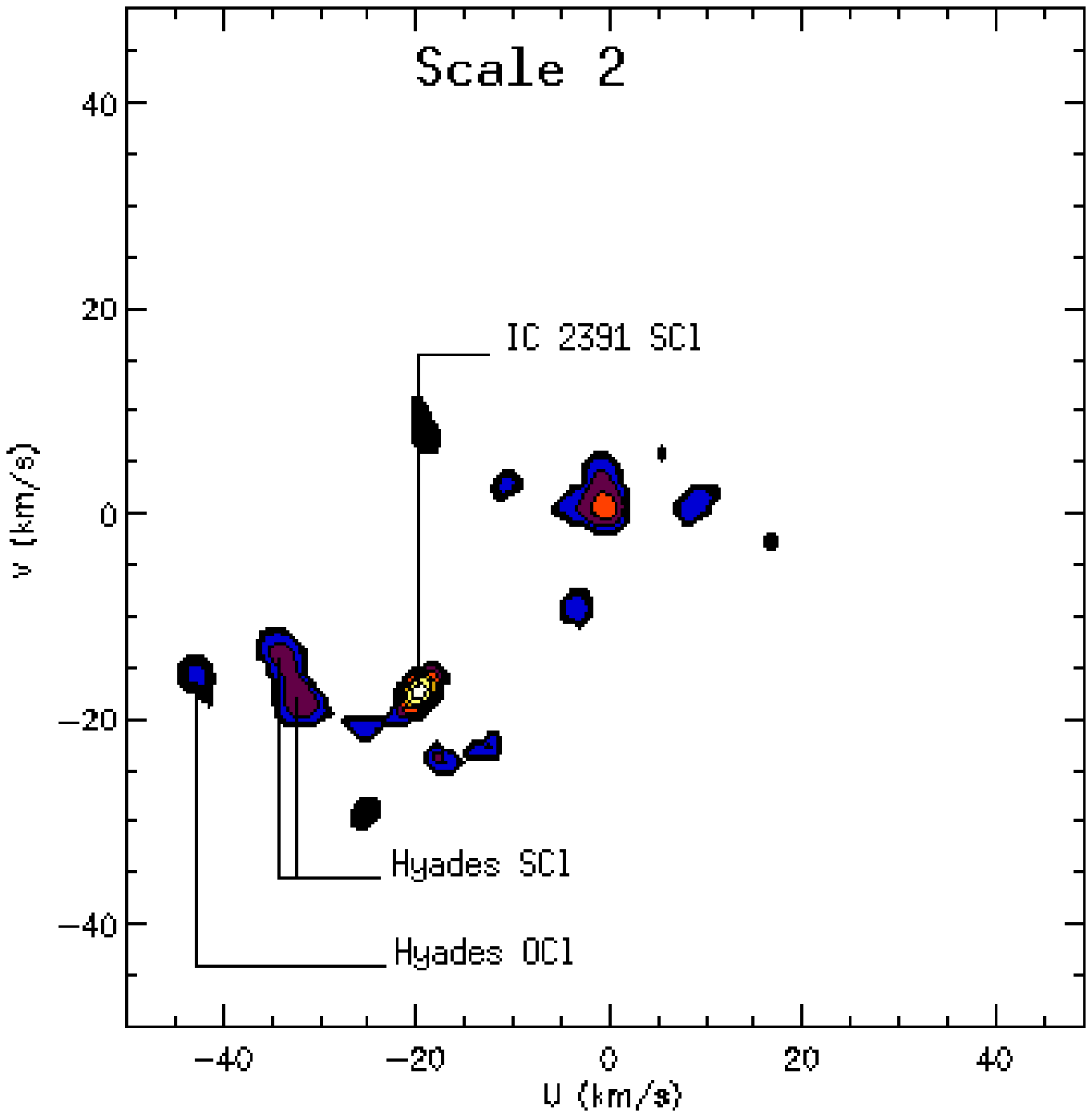,height=7cm,width=7cm,angle=0.}
   \epsfig{file=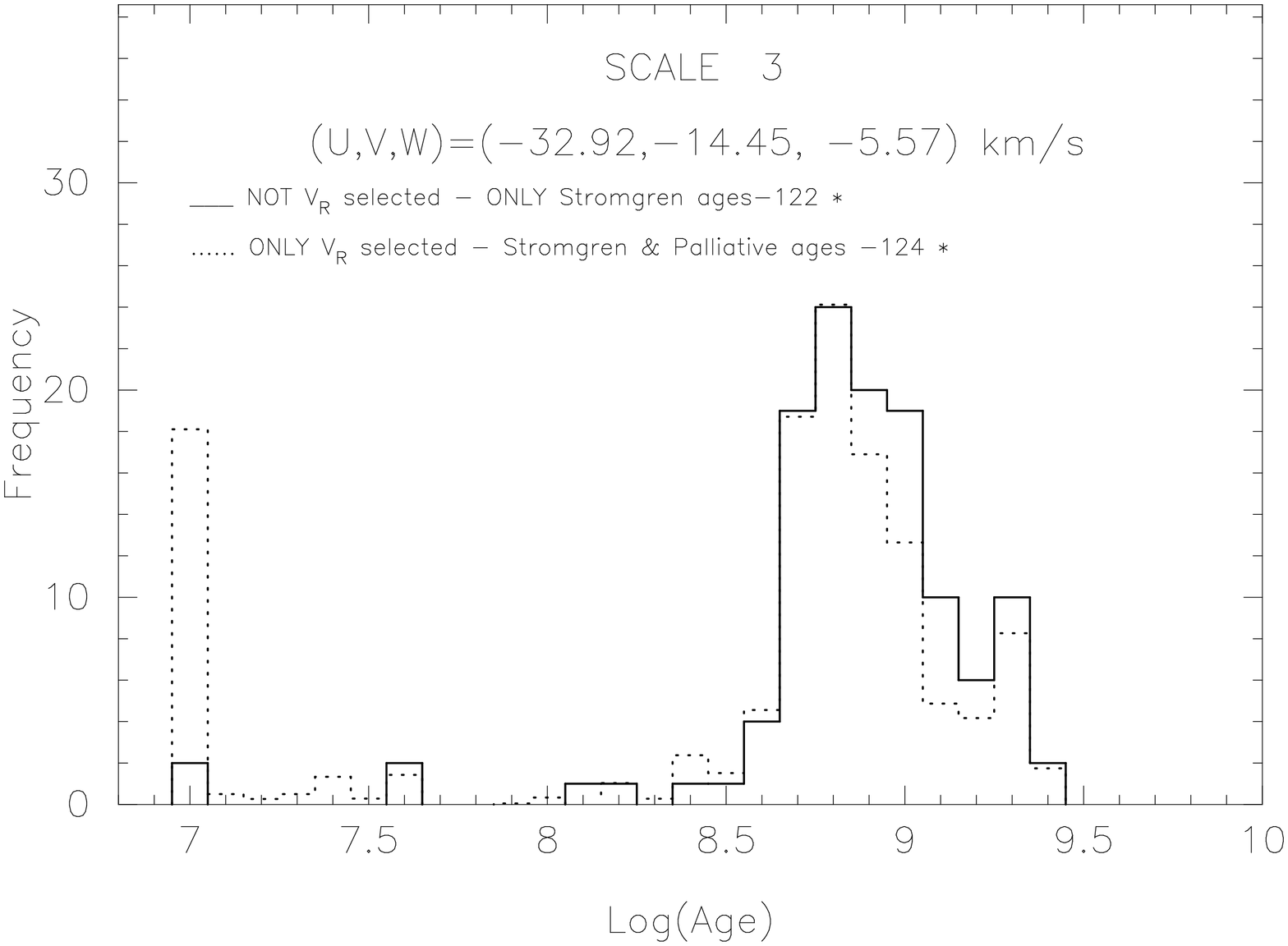,height=8.cm,width=7.cm,angle=-90.}
  \epsfig{file=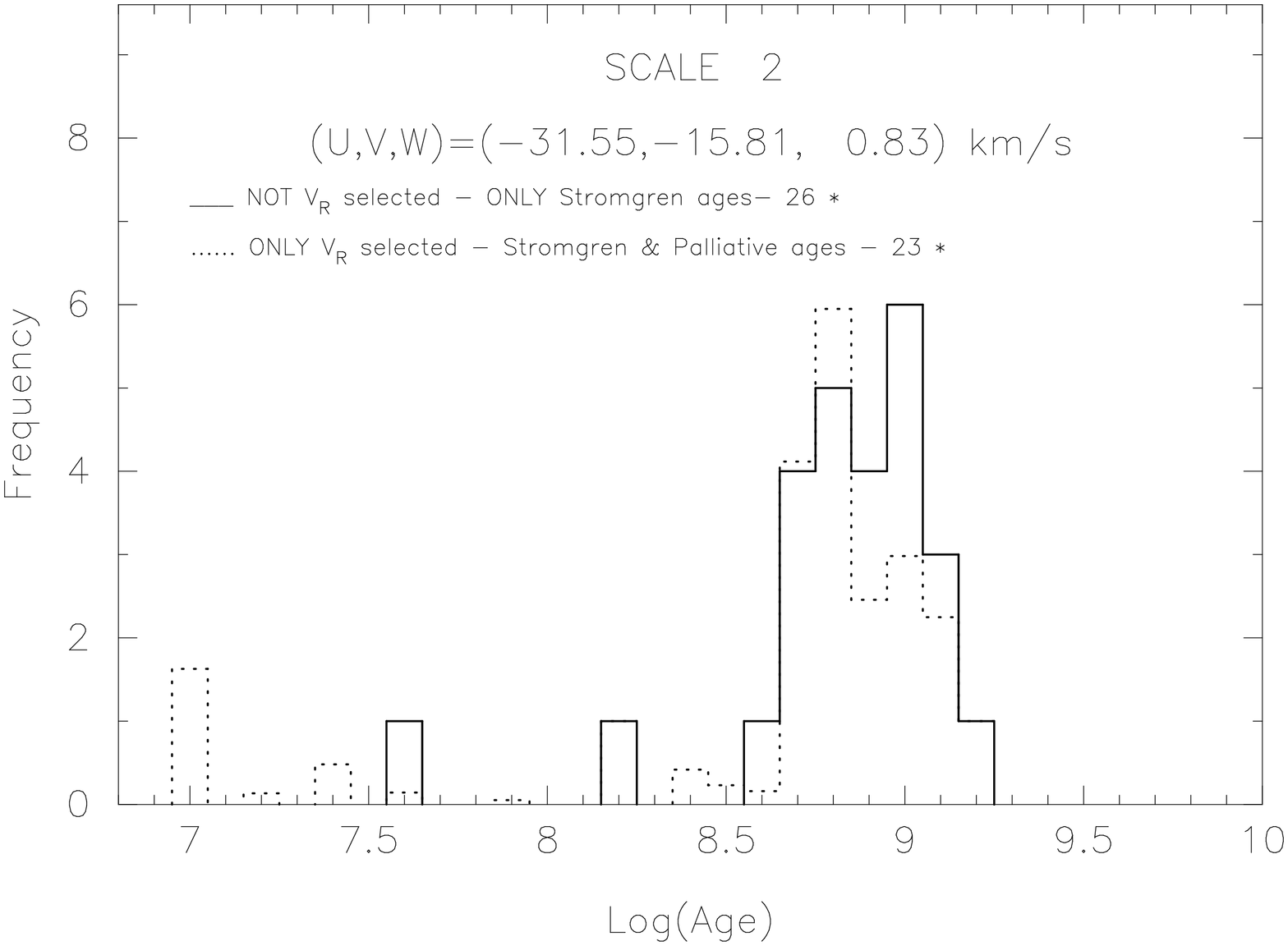,height=8.cm,width=7.cm,angle=-90.}
  \epsfig{file=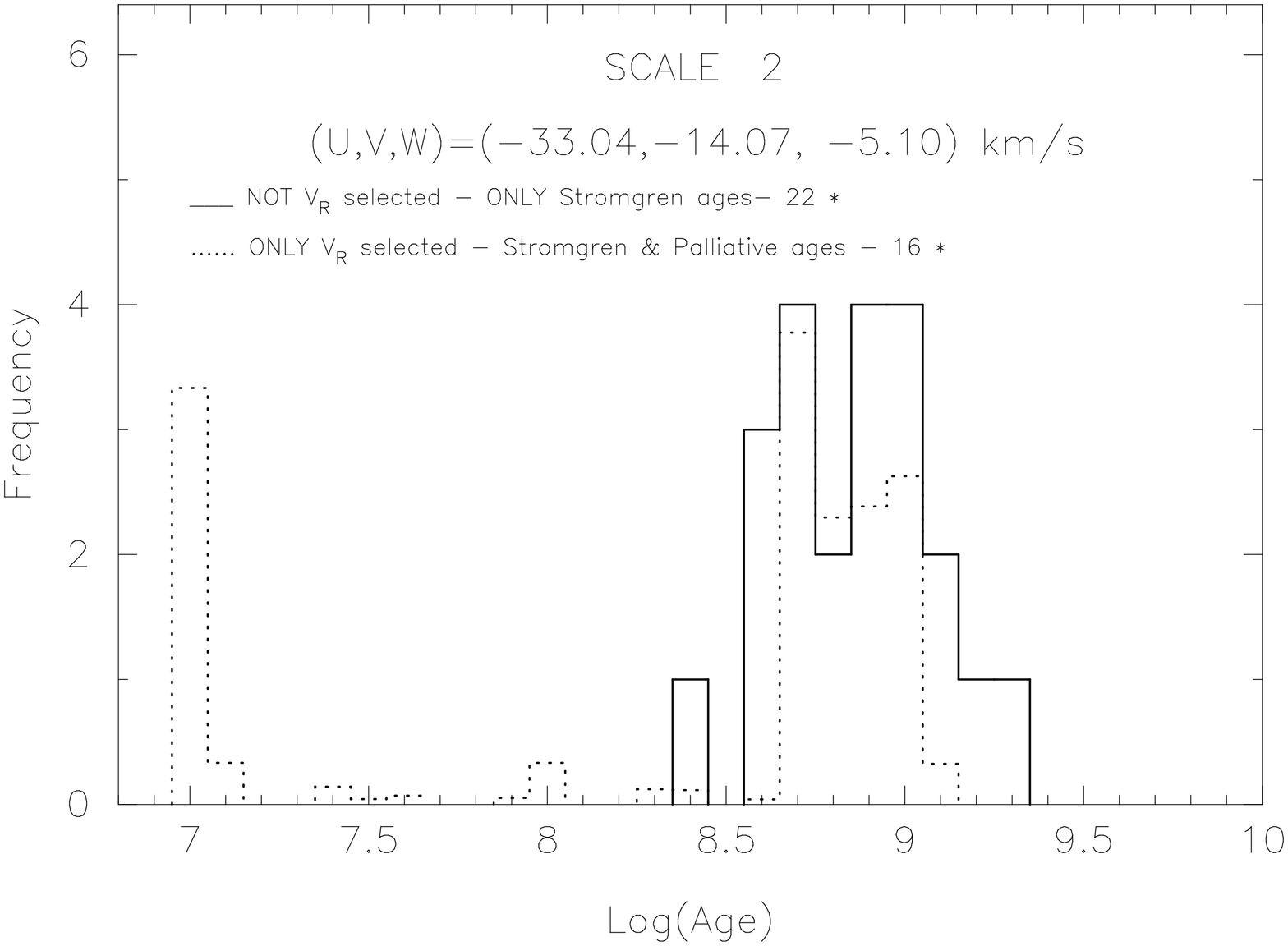,height=8.cm,width=7.cm,angle=-90.}
  \epsfig{file=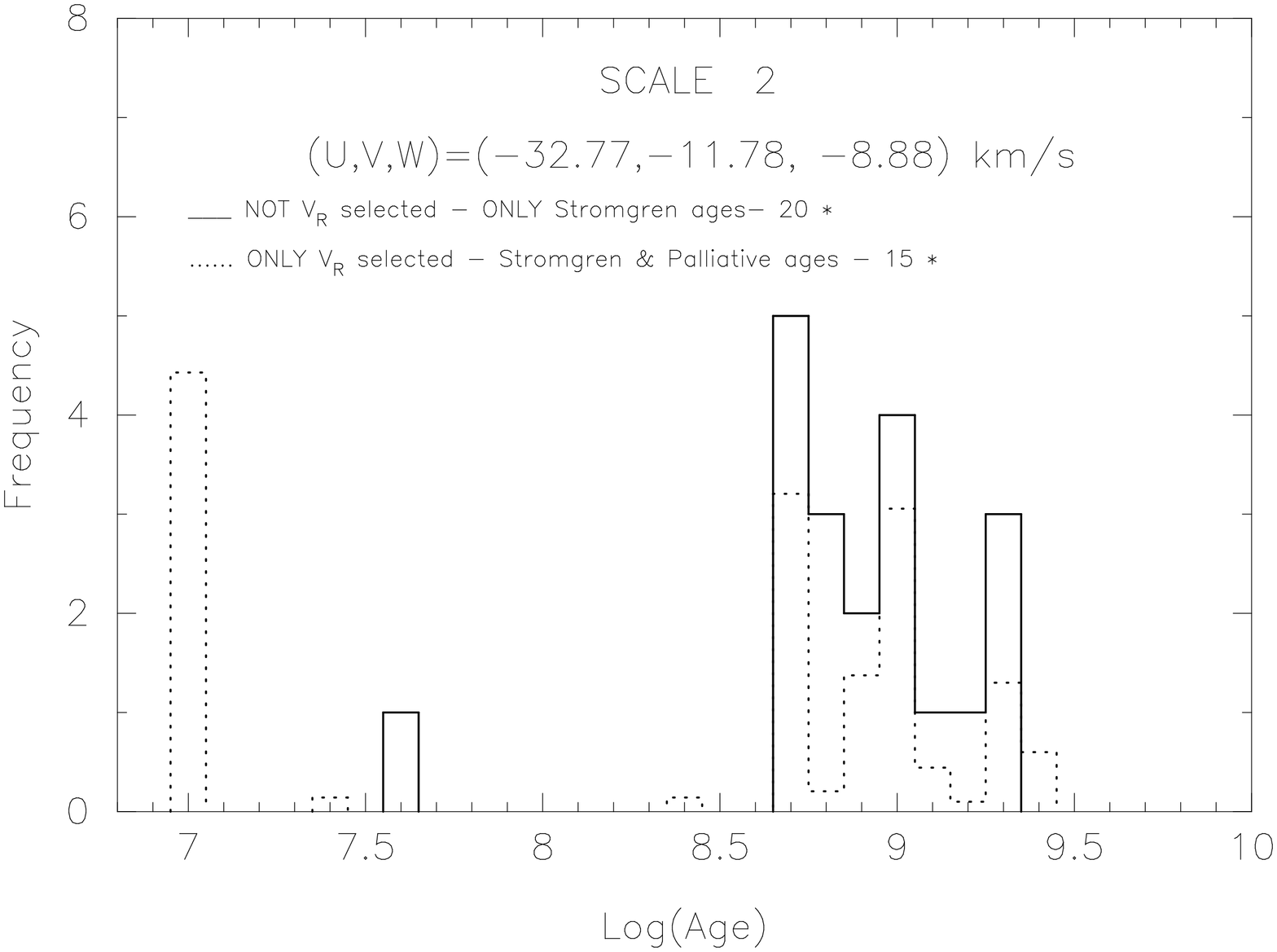,height=8.cm,width=7.cm,angle=-90.}
  \caption{\em {\bf Hyades Scl}. Thresholded wavelet coefficient isocontours at W=-2.4 $km\cdot s^{-1}$ of the velocity field at scale 3 ({\bf top left}) and scale 2 ({\bf top right}). This slice in wavelet coefficients at scale 2 reveals two of the three clumps composing the whole supercluster. Age distributions of the whole group (stream 3-10 in Table \ref{tab:table3}) at third scale ({\bf middle left}) and of the first sub-stream (stream 2-10 in Table \ref{tab:table4}) at second scale ({\bf middle right}). Age distributions of the two last sub-streams (stream 2-18 and 2-25 in Table \ref{tab:table4}) discovered at scale 2 ({\bf bottom}).}
  \label{fig:hyades}
  \end{center}
\end{figure*}
\begin{figure}
 \hspace*{-1.8cm}
  \epsfig{file=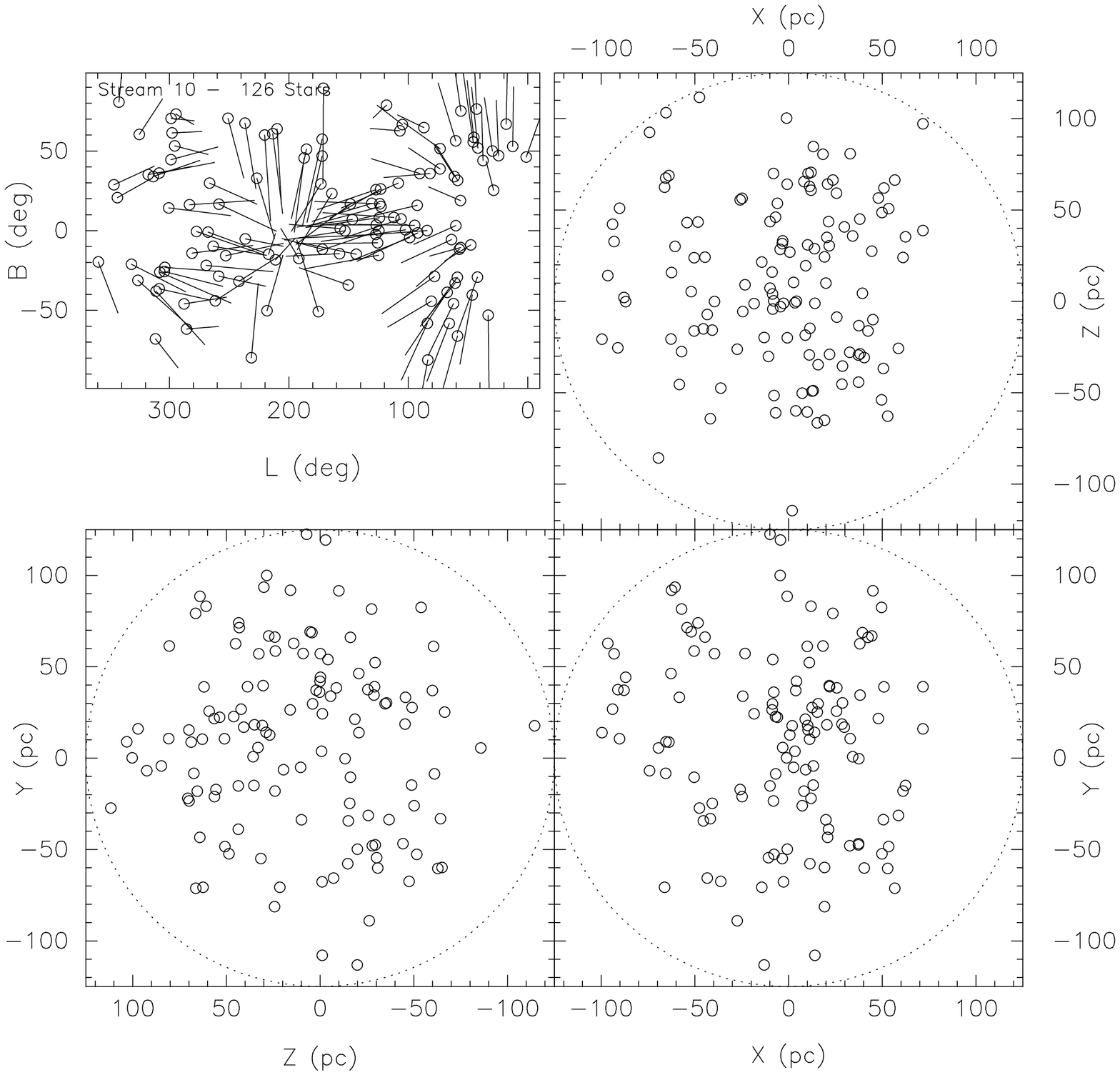,height=12.cm,width=9.4cm,angle=-90.}
\caption{\em {\bf Space distribution of Hyades SCl} from the $V_{R}$ selected sub-sample at scale 3 (stream 3-10 
in Table \ref{tab:table3}).}
  \label{fig:spat_hyades_s3}
 \end{figure}
The velocity clump (stream 3-10 in Table \ref{tab:table3}) identified at scale 3 as the Hyades SCl 
(see Figures \ref{fig:hyades} for velocity and age distributions and Figures \ref{fig:spat_hyades_s3} for space 
distributions) is located at (U,V,W)=(-32.9, -14.5, -5.6) $km\cdot s^{-1}$ with velocity dispersions 
($\sigma_{U}$,$\sigma_{V}$,$\sigma_{W}$)=(6.6, 6.8, 6.5)  $km\cdot s^{-1}$. The mean velocity deviates 
slightly from the definition given by Eggen (1992b) (cf. Table \ref{tab:cross}). At this resolution the 
bulk of star ages is between $4\cdot 10^{8}$ yr and $2\cdot 10^{9}$ yr with two peaks at 
$6\cdot 10^{8}$ and $1.6\cdot 10^{9}$ yr in Str\"omgren age distribution plus a $10^{7}$ yr peak 
in the palliative age distribution. Eggen (1992b) pointed out that the {\em supercluster} contains at 
least three age groups around 3 to 4, 6 and 8$\cdot 10^{8}$ yr. \\
The velocity pattern splits into 3 groups at scale 2, namely 2-10, 2-18 and 2-25 (Table \ref{tab:table4}). 
Each stream presents a characteristic age distribution, although the velocity separation (centers deviates 
from each other by several $km\cdot s^{-1}$ in W) does not produce a neat age separation.  
Three different main components of  $10^{7}$, $5-6\cdot 10^{8}$ and $10^{9}$ yr are mixed in the 3 clumps. 
The first clump peaks at $10^{9}$ yr in Str\"omgren ages but contains a $6\cdot 10^{8}$ yr old 
component also revealed by palliative ages. The second clump peaks at $5\cdot 10^{8}$ yr and  $10^{9}$ yr 
in Str\"omgren ages. Palliative ages produce a $10^{7}$ yr peak which is probably a statistical ghost 
of the $5\cdot 10^{8}$ year old component (see explanation of ghost at the end of Section \ref{sec:corage}). 
The third clump is dominated by a $5\cdot 10^{8}$ yr old component with two older groups of $10^{9}$ 
and $2\cdot 10^{9}$ yr. One more time the very young peak in palliative ages is also probably due to 
the $5\cdot 10^{8}$ year old component.\\
The presence of older {\em supercluster} members around $1.6\cdot 10^{9}$ yr as stipulated by Eggen 
and stressed by Chen et al (1997) is detected in the third velocity clump. Scale 1 does not reveal more 
information so that we cannot obtain one age for each stream.\\
So, the Hyades SCl contains probably three groups of  $5-6\cdot 10^{8}$ yr, $10^{9}$ and 
$1.6-2\cdot 10^{9}$ yr which are in an advanced stage of dispersion in the same velocity volume. 
Only part of these 3 streams can be linked to the evaporation of 
known open clusters. The Hyades OCl recent evaporation is clearly found separately in stream 2-15. 
The Praesepe OCl mean velocity (Table \ref{tab:cross}) accurately match none of the 3 stream velocities 
but could explain the stream 2-18 despite a difference of $\sim$ 9 km$\cdot$s$^{-1}$ in the V component. 
The NGC 1901 {\em supercluster} described in \cite{Eg96} and assumed to be a Hyades SCl component is found 
separately at scale 2 (stream 2-29 in Table \ref{tab:table4}) and exhibits a  single mode in age 
distribution at $8\cdot 10^{8}$ yr. Its velocity is more dissociated from the {\em supercluster} mean 
velocity than the 3 other streams which explain a best member extraction.\\
\item {\bf Sirius SCl} \\
  \begin{figure*}
  \begin{center}
    \epsfig{file=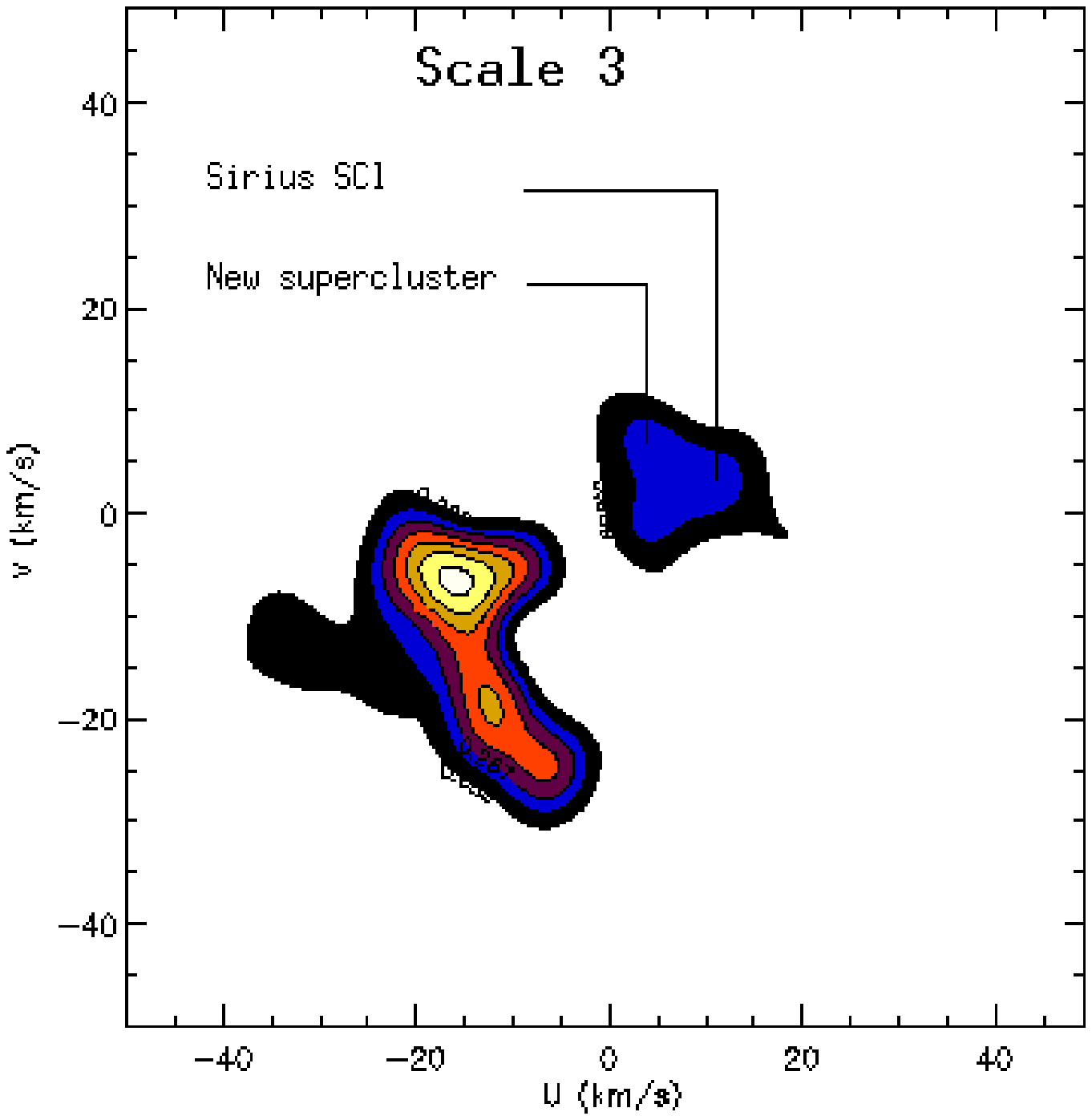,height=7cm,width=7cm,angle=0.}
  \epsfig{file=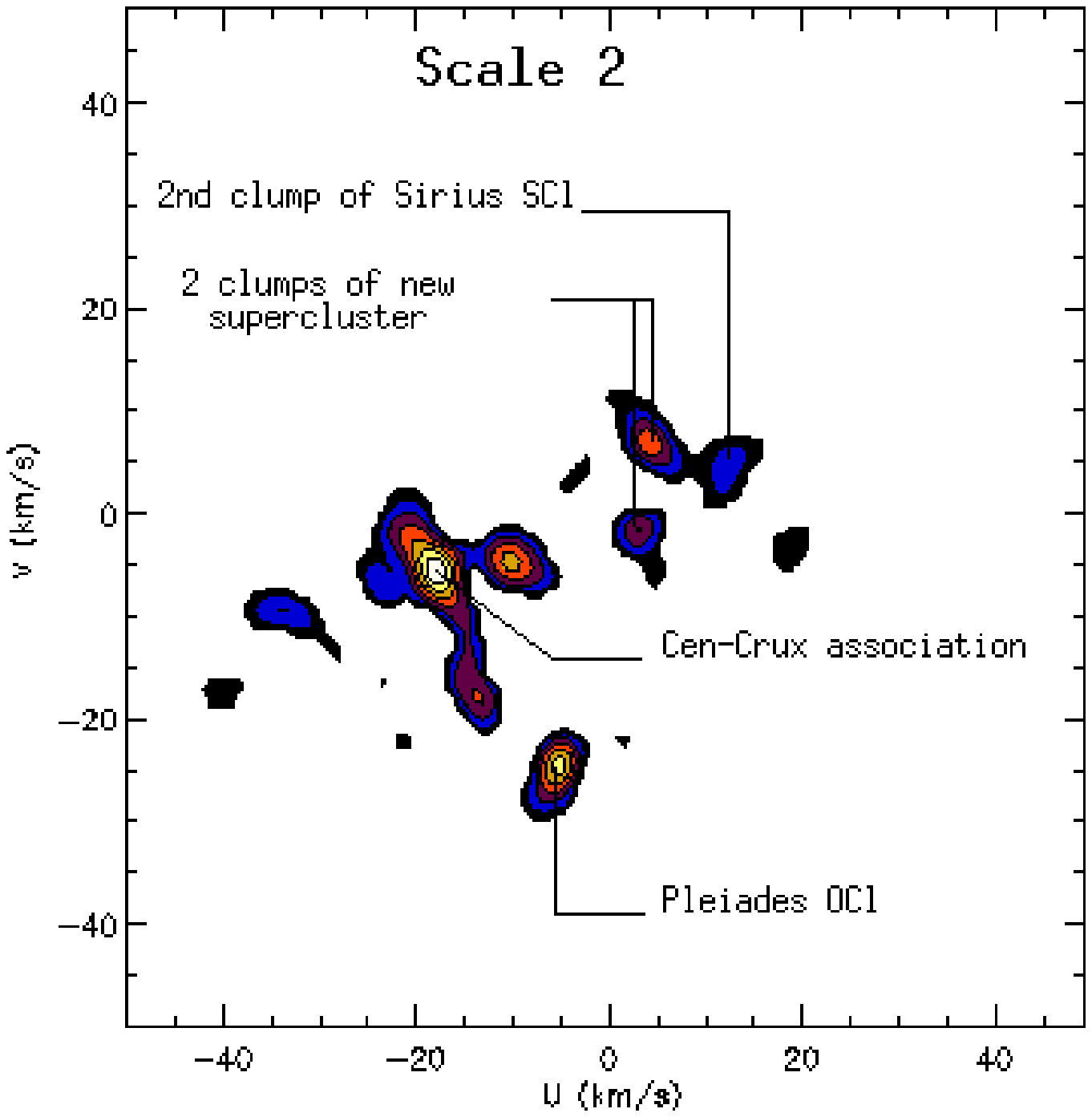,height=7cm,width=7cm,angle=0.}
   \epsfig{file=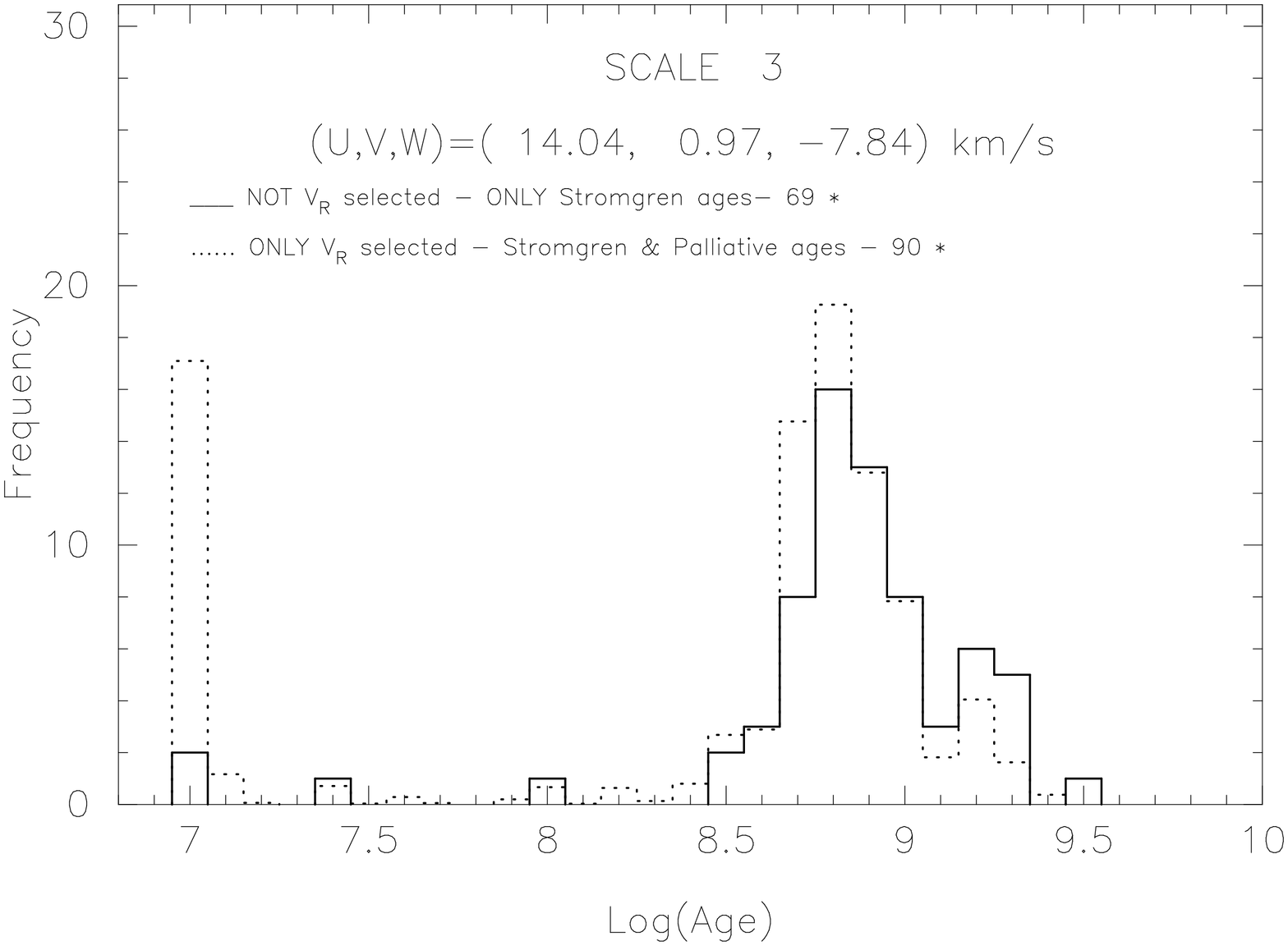,height=8.cm,width=7.cm,angle=-90.}
  \epsfig{file=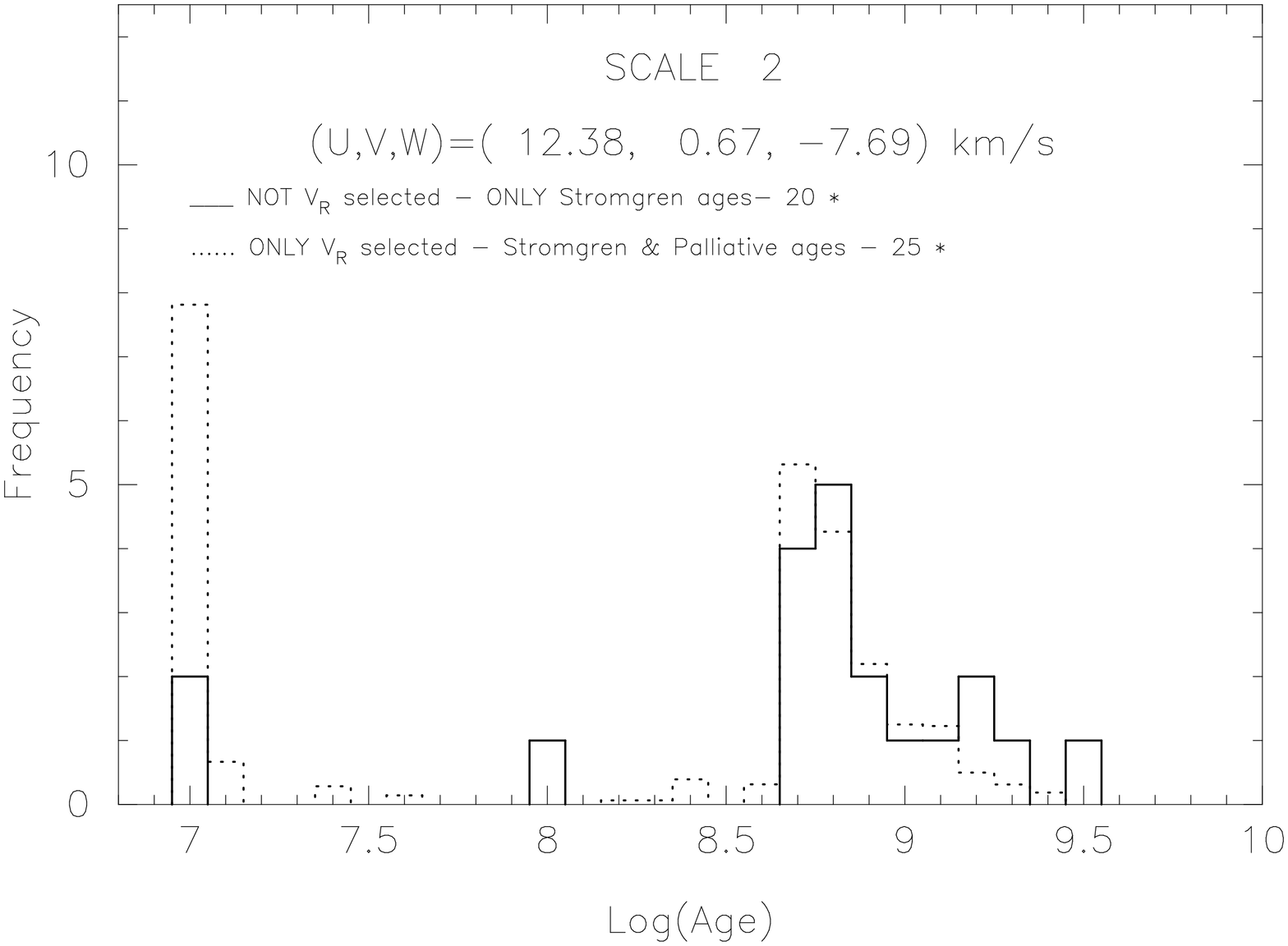,height=8.cm,width=7.cm,angle=-90.}
  \epsfig{file=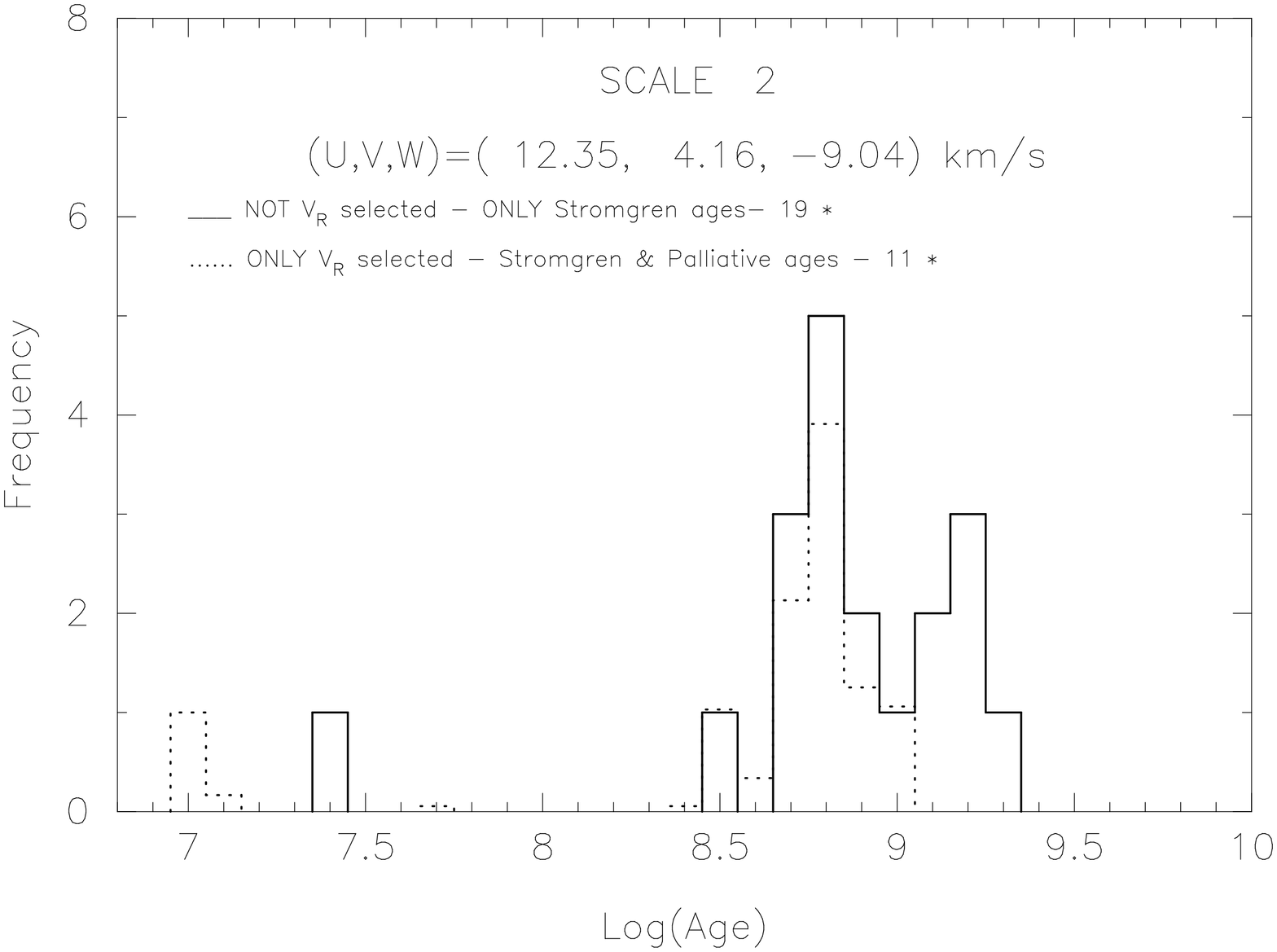,height=8.cm,width=7.cm,angle=-90.}
  \epsfig{file=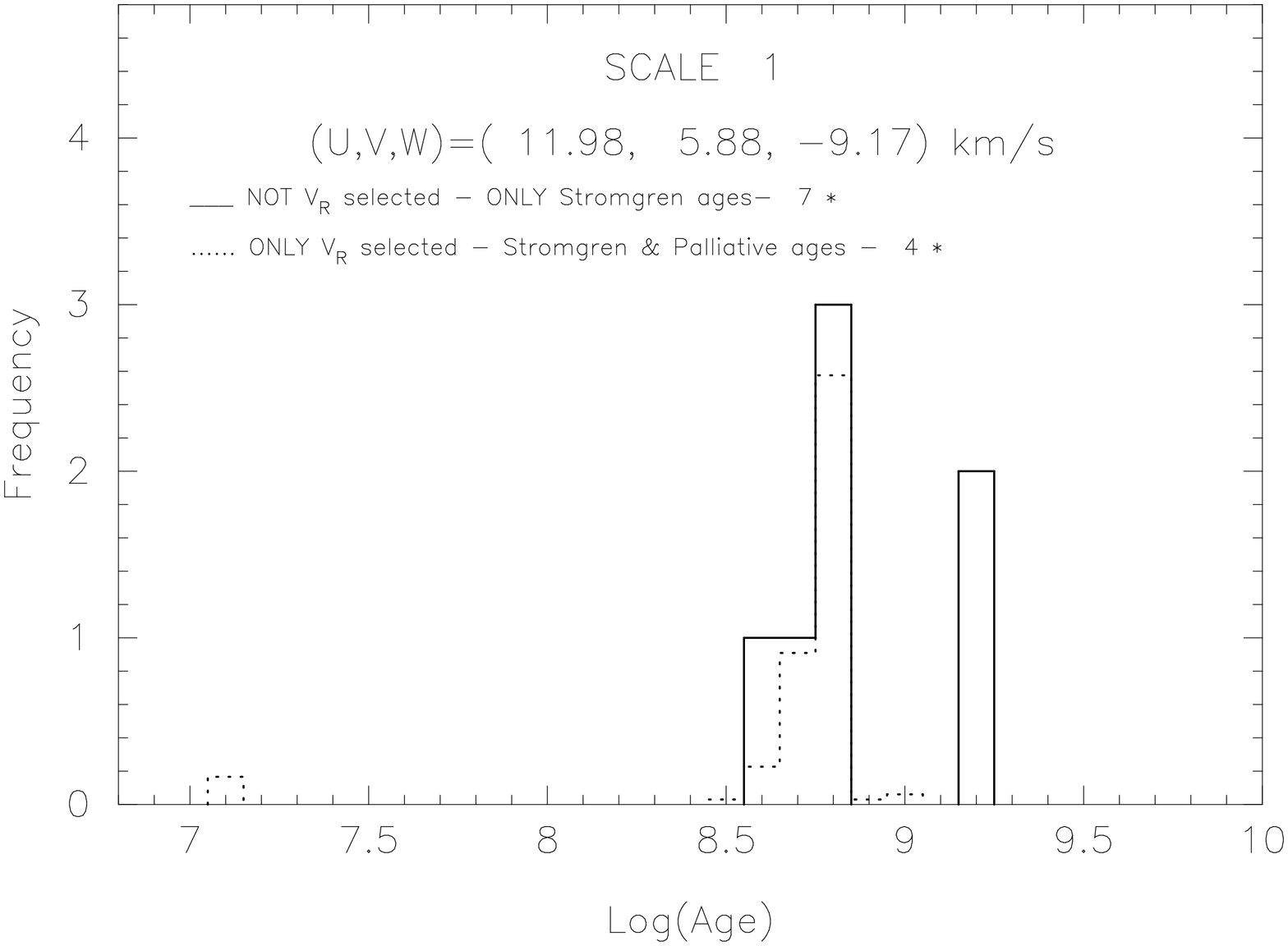,height=8.cm,width=7.cm,angle=-90.}
  \caption{\em {\bf Sirius Scl}. Thresholded wavelet coefficient isocontours at W=-10.1$km\cdot s^{-1}$ of the velocity field at scale 3 ({\bf top left}) and scale 2 ({\bf top right}). At scale 2, Sirius SCl is composed of 2 main streams (stream 2-37 and 2-41 in Table \ref{tab:table4}). The  2$^{nd}$ is shown on this W slice of wavelet coefficients.  Age distributions of the whole Sirius SCl at third scale ({\bf middle left}) and stream 2-37 at second scale ({\bf middle right}). Age distributions of stream 2-41 at scale 2 ({\bf bottom left}) and stream 1-56 (in Table \ref{tab:table5}) at scale 1 ({\bf bottom right}). This latest figure (highest resolution) shows that separating oldest populations is out of reach.}
  \label{fig:sirius}
  \end{center}
 \end{figure*}
\begin{figure}
\vspace*{-0.4cm}
  \begin{center}
\hspace*{-1.8cm}
  \epsfig{file=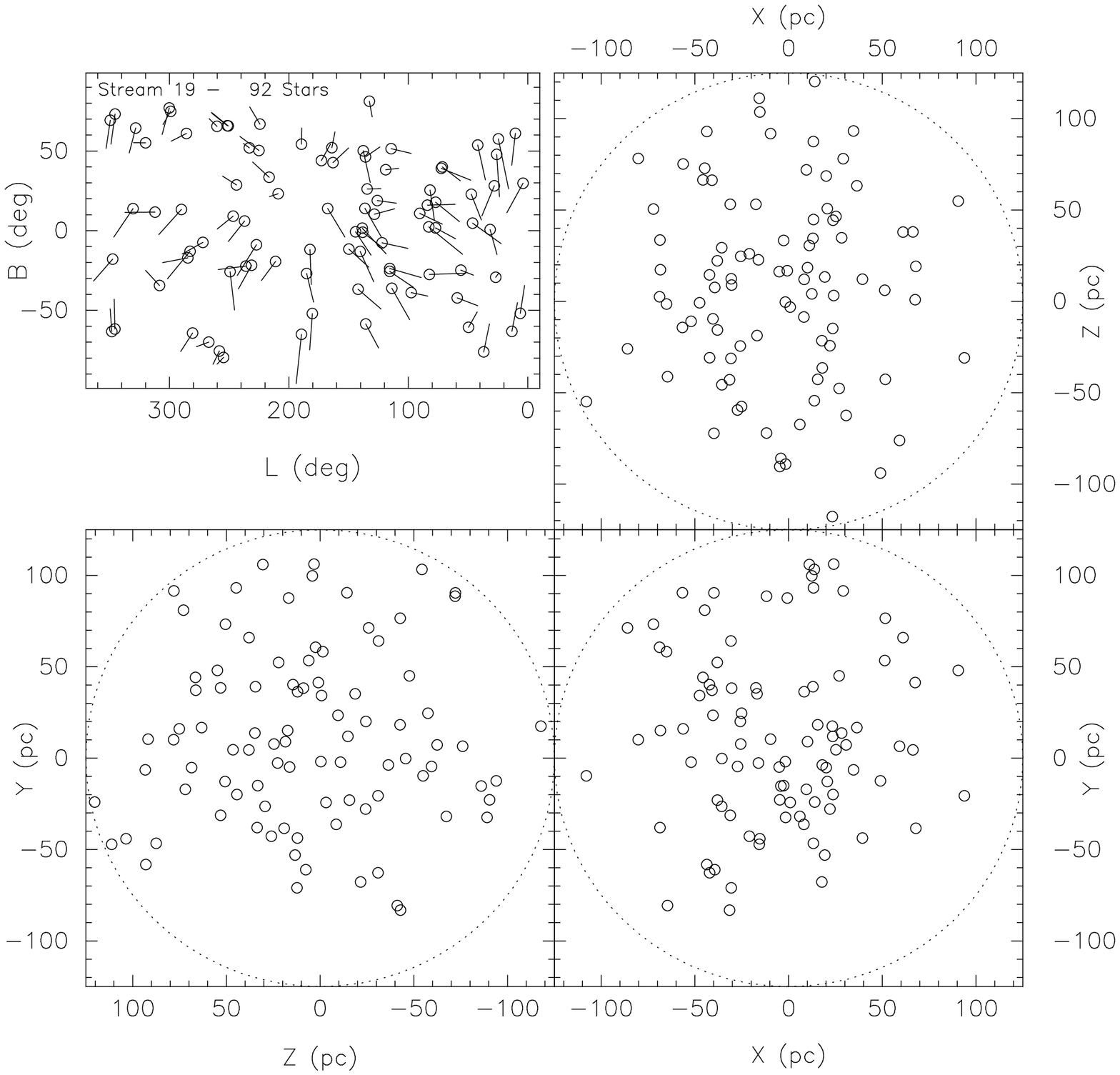,height=12.cm,width=9.4cm,angle=-90.}
  \caption{\em {\bf Space distribution of Sirius SCl} from the V$_{R}$ selected sub-sample at scale 3 
(stream 3-19 in Table \ref{tab:table3}). }
  \label{fig:spat_sirius_s3}
  \end{center}
\end{figure}
\begin{figure}
  \begin{center}
\hspace*{-1.8cm}
  \epsfig{file=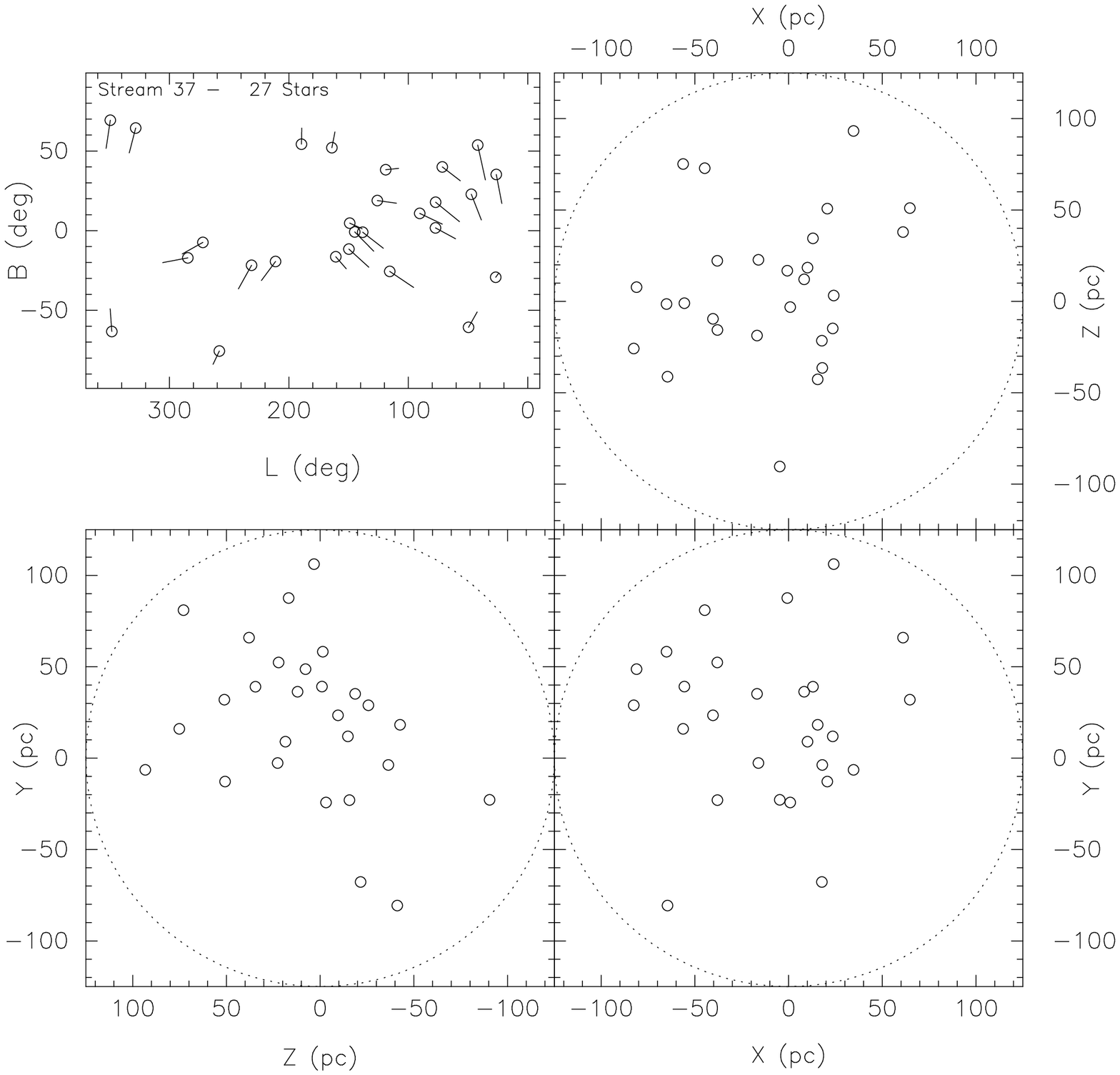,height=12.cm,width=9.4cm,angle=-90.}
\hspace*{-1.8cm}
  \epsfig{file=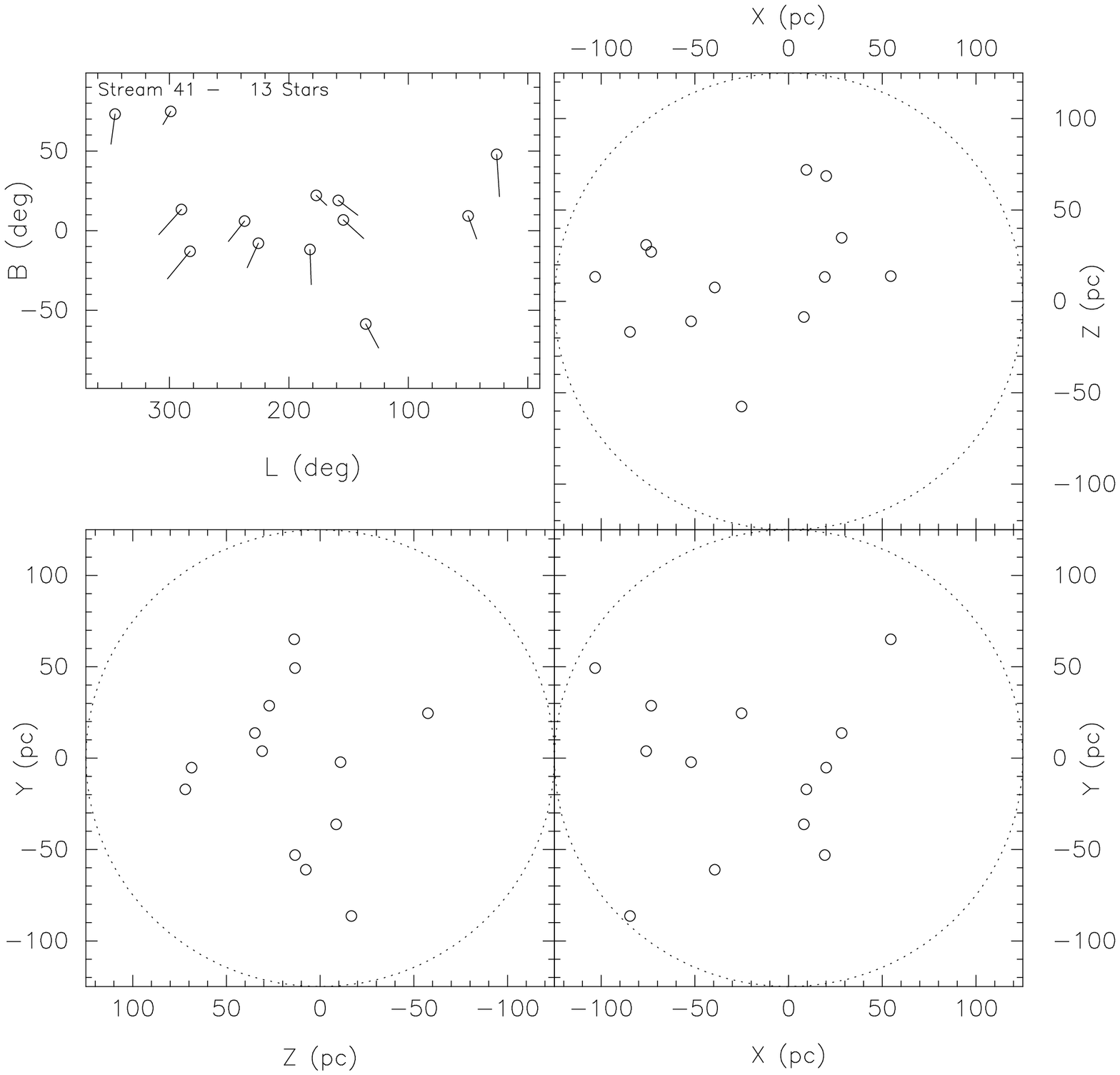,height=12.cm,width=9.4cm,angle=-90.}
  \caption{\em {\bf Space distributions of the 2 sub-streams of Sirius SCl} from the V$_{R}$ selected sub-sample at scale 2: stream 2-37 ({\bf top}) and 2-41 in Table \ref{tab:table4} ({\bf bottom}).}
  \label{fig:spat_sirius_s2}
  \end{center}
\end{figure}
The Sirius {\em supercluster} (see Figures \ref{fig:sirius} for velocity and age distributions and Figures \ref{fig:spat_sirius_s3}, \ref{fig:spat_sirius_s2} for space distributions) is found on scale 3 (stream 3-19 in Table \ref{tab:table3}) 
at mean velocity (U,V,W)=(+14.0, +1.0, -7.8) $km\cdot s^{-1}$ with velocity dispersions 
($\sigma_{U}$,$\sigma_{V}$,$\sigma_{W}$)=(7.3, 6.4, 5.5) $km\cdot s^{1}$. Eggen (1992c) identifies 
two age groups, $6.3\cdot 10^{8}$ and $10^{9}$ yr and notices that there are also younger 
($2.5\cdot 10^{8}$ yr) and older members ($1.5\cdot 10^{9}$ yr). At the coarse resolution (scale 3), 
the age distribution is in relative good agreement with this description: Str\"omgren ages peak 
at $6 \cdot 10^{8}$ yr and there is a significant proportion of stars between $10^{9}$ and $2\cdot 10^{9}$ yr. 
Stars younger than $2.5\cdot 10^{8}$ yr are probably not a statistical ghost of the $6\cdot 10^{8}$ year old
component since some stars with Str\"omgren ages are also present. \\
At scale 2, the {\em supercluster} splits into two distinct streams 
(stream 2-37 and 2-41 in Table \ref{tab:table4}) at respectively (U,V,W)= (+12.4, +0.7, -7.7) $km\cdot s^{-1}$ with ($\sigma_{U}$,$\sigma_{V}$,$\sigma_{W}$)=(4.0, 4.6, 4.7)  $km\cdot s^{-1}$ 
and (U,V,W)=(+12.4, +4.2, -9.0) $km\cdot s^{-1}$ with ($\sigma_{U}$,$\sigma_{V}$,$\sigma_{W}$)=(3.7, 3.3, 2.9)  
$km\cdot s^{-1}$ producing a very clear age separation: the very young stars are separated from a part 
of the oldest components ($6\cdot 10^{8}$ and $1.6\cdot 10^{9}$ yr). The very young component 
appears exclusively in stream 2-37 (middle right of Figure \ref{fig:sirius}). At the highest resolution, 
on scale 1 (bottom of Figure \ref{fig:sirius}), the stream 2-41 contains oldest components still 
interpenetrated. Space distributions (Figure \ref{fig:spat_sirius_s2}) show that the first stream, 
which contains the $10^{7}$ year old component is still concentrated. There are too few members 
in the second clump to make conclusions. \\
The Sirius SCl is composed by three age components of $\sim 10^{7}$, $6 \cdot 10^{8}$ and 
$1.5 \cdot 10^{9}$. The younger stream is still concentrated both kinematically and spatially 
while the two oldest streams are mixed in a larger volume of the phase space.\\
\item {\bf IC 2391 SCl}\\
\begin{figure}
  \begin{center}
  \epsfig{file=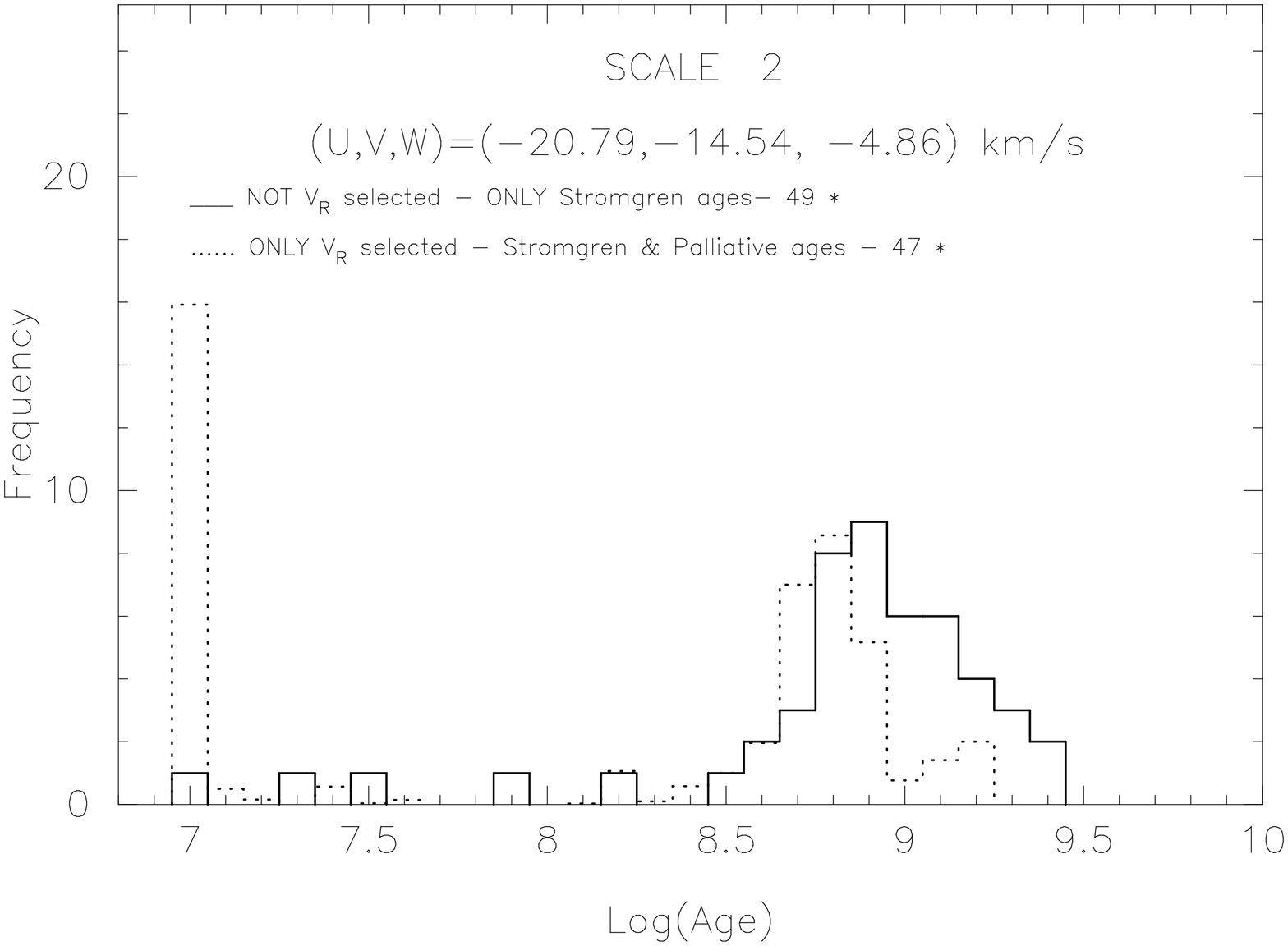,height=8.cm,width=7.cm,angle=-90.}	
  \epsfig{file=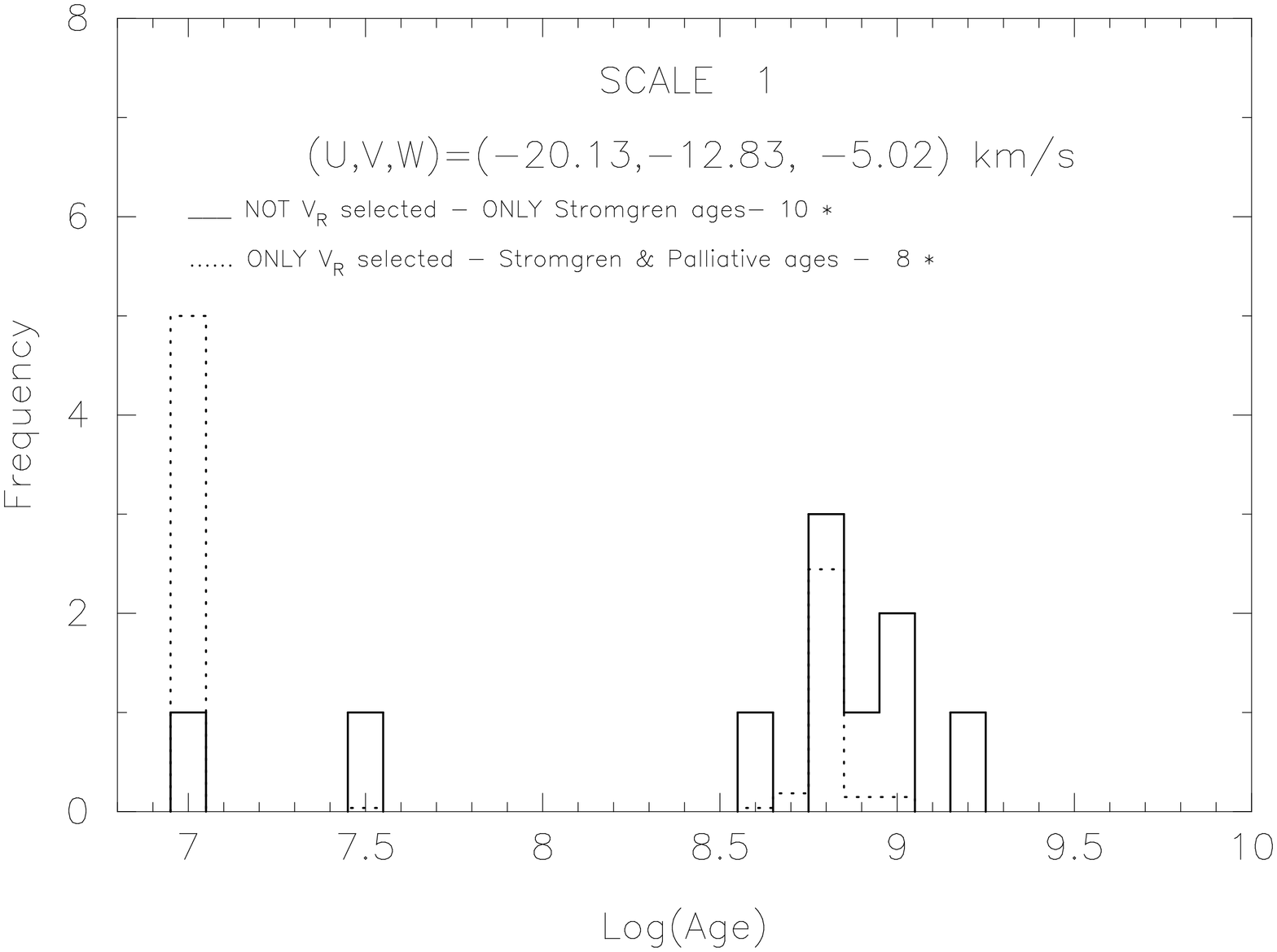,height=8.cm,width=7.cm,angle=-90.}
  \epsfig{file=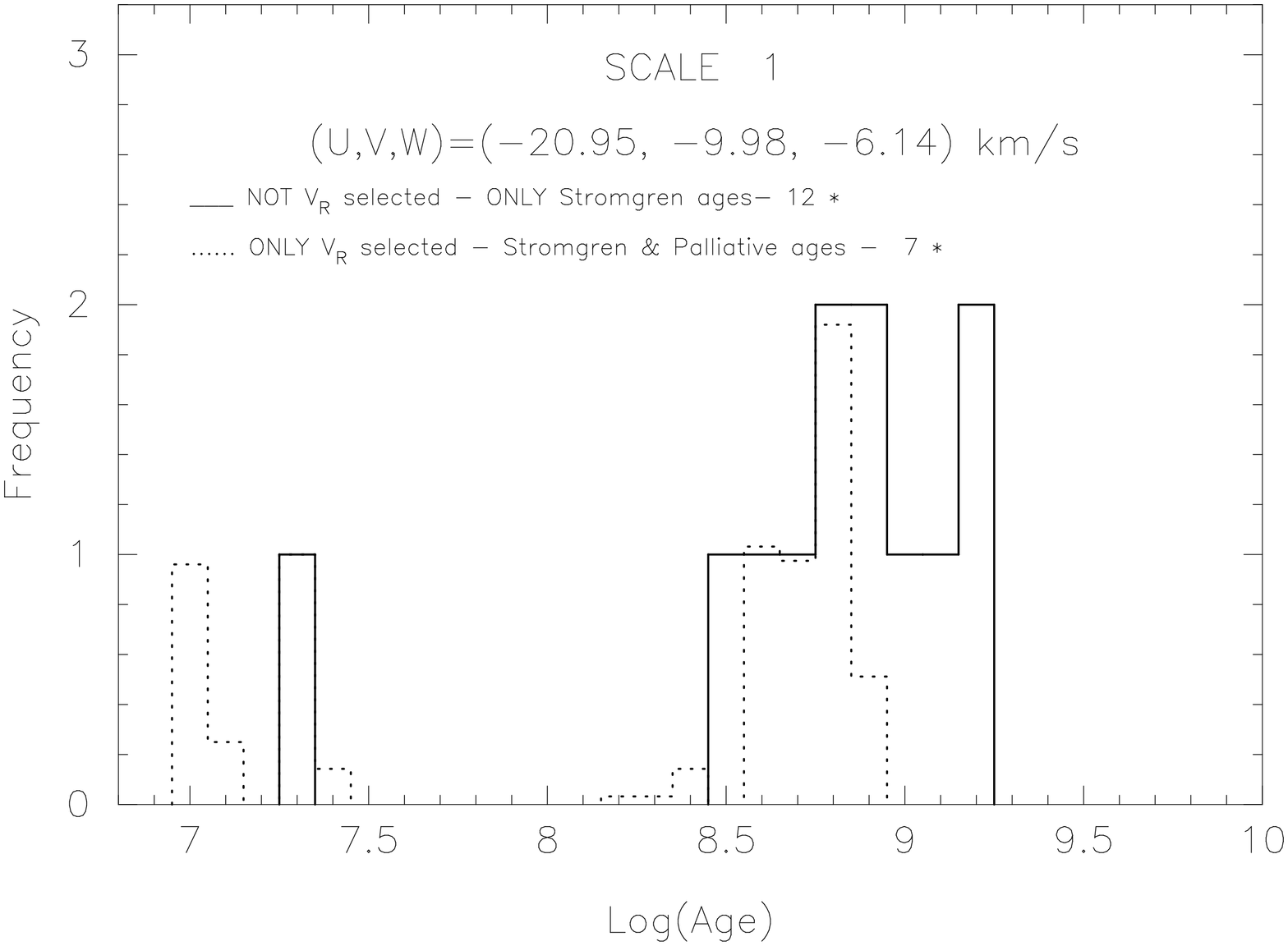,height=8.cm,width=7.cm,angle=-90.}
  \caption{\em {\bf IC 2391 SCl}. Age distributions for the IC 2391 SCl (stream 2-14 in Table \ref{tab:table4}) at scale 2 ({\bf top}), for sub-stream 1-20 (in Table \ref{tab:table5}) at scale 1 ({\bf middle}) and for sub-stream 1-25 (in Table \ref{tab:table5}) at scale 1 ({\bf bottom}).}
  \label{fig:ic2391}
  \end{center}
\end{figure}
\begin{figure}
  \begin{center}
\hspace*{-1.8cm}
  \epsfig{file=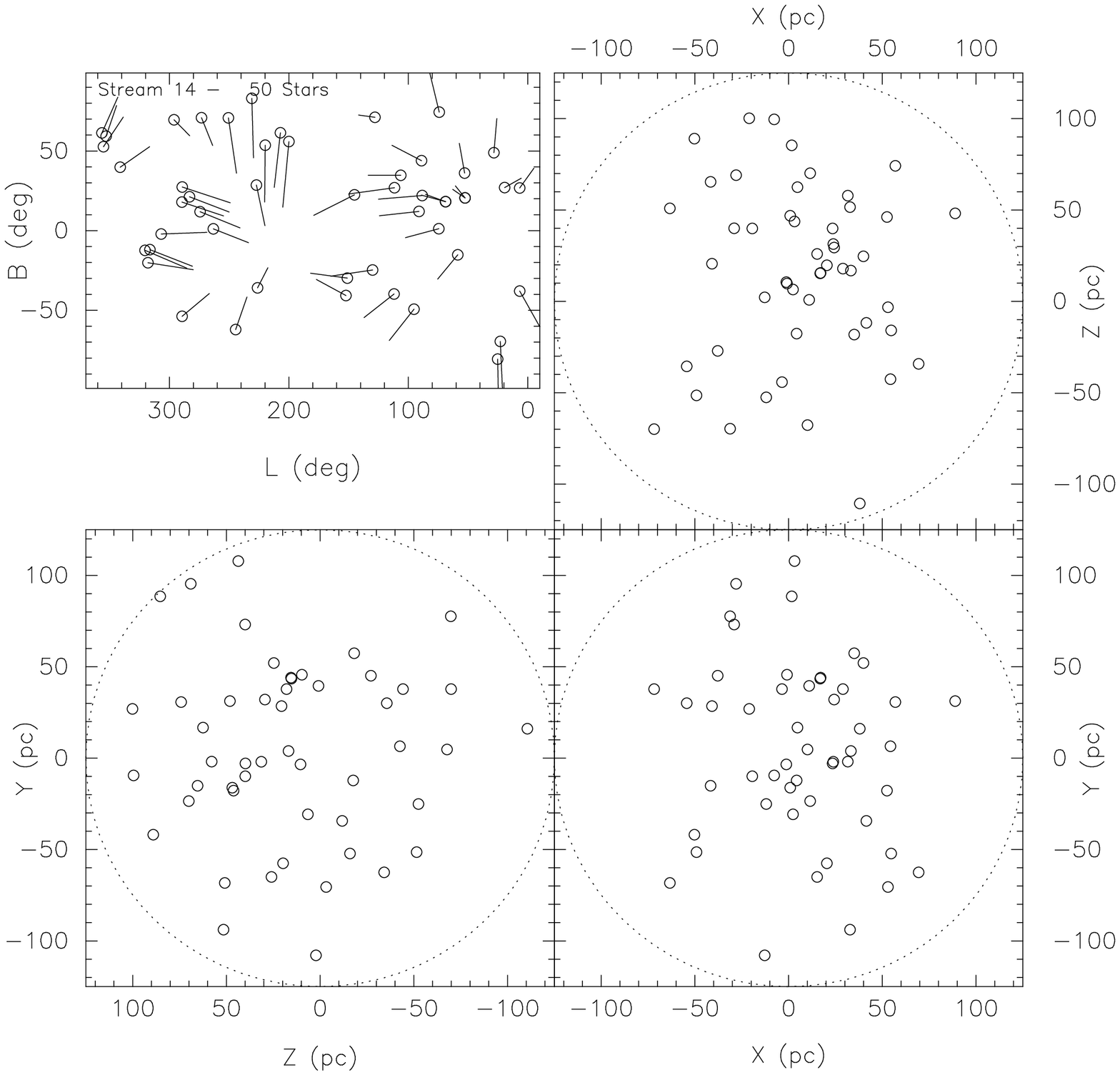,height=12.cm,width=9.4cm,angle=-90.}
  \caption{\em {\bf Space distribution of IC 2391 SCl} from the V$_{R}$ selected sub-sample at scale 2 (stream 2-14 in Table \ref{tab:table4}).}
  \label{fig:spat_ic2391_s2}
  \end{center}
\end{figure}
The IC 2391 SCl (see Figures \ref{fig:ic2391} for age distributions and Figure \ref{fig:spat_ic2391_s2} 
for space distributions) is not found at large scale because it may have been merged into the Centaurus 
association velocity group. It appears separately at scale 2 (stream 2-14 in Table \ref{tab:table4}) at 
(U,V,W)=(-20.8, -14.5, -4.9) $km\cdot s^{-1}$ with velocity dispersions 
($\sigma_{U}$,$\sigma_{V}$,$\sigma_{W}$)=(4.3, 4.9, 5.0)  $km\cdot s^{-1}$ (see Figure 15 of Paper II). 
Eggen (1991) states that IC 2391 SCl contains two ages: $8\cdot 10^{7}$ and $2.5\cdot 10^{8}$ yr 
while Chen et al (1997) found a mean age of $4.6\pm1.6\cdot 10^{8}$ yr. The Str\"omgren age 
distribution is quite different from the palliative age distribution at coarser resolution. Str\"omgren 
ages peak at $8\cdot 10^{8}$ but with ages up to  $2\cdot 10^{9}$ yr. Palliative ages exhibit a peak 
at $6\cdot 10^{8}$ yr and a $10^{7}$ year old component (Figure \ref{fig:ic2391}). This last peak is 
certainly real because its proportion is too high to be a statistical ghost of palliative ages from a 
$6\cdot 10^{8}$ year old component and moreover Str\"omgren ages show the presence of young stars. 
Two sub-streams are found at scale 1 (stream 1-20 and 1-25 in Table \ref{tab:table5}) at 
(U,V,W)=(-20.1, -12.8, -5.0) with ($\sigma_{U}$,$\sigma_{V}$,$\sigma_{W}$)= ( 2.9, 3.1, 1.8)  $km\cdot s^{-1}$ 
and (U,V,W)=( -20.9, -10.0, -6.1) with ($\sigma_{U}$,$\sigma_{V}$,$\sigma_{W}$)=(4.5, 3.7, 2.9)  $km\cdot s^{-1}$. 
The stream 1-20 contains all the youngest stars while the sub-stream 1-25 is only constituted of 
the $6\cdot 10^{8}$ year old population. Velocity dispersions of the two streams are in agreement with 
this view: they are smaller for the stream with the younger component. This configuration is exactly the 
opposite of the Pleiades' one: in this case the youngest population is more concentrated in the velocity 
space and is entirely detected in one sub-stream while the oldest span over the two streams.\\
\item {\bf Centaurus Associations} and {\bf the Gould belt}\\
\begin{figure}
  \begin{center}
  \epsfig{file=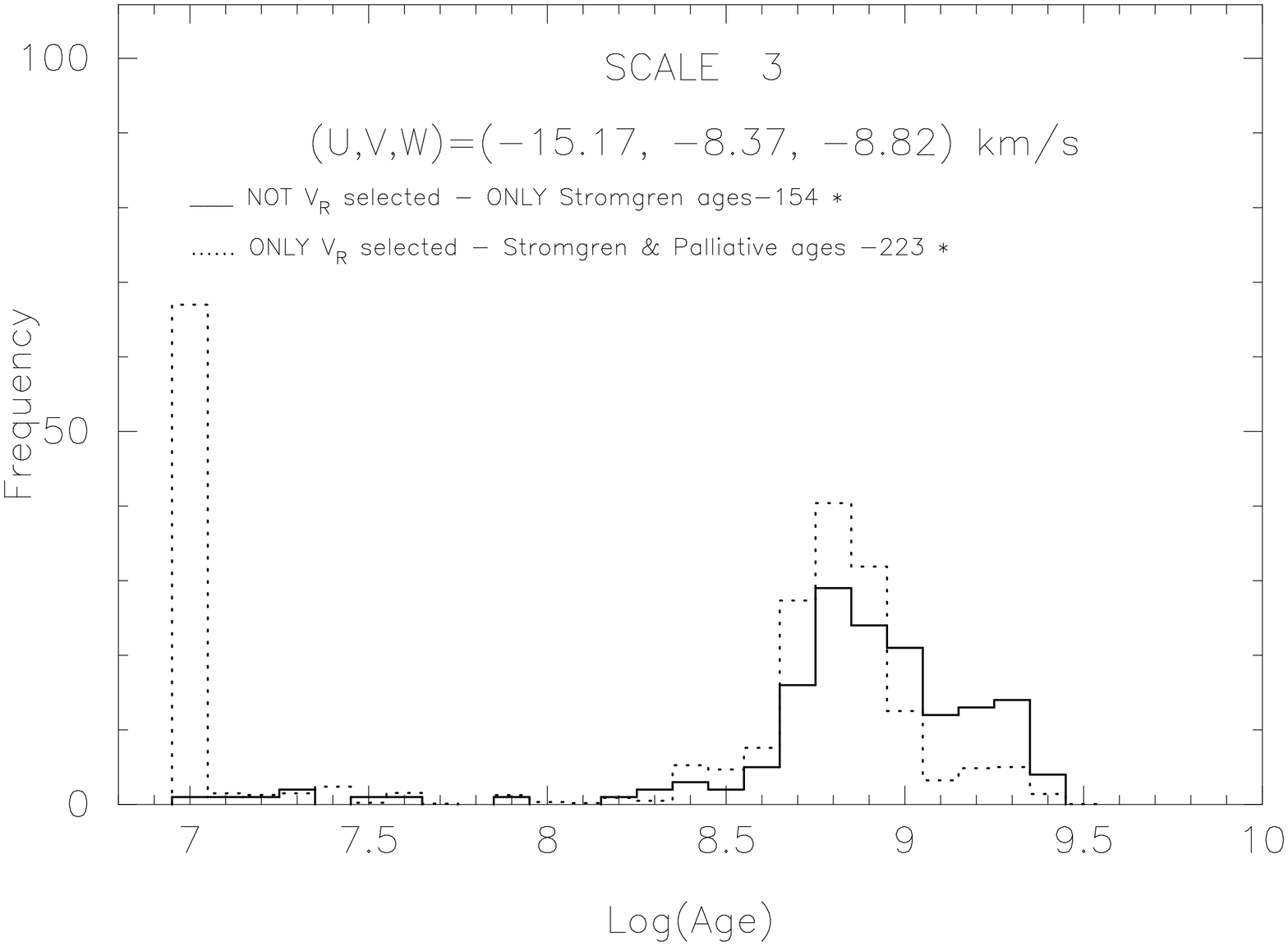,height=8.cm,width=7.cm,angle=-90.}
  \epsfig{file=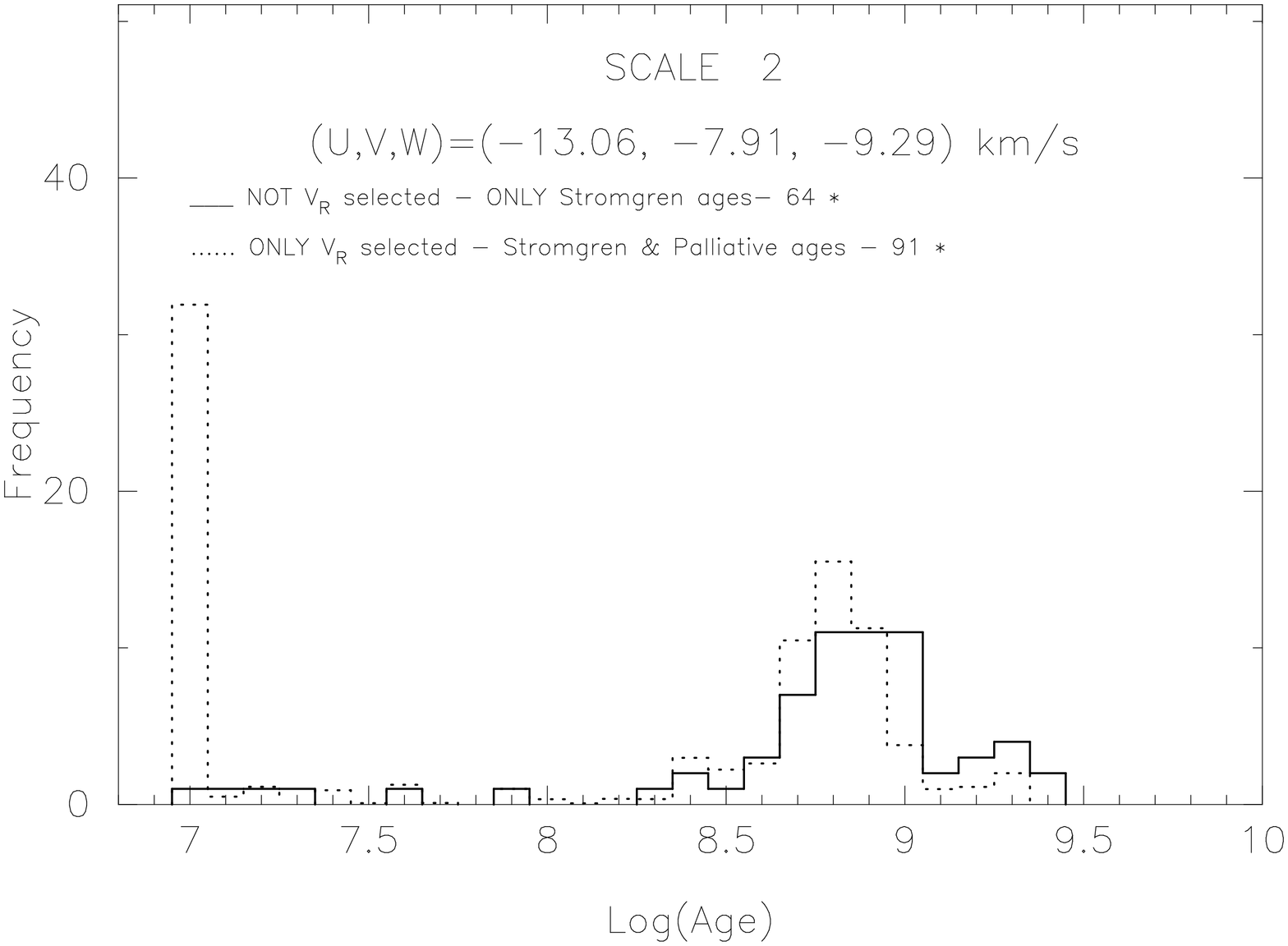,height=8.cm,width=7.cm,angle=-90.}
  \epsfig{file=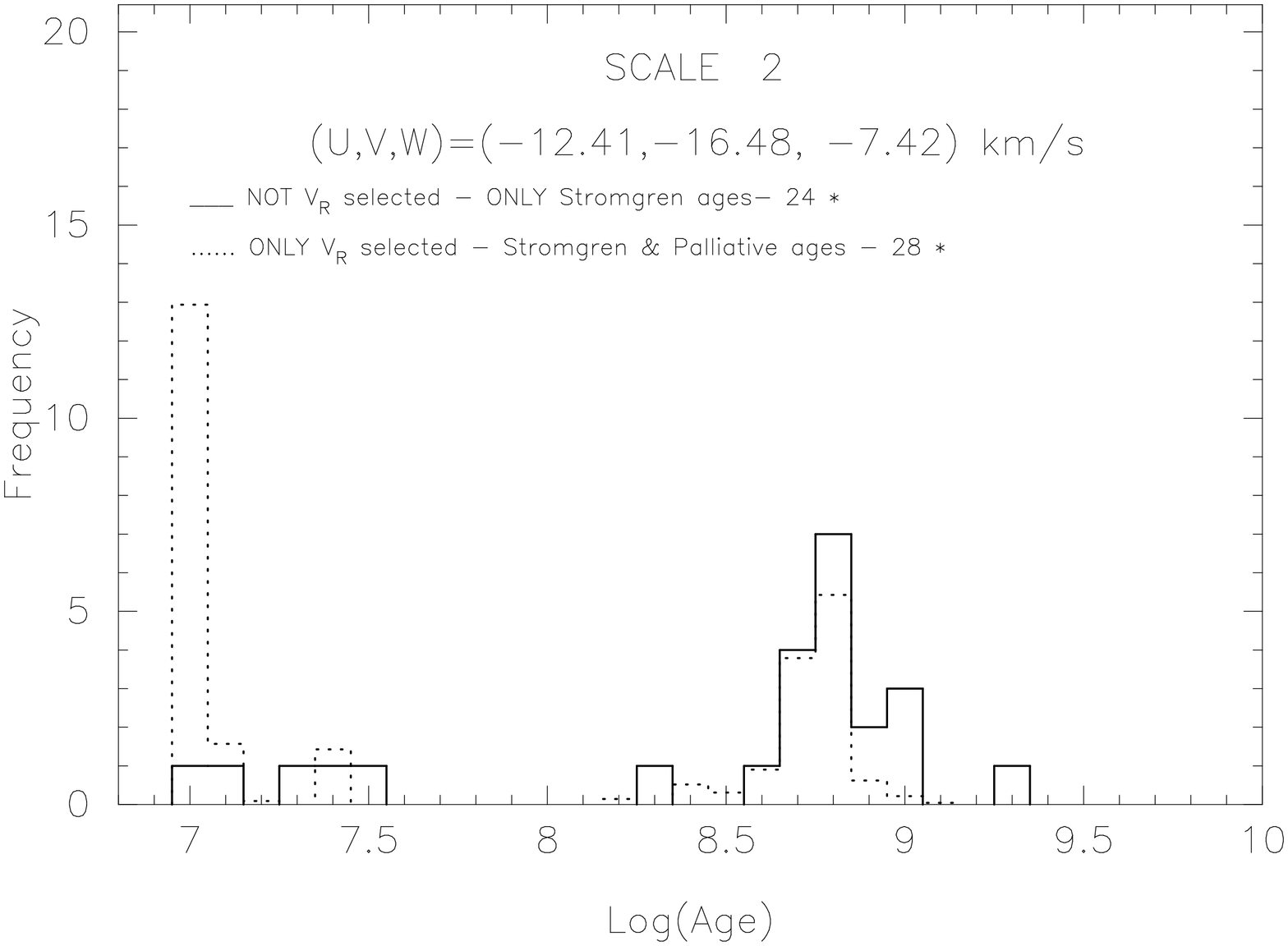,height=8.cm,width=7.cm,angle=-90.}
  \caption{\em {\bf Streams associated with Centaurus Associations}. Age distributions of the overall association (stream 3-15 in Table \ref{tab:table3}) at scale 3 ({\bf top}). Age distributions of Centaurus-Crux (stream 2-26 in Table \ref{tab:table4}) at scale 2 ({\bf middle}). Age  distributions of Centaurus-Lupus (stream 2-12 in Table \ref{tab:table4}) at scale 2 ({\bf bottom}).}
  \label{fig:centaurus}
  \end{center}
 \end{figure}
\vspace{-0.3cm}
\begin{figure}
  \begin{center}
\hspace*{-1.8cm}
\epsfig{file=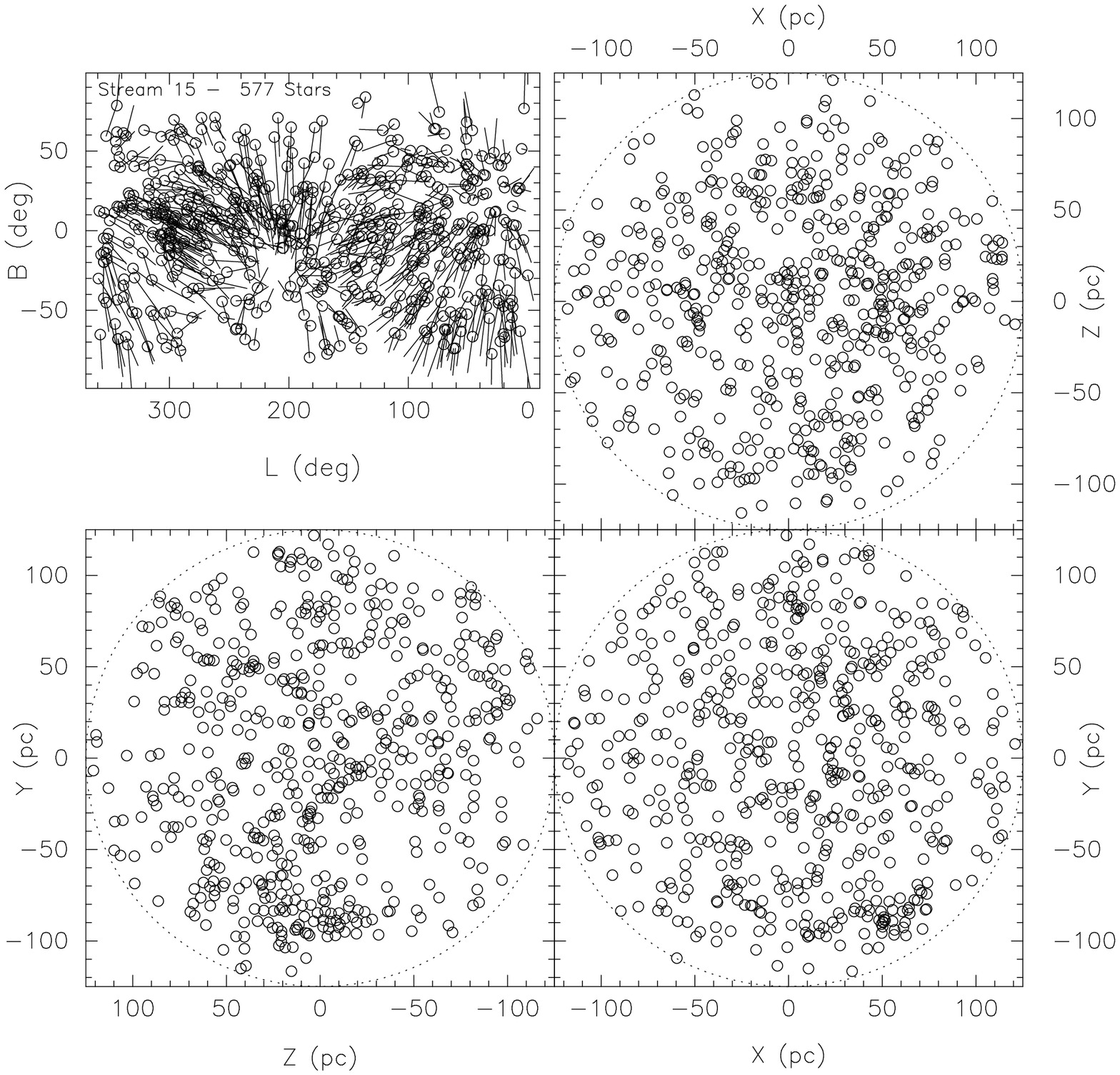,height=12.cm,width=9.4cm,angle=-90.}
\hspace*{-1.8cm}
  \epsfig{file=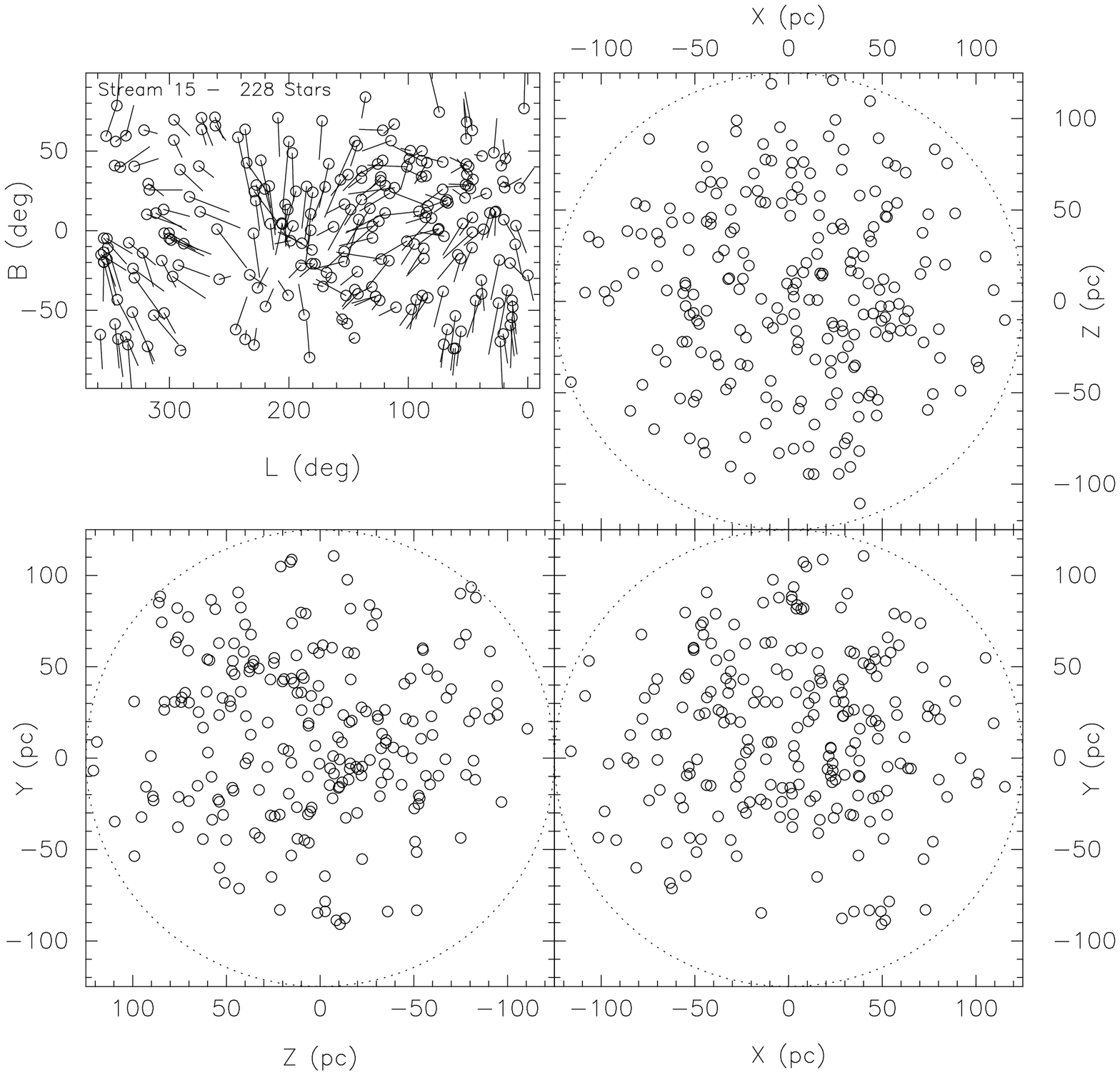,height=12.cm,width=9.4cm,angle=-90.}
  \caption{\em {\bf Space distributions of the stream associated with Centaurus Associations} (stream 3-15 in Table \ref{tab:table3}) for all the stars ({\bf top}) and for the V$_{R}$ selected sub-sample ({\bf bottom}) at scale 3. Stars belonging to spatial clumps in the upper figure disappear because of the lack of radial velocities.}
  \label{fig:spat_cen_s3}
  \end{center}
\end{figure}
\begin{figure}
  \begin{center}
\hspace*{-1.8cm}
  \epsfig{file=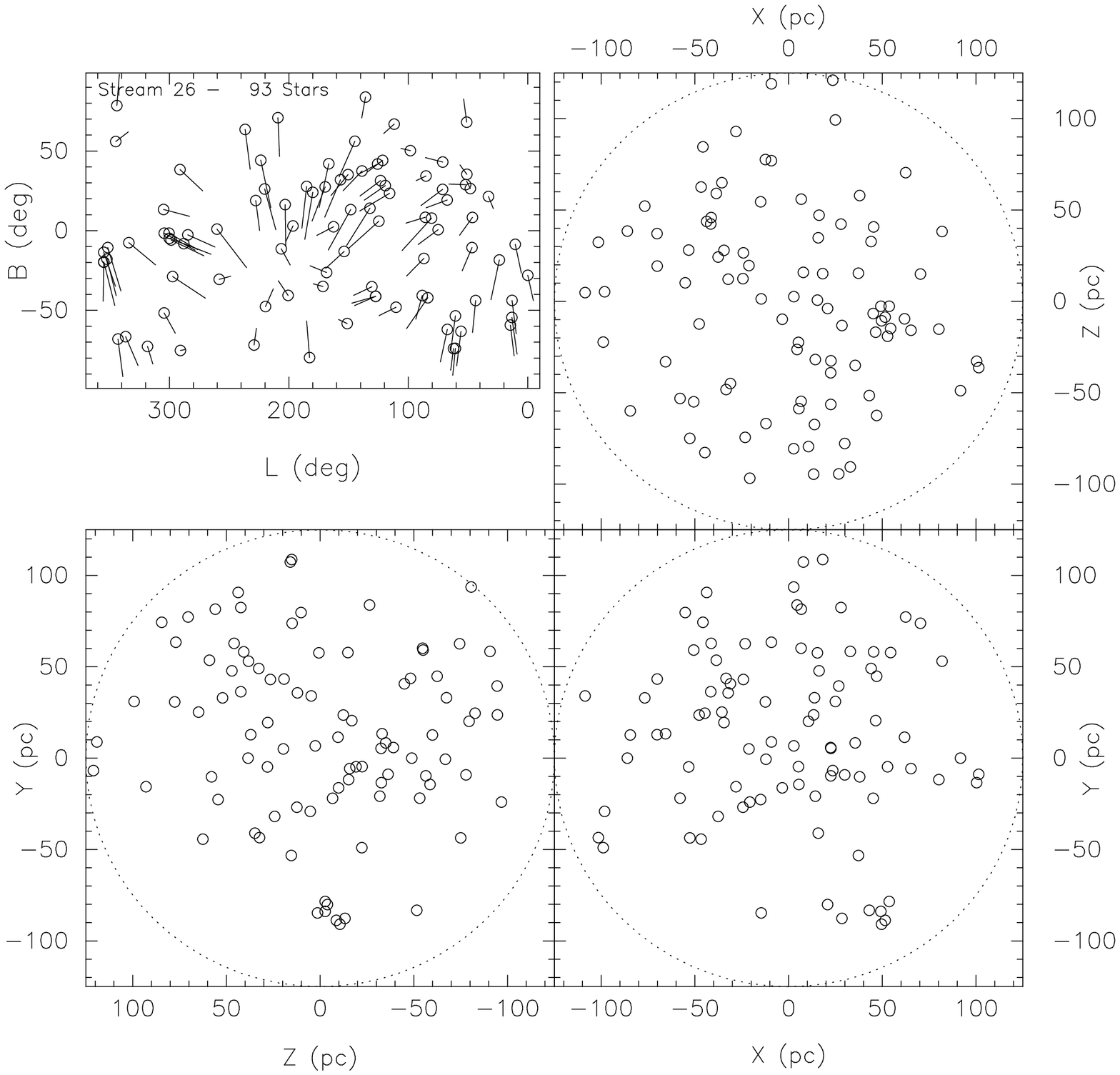,height=12.cm,width=9.4cm,angle=-90.}
\hspace*{-1.8cm}
  \epsfig{file=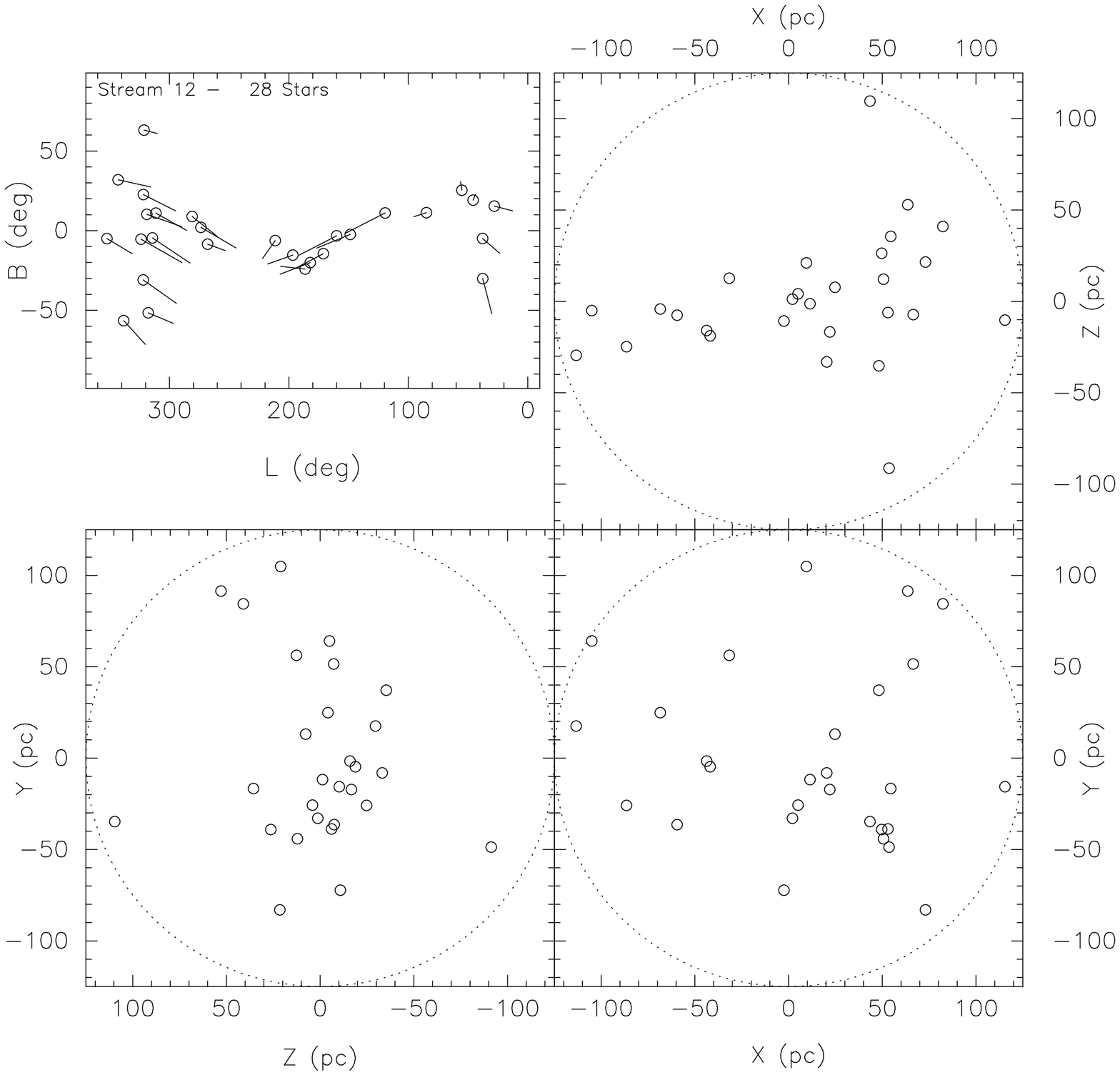,height=12.cm,width=9.4cm,angle=-90.}
  \caption{\em {\bf Space distributions of sub-streams associated with Centaurus Associations}. Centaurus-Crux (stream 2-26 in Table \ref{tab:table4}) ({\bf top}) and Centaurus-Lupus (stream 2-12 in Table \ref{tab:table4}) ({\bf bottom}) associations from the V$_{R}$ selected sub-sample at scale 2. A disklike structure appears with the Centaurus-Lupus stream, on the (X,Z) projection, reflecting the Gould Belt.}
  \label{fig:spat_cencruxlup_s2}
  \end{center}
\end{figure}
Lower Centaurus-Crux and upper Centaurus-Lupus associations (see Figures \ref{fig:centaurus} 
for age distributions and Figures \ref{fig:spat_cen_s3} for space distributions) 
which are the main components of the entire Centaurus association are detected as one velocity 
clump at scale 3 (stream 3-15 in Table \ref{tab:table3}) with (U,V,W)=(-15.2, -8.4, -8.8) $km\cdot s^{-1}$ 
with velocity dispersion  ($\sigma_{U}$,$\sigma_{V}$,$\sigma_{W}$)=(8.6, 6.7, 6.1)  $km\cdot s^{-1}$. 
Scale 3 is a too coarse resolution and the age distributions reflect the overall distribution. 
The whole Centaurus association is splitted into two parts at scale 2 and does not evolve at scale 1 
(see Paper II, Figure 15).\\
At scale 2, Centaurus-Crux  (stream 2-26 in Table \ref{tab:table4}) and Centaurus-Lupus 
(stream 2-12 in Table \ref{tab:table4}) are identified at (U,V,W)=(-13.1, -7.9, -9.3) $km\cdot s^{-1}$ 
with ($\sigma_{U}$,$\sigma_{V}$,$\sigma_{W}$)=(6.2, 6.1, 5.5)  $km\cdot s^{-1}$ and 
(U,V,W)=(-12.4, -16.5, -7.4) $km\cdot s^{-1}$ with ($\sigma_{U}$,$\sigma_{V}$,$\sigma_{W}$)=(6.1, 4.6, 3.1)  
$km\cdot s^{-1}$ respectively. 
Unfortunately, as for the Pleiades SCl, a lot of young stars have not Str\"omgren photometry. 
Str\"omgren age distributions peak at  $6\cdot 10^{8}$ yr for both sub-streams but palliative age 
distributions show the predominance of the very young population ($10^{7}$ yr) in each case.\\
There is a crucial lack of radial velocities for the spatially clustered structures Centaurus-Crux  and 
Centaurus-Lupus: one fifth of stars have observed V$_{R}$. That is why these space clumps are 
visible on space distributions when taking into account all the stars of the detected streams but 
disappear with the sub-sample selected on observed V$_{R}$ (Figure \ref{fig:spat_cen_s3}). 
At scale 2, space distributions show that stars of the velocity substructure identified as 
Centaurus-Lupus association belong to a disk-like structure (see XZ projection on 
Figure \ref{fig:spat_cencruxlup_s2}) tilted with respect to the Galactic disk. Eigenvectors of the 
spatial ellipsoid are calculated.  The two vectors associated with the largest eigenvalues allow to 
define the plane of the structure, assuming it passes through the Sun. The ascending node of the 
intersection between this disk-like structure and the Galactic plane is $l_{\Omega}=317^{o}$ which 
differs slightly from usual values ($l_{\Omega}\sim 275^{o}$ to $290^{o}$, \cite{Poppel97}). 
The angle between the two planes is $i=18.8^{o}$ in agreement with previous study ($i=18^{o}$ - $23^{o}$)\\
\item {\bf A new {\em supercluster}}\\
 \begin{figure*}
  \begin{center}
  \epsfig{file=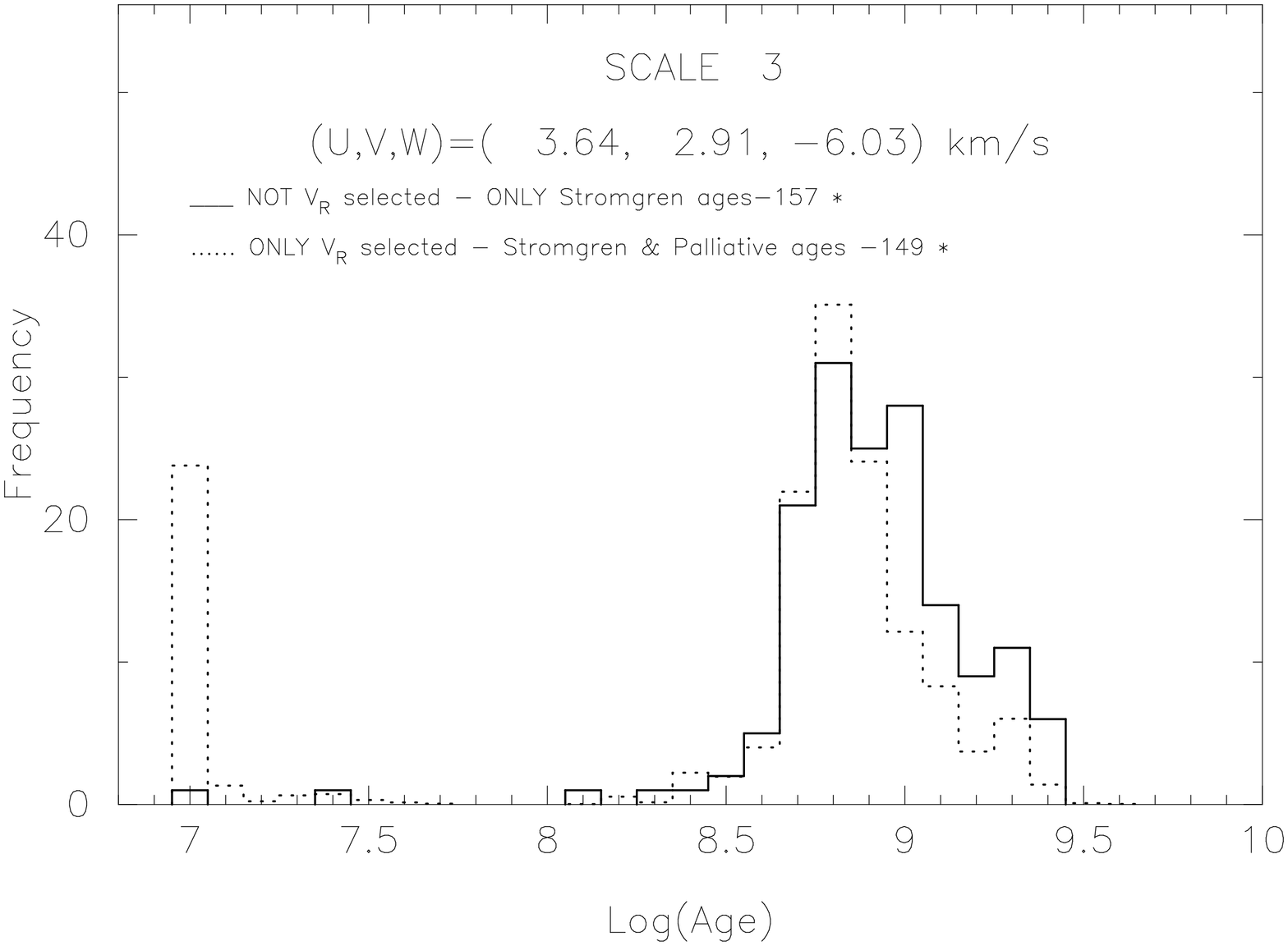,height=8.cm,width=7.cm,angle=-90.}
  \epsfig{file=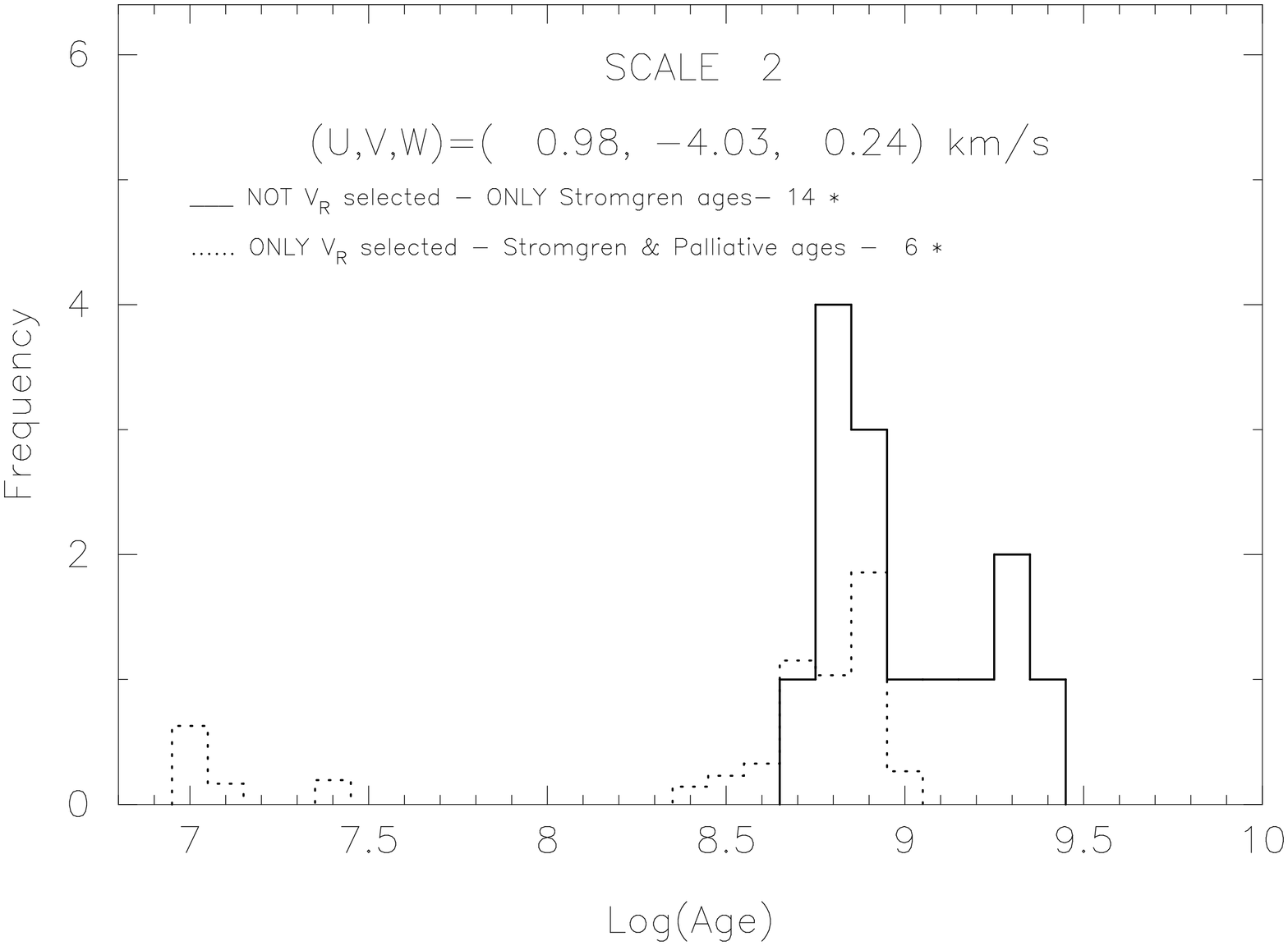,height=8.cm,width=7.cm,angle=-90.}
  \epsfig{file=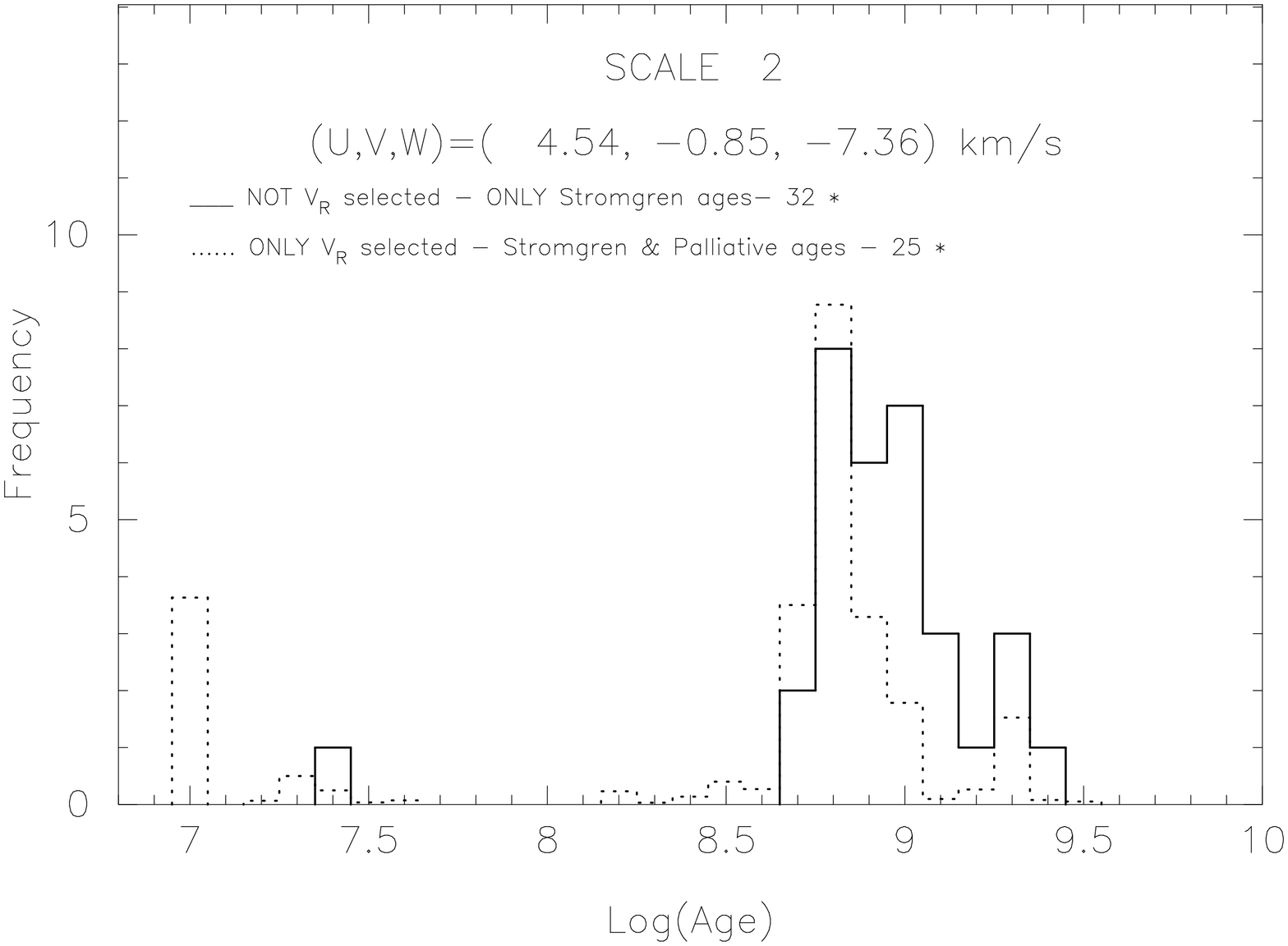,height=8.cm,width=7.cm,angle=-90.}
  \epsfig{file=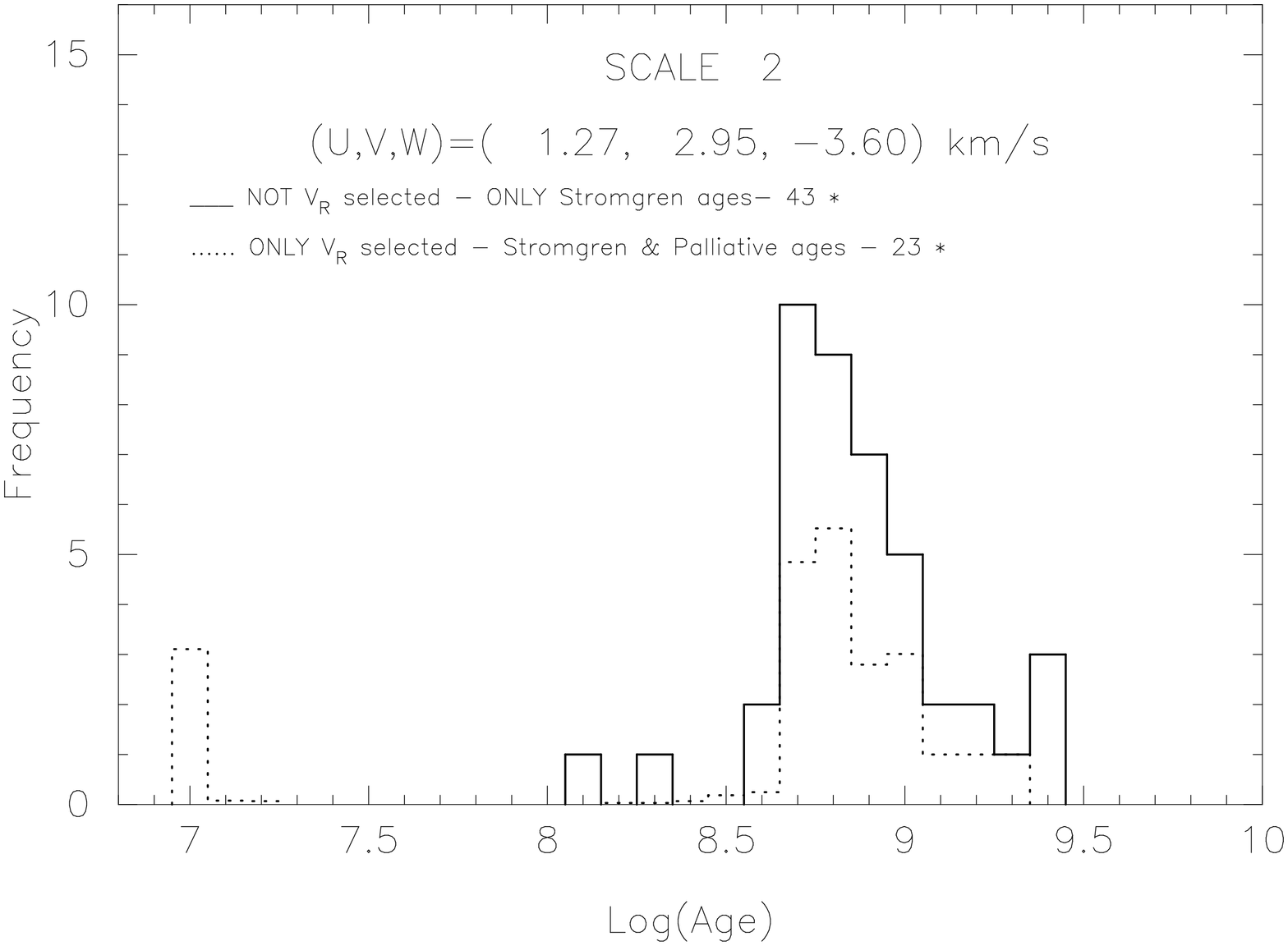,height=8.cm,width=7.cm,angle=-90.}
  \epsfig{file=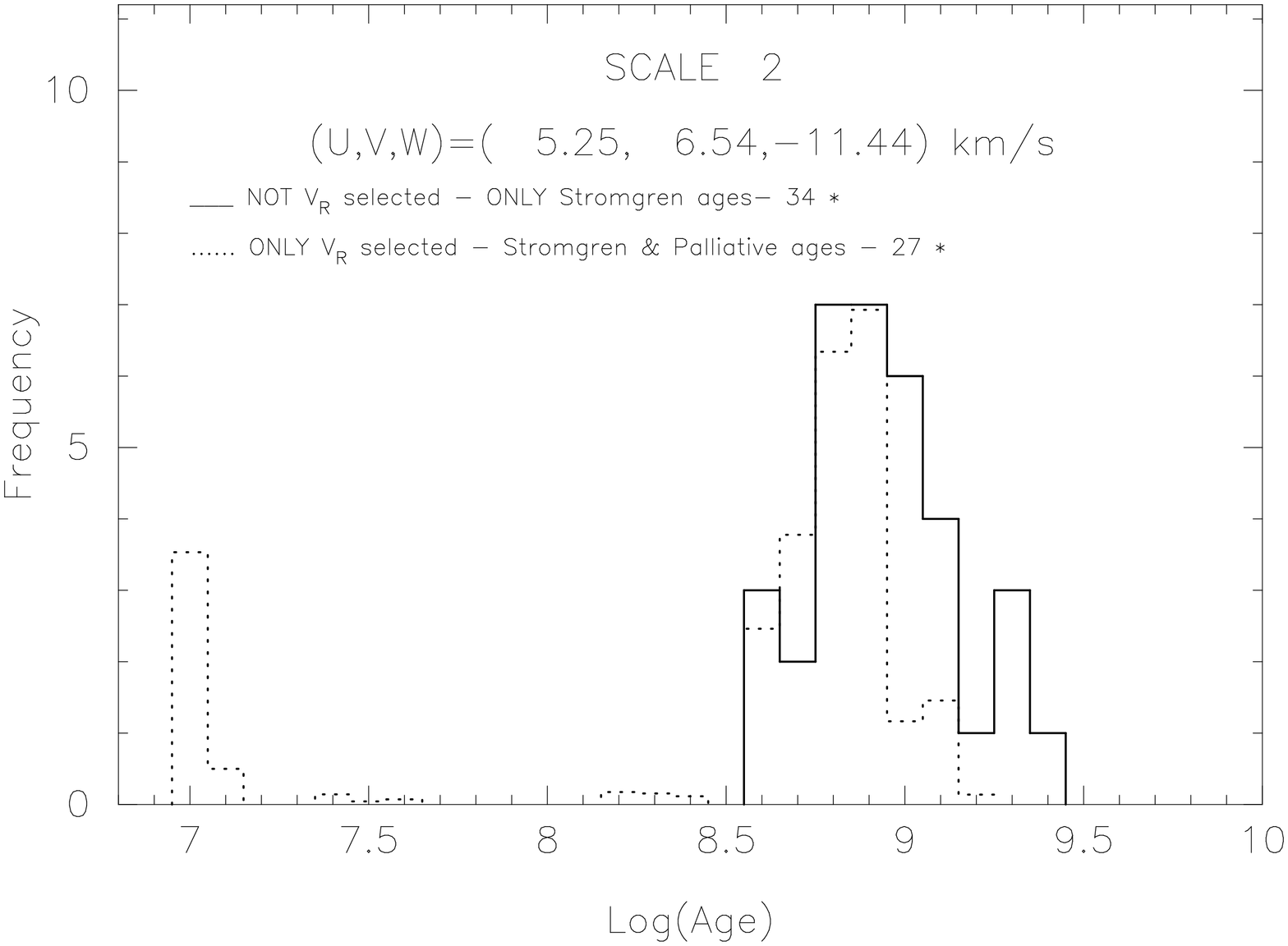,height=8.cm,width=7.cm,angle=-90.}
  \epsfig{file=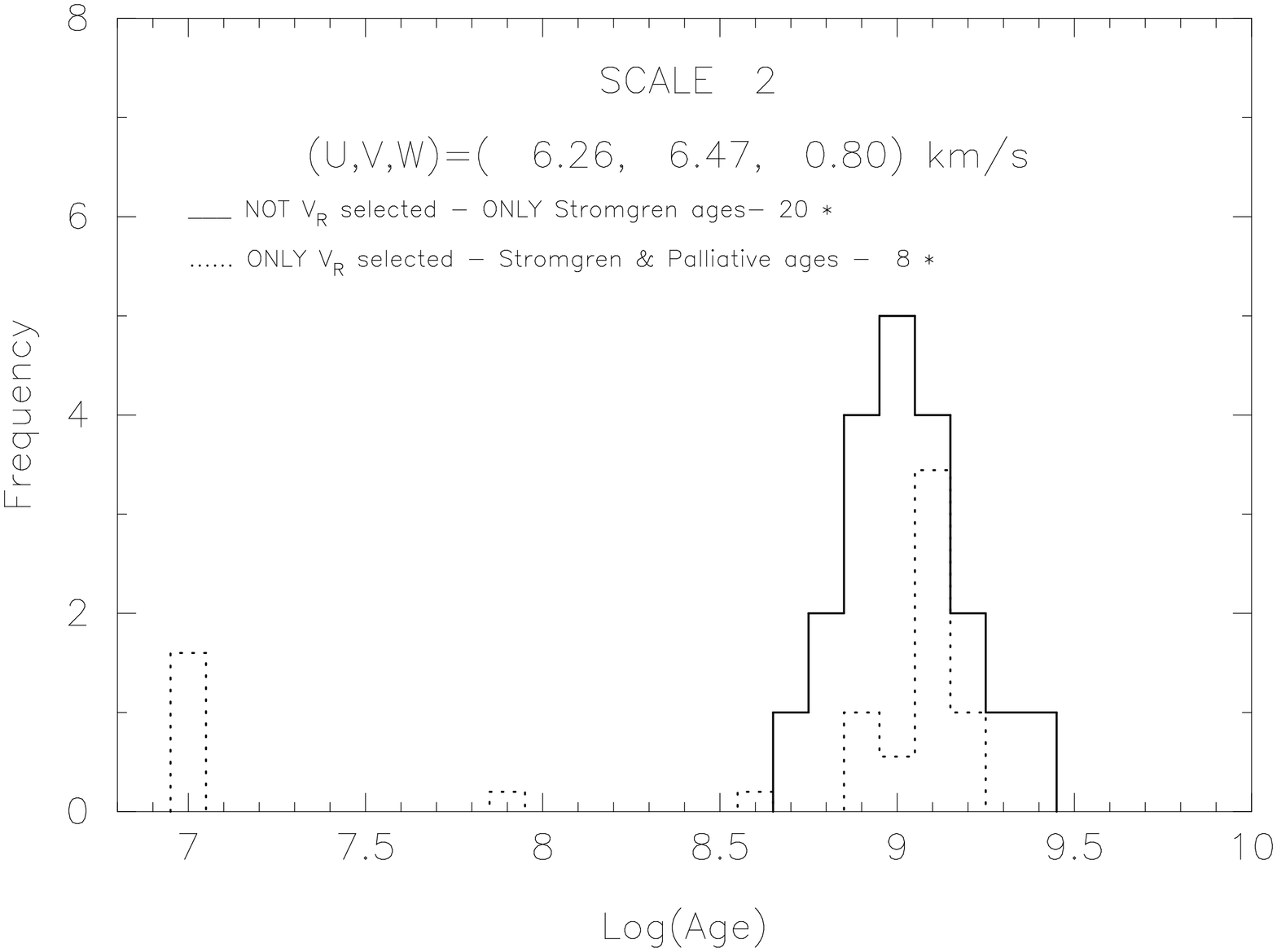,height=8.cm,width=7.cm,angle=-90.}
  \caption{\em {\bf New supercluster}. Age distributions of the new moving group (stream 3-18 in Table \ref{tab:table3}) at scale 3 ({\bf top left}) and the sub-stream 2-27 (in Table \ref{tab:table4}) at scale 2 ({\bf top right}). Age distributions of the sub-streams 2-35 ({\bf middle left}) and 2-38  ({\bf middle right}) at scale 2. Age distributions of the sub-streams 2-43 ({\bf bottom left}) and 2-44 ({\bf bottom right}) at scale 2.}
  \label{fig:newmov}
  \end{center}
\end{figure*}
\begin{figure}
  \begin{center}
\hspace*{-1.8cm}
  \epsfig{file=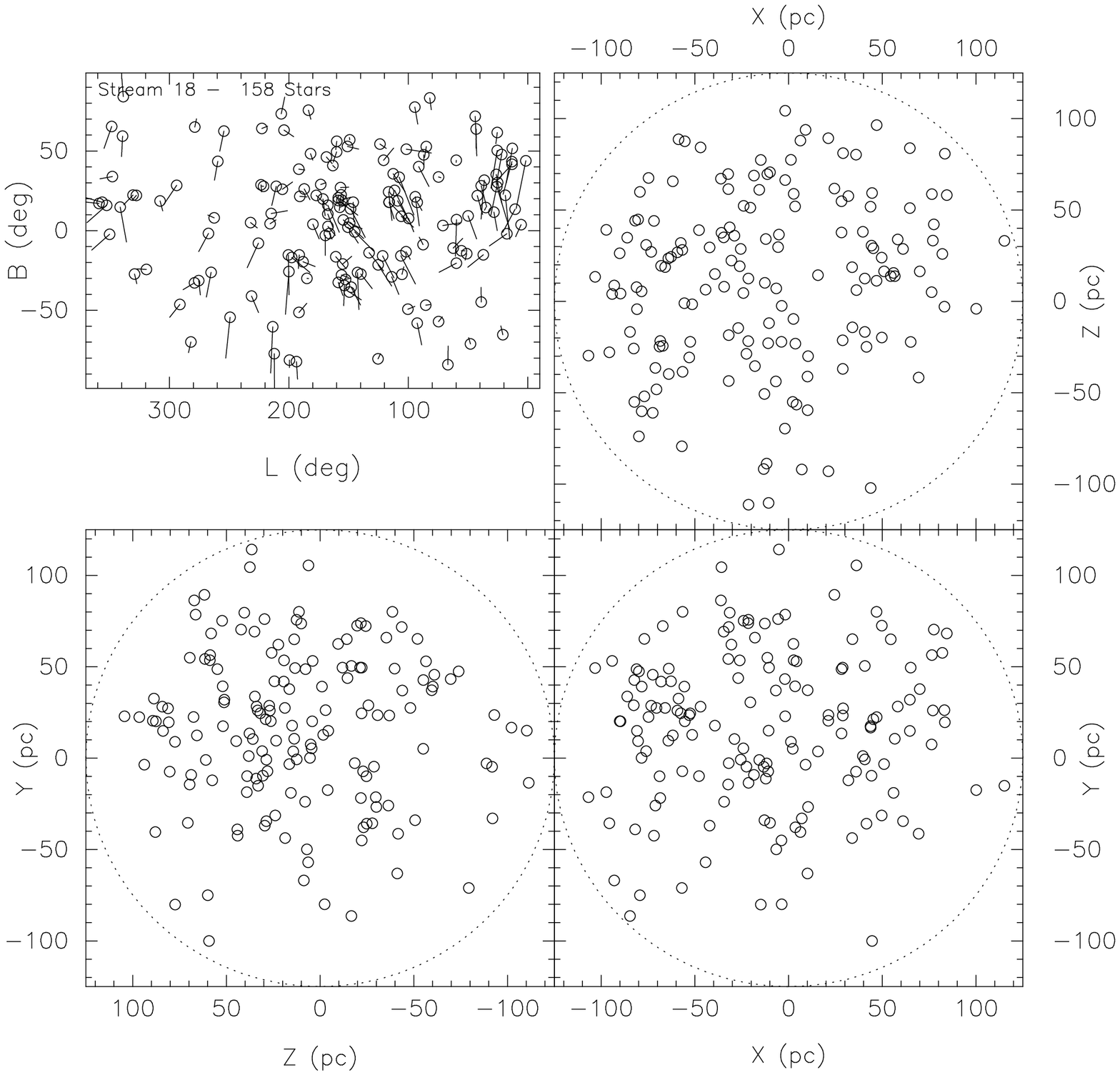,height=12.cm,width=9.4cm,angle=-90.}
  \caption{\em {\bf Space distributions of New SCl} (stream 3-18 in Table \ref{tab:table3}) from the V$_{R}$ selected 
sub-sample at scale 3.}
  \label{fig:spat_new_s3}
  \end{center}
\end{figure}
Close to the Sirius SCl in velocity space, located at the mean velocity (U,V,W)=(+3.6,+2.9,-6.0) $km\cdot s^{-1}$ 
with velocity dispersions ($\sigma_{U}$,$\sigma_{V}$,$\sigma_{W}$)=(6.8,5.0,6.3)  $km\cdot s^{-1}$, a new massive 
{\em supercluster} (stream 3-18 in Table \ref{tab:table3}) is detected at scale 3 (see Figures \ref{fig:newmov} for age 
distributions and Figure \ref{fig:spat_new_s3} for spatial distribution). It contains 
almost twice as many members as the Sirius SCl. None of the previously known {\em superclusters} corresponds 
to this velocity definition. Figueras et al (1997) indicate the presence of a velocity structure at (U,V)=(+7,+6) 
which they cannot confirm without doubt by their analysis and interpreted it as a possible sub-structure of 
Sirius SCl with a mean age of $10^{9}$ yr. We confirm the existence of a {\it supercluster} like structure, 
probably never detected before because of its low velocity with respect to the Sun. On a kinematics basis it is 
clearly dissociated from the Sirius SCl.\\
Age distributions at coarser scale are similar to the whole sample ones with ages ranging from $10^{7}$ 
to $2.5\cdot 10^{9}$ yr. But at least 5 sub-streams at scale 2 (see Table \ref{tab:table4}) are found to form 
this structure. These streams show age distributions of relatively old components. Stream 2-27 shows an 
unambiguous peak at $6\cdot 10^{8}$ yr with few $1.6\cdot 10^{9}$ year old stars. On the basis of velocity 
and age content, this stream could originate from the evaporation of the Coma OCl. Stream 2-35 has stars 
which are $6\cdot 10^{8}$, $10^{9}$ and $1.6\cdot 10^{9}$ year old on the basis of Str\"omgren photometry 
but palliative ages exhibit only one peak at $6\cdot 10^{8}$ yr. Stream 2-38 shows a peak at $5\cdot 10^{8}$ yr. 
Stream 2-43 has Str\"omgren ages between $6-8\cdot 10^{8}$ and few $1.6\cdot 10^{9}$ year old stars 
but the palliative ages exhibit only one peak at $8\cdot 10^{8}$. Stream 2-44 is clearly a $10^{9}$ year old 
group. All the few very young palliative ages in each stream are probably statistical ghost because very young 
Str\"omgren ages are never present.\\
Age distributions at the highest resolution (scale 1), not shown here, give exactly the same results as scale 2 
but the number of stars dramatically decreases.\\
This structure has the same features as the previously known {\em superclusters}: a juxtaposition of several little 
star formation bursts at different epochs in adjacent cells of the velocity field. The correlation between 
velocity and age is not always obvious because these bursts ($5-6\cdot 10^{8}$, $8\cdot10^{8}$ and
$10^{9}$ yr) are not very recent. As in the Hyades SCl case, stream velocity volumes, defined by their 
velocity dispersions, are substantially recovering.\\
\end{enumerate}
Implications of these results on the understanding of the {\em supercluster} concept are discussed in Paper II.
\subsubsection{Outside {\em superclusters}}
\begin{itemize}
\item Small scale streams\\
\label{sec:smallstream}
While {\em superclusters} are found to split into smaller scale streams most of them corresponding to 
well defined age, a number of other streams are detected only at small scales. Stream 2-13 in 
Table \ref{tab:table4} (Figure \ref{fig:spat_g13_s2}) is a typical example of such a stream. 
Its age distribution shows a mono-age component of  $10^{9}$ year old and its vertical velocity 
is high (W=-15.3 $km\cdot s^{-1}$). Space distribution does not fill the 125 pc radius sphere 
and Figure \ref{fig:spat_g13_s2} probably shows on (X,Z) and (Y,Z) projections the orbit tube in 
which stars are confined. Stream 2-7 in Table \ref{tab:table4} shows also the same features 
with a $6\cdot 10^{8}$ year old component and a lower vertical velocity (W=-8.5 $km\cdot s^{-1}$).\\
\begin{figure}
  \begin{center}
\hspace*{-1.8cm}
  \epsfig{file=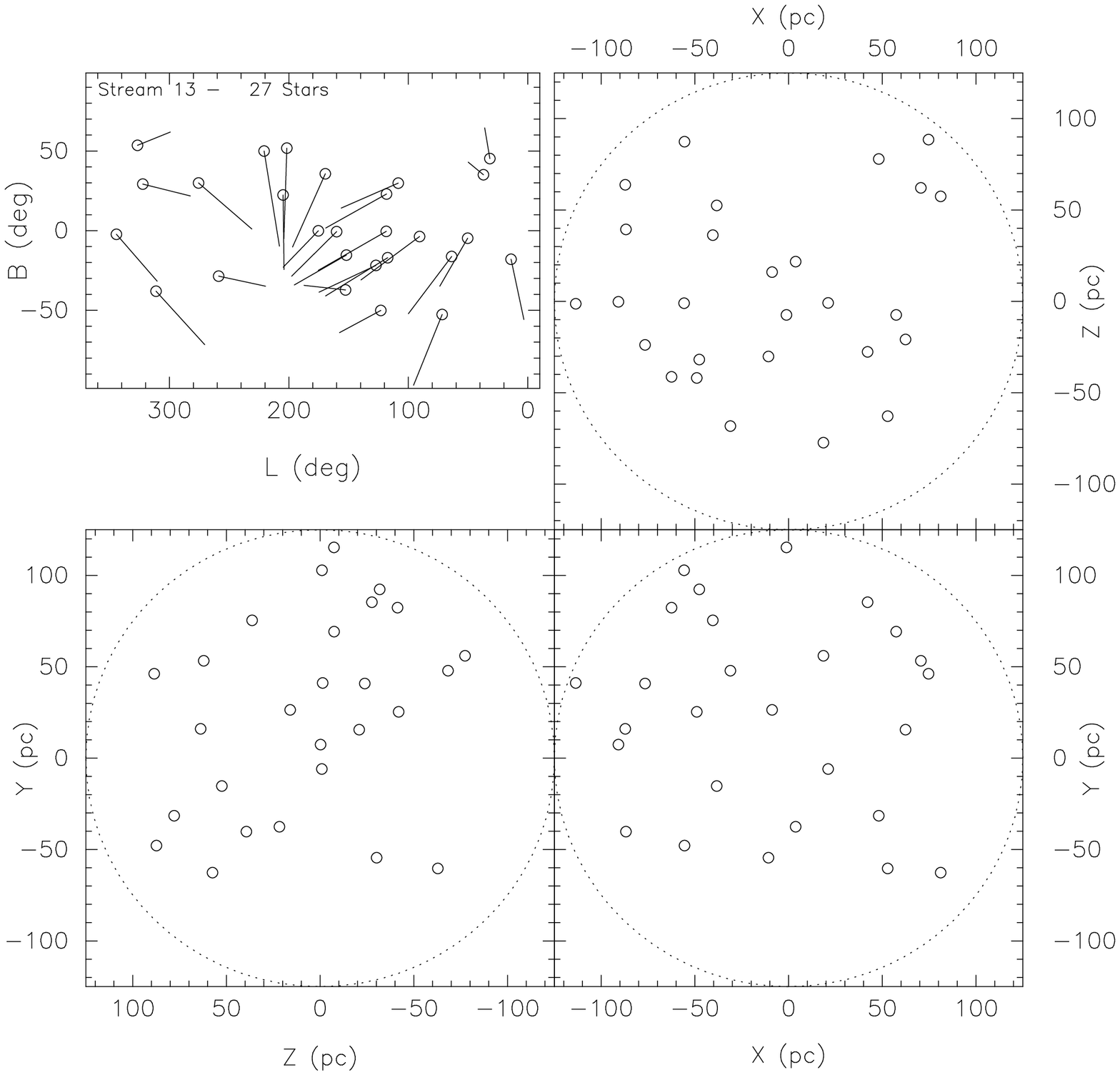,height=12.cm,width=9.4cm,angle=-90.}
  \epsfig{file=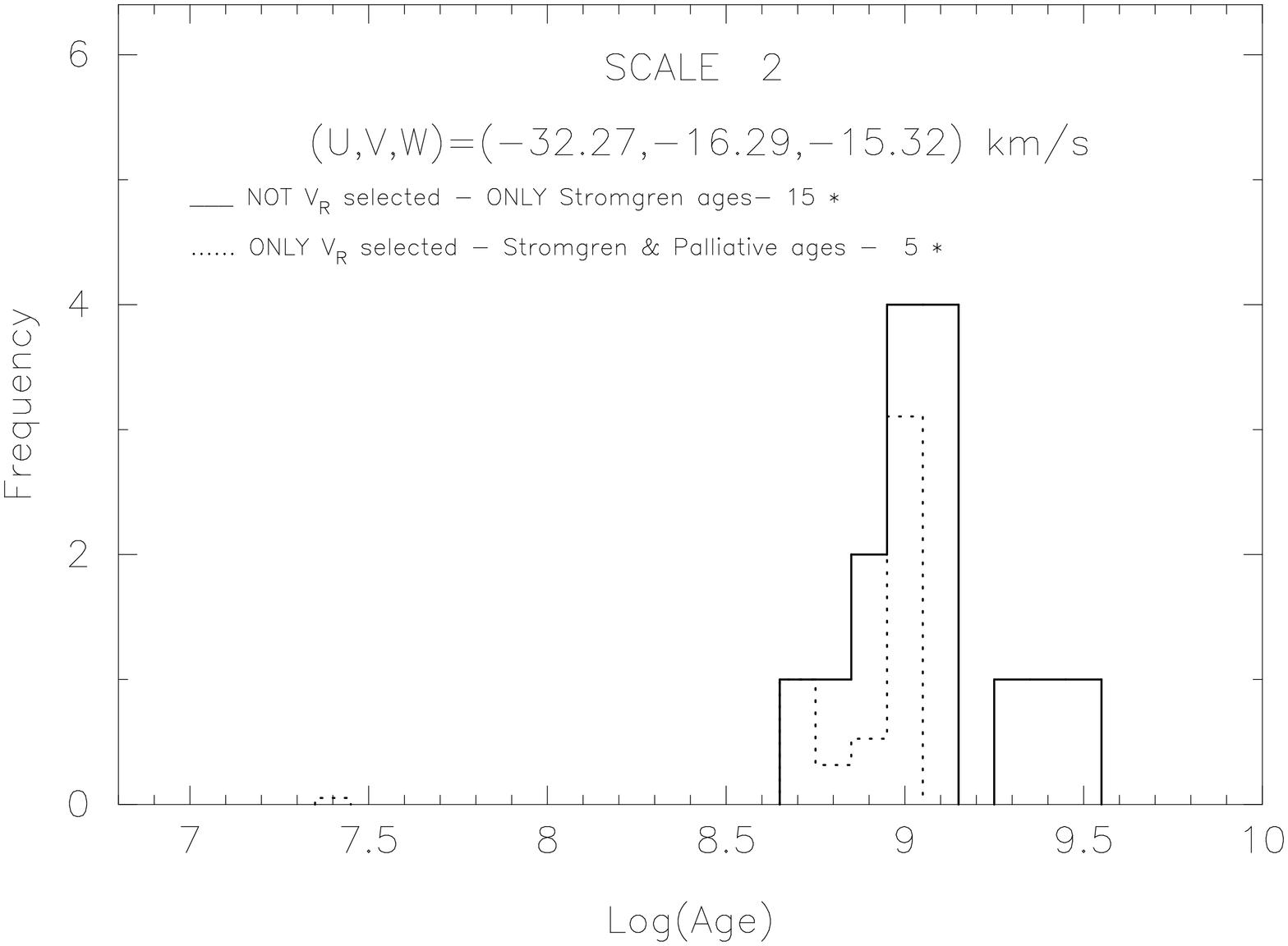,height=8.cm,width=7.cm,angle=-90.}
  \caption{\em {\bf Stream 2-13} (in Table \ref{tab:table4}). Space ({\bf top}) and age ({\bf bottom}) distributions from the sub-set without V$_{R}$ selection at scale 2.}
  \label{fig:spat_g13_s2}
  \end{center}
\end{figure}
\item Particulars on oldest groups\\
\label{sec:oldstream}
Another striking feature of the velocity field is the existence of a $2\cdot 10^{9}$ year old 
population still in velocity  structures. These streams are only detected at the coarser resolution 
(scale 3) in agreement with an intrinsically large velocity dispersion. They have much less members 
(between 20 and 30 initial members) than the previously  investigated streams and have very few 
observed $V_{R}$. Two main old groups (stream 3-9 and 3-14 in Table \ref{tab:table3} and Figures \ref{fig:spat_g9_s3}, 
\ref{fig:spat_g14_s3} ) are clearly detected and have similar characteristics:\\
\begin{itemize}
\item Age distributions peak between $1.6$ and $2\cdot 10^{9}$ yr
\item Their velocity dispersion obtained from the few selected stars on observed $V_{R}$ are of order 6 $km \cdot s^{-1}$ (stream 3-14 seems to have a lower velocity dispersion but the result is obtained for only 3 stars).
\item U-component is positive (towards the galactic center) and often larger than 20 $km\cdot s^{-1}$
\item Space distributions may still be clumpy\\
\end{itemize}
  \begin{figure}
  \begin{center}
 \hspace*{-1.8cm}
  \epsfig{file=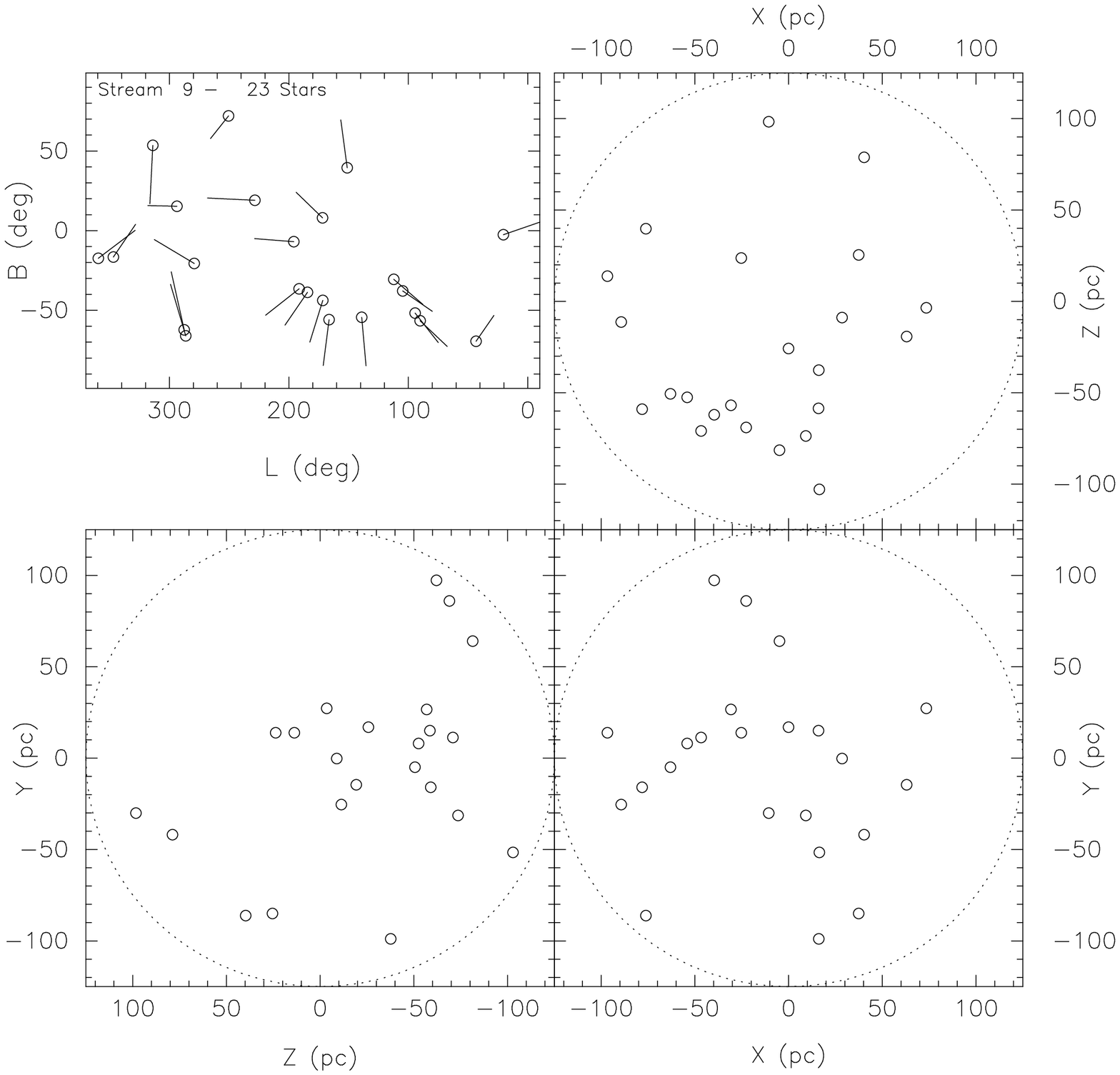,height=12.cm,width=9.4cm,angle=-90.}
  \epsfig{file=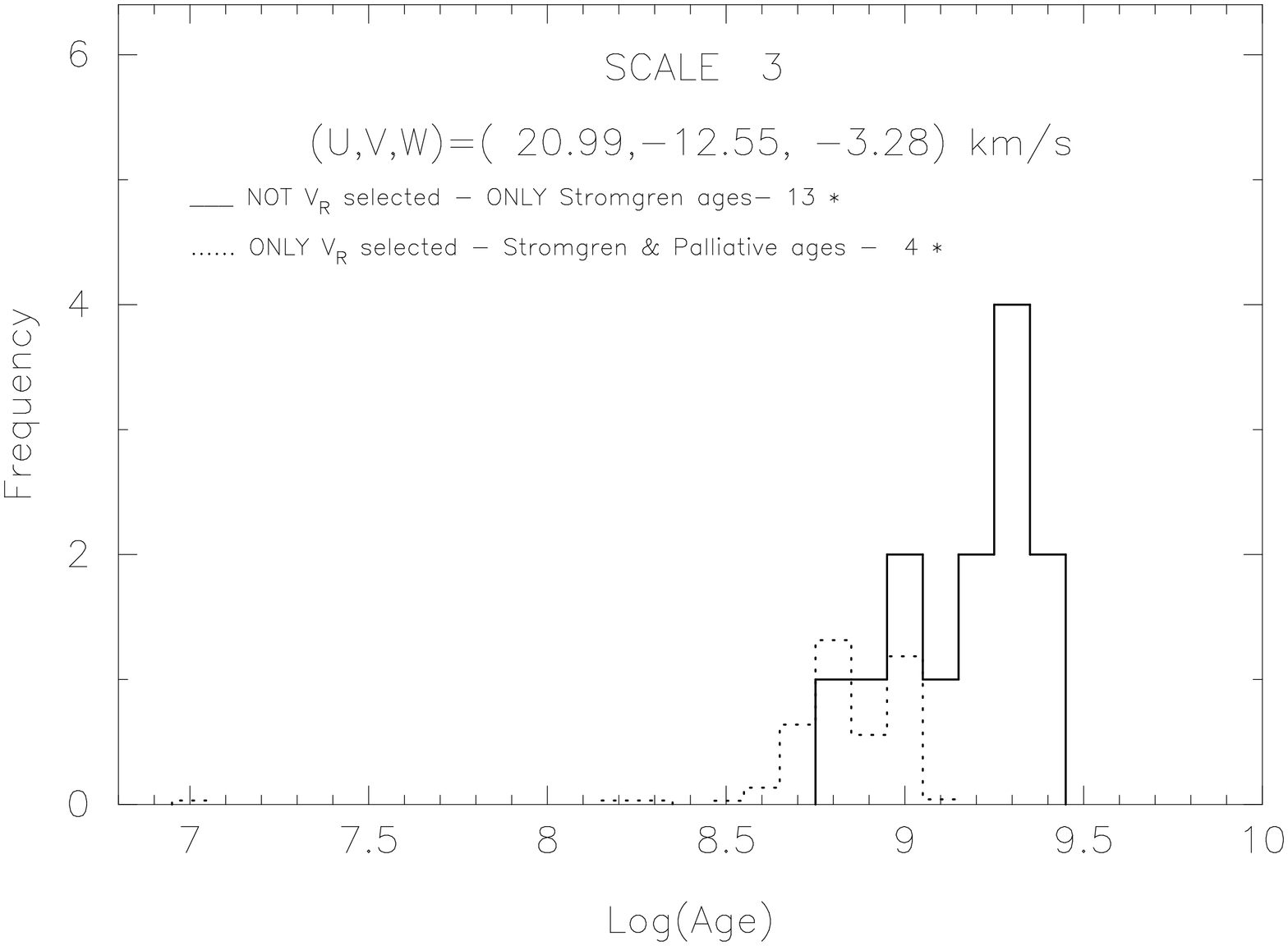,height=8.cm,width=7.cm,angle=-90.}
  \caption{\em {\bf Stream 3-9} (in Table \ref{tab:table3}). Space ({\bf top}) and age ({\bf bottom}) distributions from the sub-set without V$_{R}$ selection at scale 3.}
  \label{fig:spat_g9_s3}
  \end{center}
  \end{figure}
  \begin{figure}
  \begin{center}
 \hspace*{-1.8cm}
  \epsfig{file=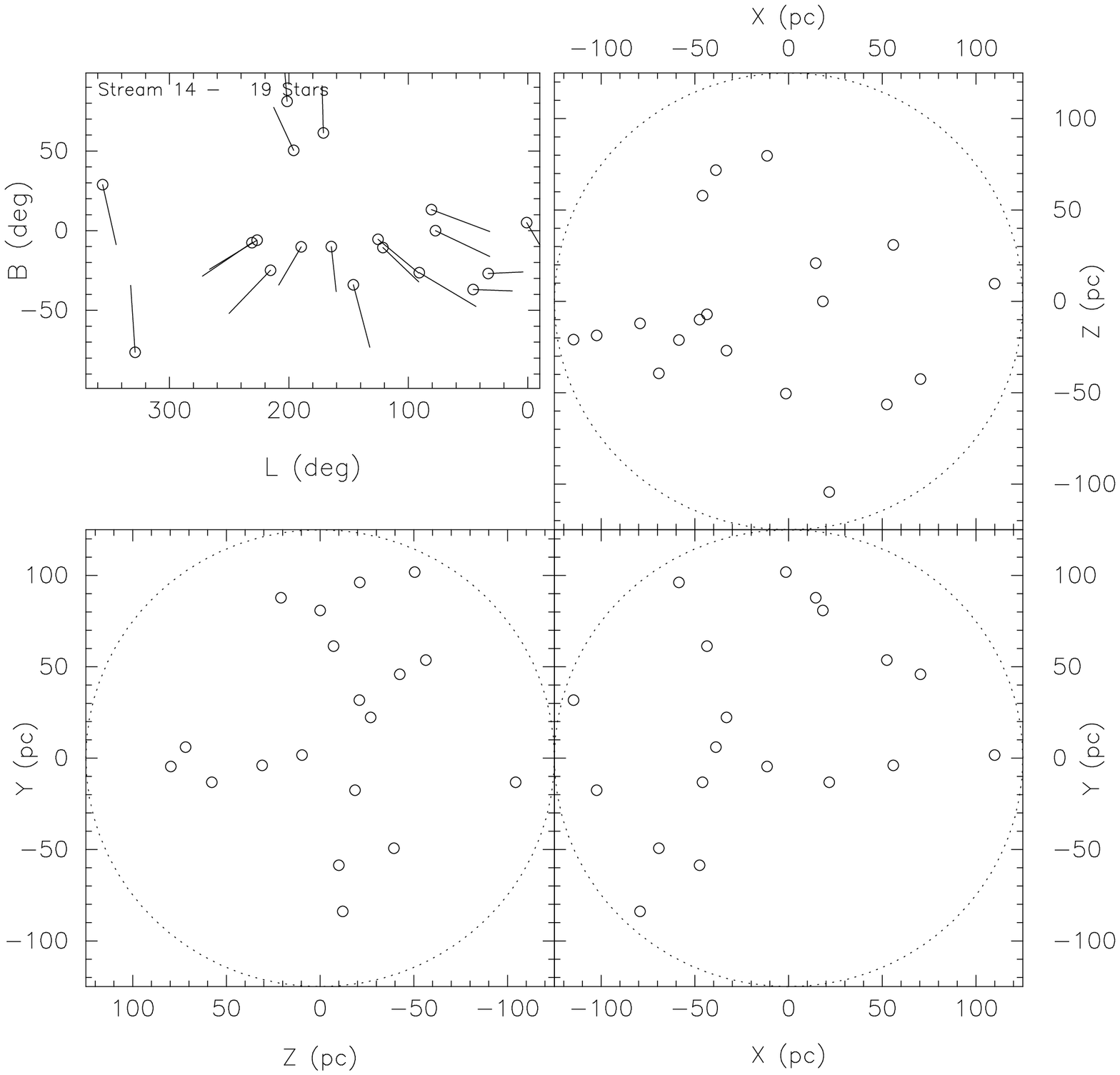,height=12.cm,width=9.4cm,angle=-90.}
  \epsfig{file=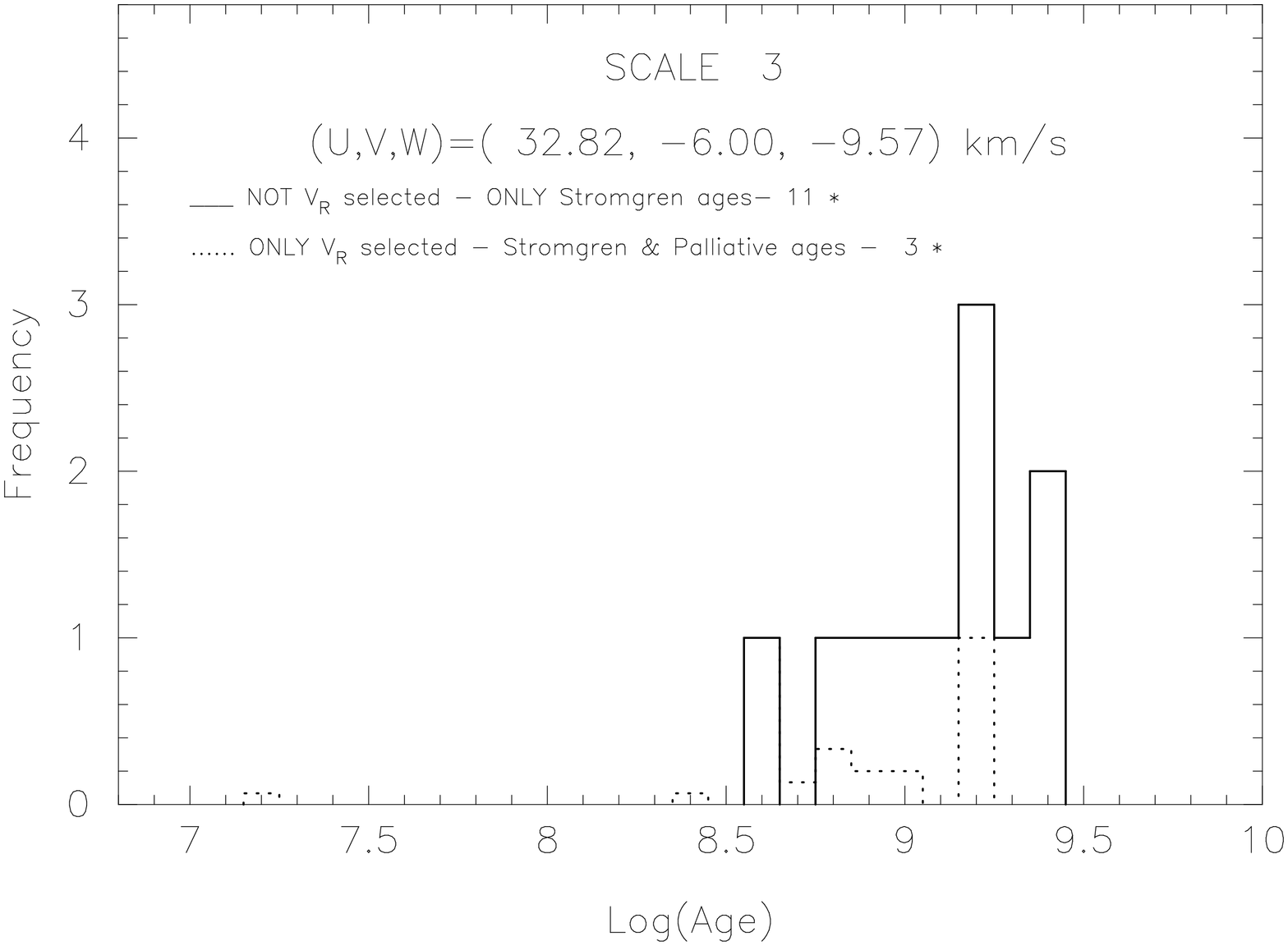,height=8.cm,width=7.cm,angle=-90.}
  \caption{\em {\bf Stream 3-14} (in Table \ref{tab:table3}). Space ({\bf top}) and age ({\bf bottom}) distributions from the sub-set without V$_{R}$ selection  at scale 3.}
  \label{fig:spat_g14_s3}
  \end{center}
  \end{figure}
We should notice at this point that these old star streams are both affected by two biases with 
respect to the whole sample characteristics. They have much less observed radial velocities 
(16$\%$ vs. 46$\%$) and have much more Str\"omgren photometry (57$\%$ vs. 37$\%$). For 
this reason, space distributions are displayed without radial velocity selection.\\
Streams are rather fragile structures originating from cluster evaporation. The survival of these 
relatively old moving groups with coherent ages can be explain by two non exclusive mechanisms. 
On one hand, recent simulations (Zwart et al, 1998) of heavy cluster dynamical evolution 
(with $\sim$ 32000 particles) have shown that contrary to lighter open clusters whose typical 
lifetime is $1-2\cdot 10^{8}$ (\cite{Wielen71} and \cite{Lyn82}), heavy clusters can survive up to 4 Gyr 
in the Galactic gravitational field. This long lifetime can authorize to maintain old streams but such heavy 
open clusters are probably very rare in the disc. 
On the other hand, as sketched in Dehnen (1998) to explain their eccentric orbits, stars of these moving 
groups, provided that they were formed in the inner part of the disc, could have been trapped into 
resonant orbits with the non-axisymmetric force created by the gravitational potential of the Galactic bar. 
This latest explanation does not require any minimum lifetime to the initially bound structure from which 
streams originate.\\ 
\end{itemize}
\begin{table*}
 \vspace{-0.1cm}
  \caption{\em Main characteristics of detected streams at scale 3 with filter size of  8.6 $km\cdot s^{-1}$ ($\overline{\sigma}_{stream} \sim $ 6.3 $km\cdot s^{-1}$). $\overline{\bf U} \pm \sigma_{u}$, $\overline{\bf V}\pm \sigma_{v}$ and $\overline{\bf W} \pm \sigma_{w}$ are the mean velocity components and their dispersions calculated, after selection on $V_{R}$,  with the true radial velocity data - {\bf N$_{init}$} is the initial number of stars belonging to the structure - {\bf N$_{V_{R}}$} is the number of stars with observed radial velocity - {\bf N$_{sel}$} is the number of stars among N$_{Vr}$ after selection 5.2.1. The selection has not been done in case where there is less than 3 observed radial velocities.  In this last case and also when none of the stars are selected we do not present the line in the table. However the same stream identifiers are conserved: the first digit is the scale followed by the stream number. Column {\bf $\%_{field}$} is the percentage of field stars estimated in each structure with the procedure 5.2.2. Cross-identification is done with Eggen's superclusters (SCl) and open cluster (OCl) data from Palo\`us and Eggen (in Gomez et al, 1990) (see Table \ref{tab:cross}) - {\bf Total 1} gives the sum of suspected stream members (col. N$_{init}$) and the sum of observed $V_{R}$ among them (col. N$_{V_{R}}$) - {\bf Total2} gives the sum of all confirmed stream stars among available observed radial velocities (col. N$_{sel}$) - {\bf Total 3} is the sum of all expected stream members in each stream (col. N$_{init}$) taking into account the confirmed/suspected ratio obtained in each stream - {\bf Total 4} gives the inferred fraction of stars in stream in the full sample of 2910 stars (col. N$_{init}$). {\bf Total 5} is the fraction of stars in stream corrected for field contamination (col. N$_{init}$). }
  \label{tab:table3}
    \footnotesize  
   \begin{center}
    \begin{tabular}[h]{lrrrrccc}
      \hline \\[-5pt]
 Stream &N$_{init}$ &N$_{V_{R}}$&N$_{sel}$ &$\%_{field}$&$\overline{U}$ $ \pm\sigma_{u}$ & $\overline{V}$ $\pm \sigma_{v}$ &$\overline{W}$ $\pm \sigma_{w}$\\
 & & & ($\mid\kappa\mid$=3.)&&($km\cdot s^{-1}$)&($km\cdot s^{-1}$)&($km\cdot s^{-1}$)\\
 \hline \\ [-5pt]
    ~3-8~Pleiades SCl& 554 & 276 & 254 &23$\%$&-14.4$\pm$ 8.4&-20.1$\pm$ 5.9& -6.2$\pm$ 6.3\\
    ~3-9 &  23 &   7 &   4 && 21.0$\pm$ 5.7&-12.6$\pm$ 3.2& -3.3$\pm$ 9.1\\
   3-10 Hyades SCl& 287 & 140 &126& 34$\%$&-32.9$\pm$ 6.6&-14.5$\pm$ 6.8& -5.6$\pm$ 6.5\\
   3-11 &  10 &   7 &   5 &&-34.8$\pm$ 3.1&-14.4$\pm$ 2.5&-17.6$\pm$ 1.7\\
   3-14 &  19 &   6 &   3 && 32.8$\pm$ 3.0& -6.0$\pm$ 2.0& -9.6$\pm$ 1.1\\
   3-15 Cen Associations& 577 & 254 & 228 &20$\%$&-15.2$\pm$ 8.6& -8.4$\pm$ 6.7& -8.8$\pm$ 6.1\\
   3-16 &  21 &   8 &   3 & &-2.0$\pm$ 1.6& -7.1$\pm$ 2.4& -2.5$\pm$ 4.8\\
   3-18 {\bf New SCl} & 406 & 197 & 158 &23$\%$&  3.6$\pm$ 6.8&  2.9$\pm$ 5.0& -6.0$\pm$ 6.3\\
   3-19 Sirius SCl& 195 &  97 &  92 &31$\%$& 14.0$\pm$ 7.3&  1.0$\pm$ 6.4& -7.8$\pm$ 5.5\\
      \hline \\[-5pt]
Total1& 2092&992 & && & & \\
      \hline \\[-5pt]
Total2& & &873&\multicolumn{4}{l}{($V_{R}$ confirmed stream members)} \\
Total3&1834.3 & & &\multicolumn{4}{l}{(Total stream members)}\\
       \hline \\[-5pt]
Total4&{\bf 63.0$\%$}&&&\multicolumn{4}{l}{(Fraction of stars in streams in the full sample)}\\
Total5&{\bf 46.4$\%$}&&&\multicolumn{4}{l}{(Fraction of stars in streams corrected for field contamination)}\\
      \hline \\[-5pt]
   \end{tabular}
   \end{center}
   \end{table*}
\begin{table*}
 \vspace{-0.1cm}
  \caption{\em Main characteristics of detected velocity structures at scale 2 with filter size of 5.5 $km\cdot s^{-1}$ ($\overline{\sigma}_{stream} \sim $ 3.8 $km\cdot s^{-1}$)- Legend is the same as Table \ref{tab:table3}}
  \label{tab:table4}
    \footnotesize
   \begin{center}
     \begin{tabular}[h]{lrrrccc}
      \hline \\[-5pt]
 Stream &N$_{init}$ &N$_{V_{R}}$&N$_{sel}$ &$\overline{U}$ $\pm \sigma_{u}$
 & $\overline{V}$ $\pm \sigma_{v}$ &$\overline{W}$ $\pm \sigma_{w}$\\
  & & & ($\mid\kappa\mid$=3.)&($km\cdot s^{-1}$) &($km\cdot s^{-1}$) &($km\cdot s^{-1}$)\\
 \hline \\ [-5pt]
    ~2-2~Pleiades OCl&  89 &  50 &  37 & -7.5$\pm$ 4.6&-24.6$\pm$ 3.5&-10.2$\pm$ 3.5\\
    ~2-5~Pleiades SCl& 161 &  85 &  67 &-12.0$\pm$ 5.3&-21.6$\pm$ 4.7& -5.3$\pm$ 5.9\\
    ~2-7 &  52 &  22 &  13 &-27.3$\pm$ 5.6&-17.8$\pm$ 1.4& -8.5$\pm$ 5.4\\
   2-10 Hyades SCl (1)&  67 &  33 &  23 &-31.6$\pm$ 2.8&-15.8$\pm$ 2.8&  0.8$\pm$ 2.7\\
   2-12 Cen-Lup& 112 &  45 &  28 &-12.4$\pm$ 6.1&-16.5$\pm$ 4.6& -7.4$\pm$ 3.1\\
   2-13 &  27 &  15 &   5 &-32.3$\pm$ 4.9&-16.3$\pm$ 2.2&-15.3$\pm$ 4.6\\
   2-14 IC 2391 SCl& 149 &  66 &  50 &-20.8$\pm$ 4.3&-14.5$\pm$ 4.9& -4.9$\pm$ 5.0\\
   2-15 Hyades OCl&  26 &  14 &   8 &-42.8$\pm$ 3.6&-17.9$\pm$ 3.2& -2.2$\pm$ 5.2\\
   2-18 Hyades SCl (2)&  59 &  27 &  17 &-33.0$\pm$ 4.2&-14.1$\pm$ 4.0& -5.1$\pm$ 3.1\\
   2-23 &  26 &  15 &   9 & -2.9$\pm$ 1.9&-10.6$\pm$ 2.4& -5.3$\pm$ 5.0\\
   2-25 Hyades SCl (3)&  49 &  29 &  17 &-32.8$\pm$ 2.8&-11.8$\pm$ 2.8& -8.9$\pm$ 2.9\\
   2-26 Cen-Crux& 273 & 116 &  93 &-13.1$\pm$ 6.2& -7.9$\pm$ 6.1& -9.3$\pm$ 5.5\\
   2-27 New SCl (1)&  45 &  20 &   6 &  1.0$\pm$ 6.0& -4.0$\pm$ 6.1&  0.2$\pm$ 4.2\\
   2-29 NGC 1901&  33 &  14 &   5 &-26.1$\pm$ 2.1& -7.6$\pm$ 3.7&  0.7$\pm$ 3.2\\
   2-31 &  28 &  16 &   7 & -5.8$\pm$ 4.5& -0.8$\pm$ 6.6&-15.6$\pm$ 4.1\\
   2-32 &  20 &  12 &   6 & 19.3$\pm$ 1.9& -5.1$\pm$ 5.3&-11.9$\pm$ 2.6\\
   2-33 &  16 &   8 &   4 & 20.1$\pm$ 1.6& -2.6$\pm$ 1.4&-15.9$\pm$ 4.8\\
   2-34 &  15 &   7 &   6 & 21.1$\pm$ 2.4&  0.4$\pm$ 3.9& -0.8$\pm$ 2.9\\
   2-35 New SCl (2)&  89 &  40 &  28 &  4.5$\pm$ 5.4& -0.9$\pm$ 4.4& -7.4$\pm$ 2.7\\
   2-37 Sirius SCl (1)&  64 &  34 &  27 & 12.4$\pm$ 4.0&  0.7$\pm$ 4.6& -7.7$\pm$ 4.7\\
   2-38 New SCl (3)& 102 &  51 &  23 &  1.3$\pm$ 4.8&  3.0$\pm$ 2.3& -3.6$\pm$ 3.1\\
   2-40 &  16 &   8 &   3 & -4.8$\pm$ 3.1&  5.2$\pm$ 6.7&-11.3$\pm$ 3.6\\
   2-41 Sirius SCl (2)&  42 &  21 &  13 & 12.4$\pm$ 3.7&  4.2$\pm$ 3.3& -9.0$\pm$ 2.9\\
   2-43 New SCl (4)&  77 &  46 &  28 &  5.3$\pm$ 2.7&  6.5$\pm$ 3.6&-11.4$\pm$ 4.0\\
   2-44 New SCl (5)&  45 &  20 &   8 &  6.3$\pm$ 2.8&  6.5$\pm$ 2.6&  0.8$\pm$ 2.7\\
   2-46 &  24 &  13 &   7 & -4.8$\pm$ 6.1& 10.8$\pm$ 2.0& -6.5$\pm$ 1.9\\
      \hline \\[-5pt]
Total1& 1706&827 & && & \\
      \hline \\[-5pt]
Total2& & &538&\multicolumn{3}{l}{($V_{R}$ confirmed stream members)} \\
Total3&1114.3 & & & \multicolumn{3}{l}{(Total stream members)}\\
       \hline \\[-5pt]
Total4&{\bf 38.3$\%$}&&&\multicolumn{3}{l}{(Fraction of stars in streams)}\\
      \hline \\[-5pt]
      \end{tabular}
      \end{center}
      \end{table*}
\begin{table*}
 \vspace{-0.1cm}
  \caption{\em Main characteristics of detected velocity structures at scale 1 with typical size of 3.2 $km\cdot s^{-1}$ ($\overline{\sigma}_{stream} \sim $ 2.4 $km\cdot s^{-1}$) - Legend is the same as Table \ref{tab:table3}}
  \label{tab:table5}
    \footnotesize
   \begin{center}
     \begin{tabular}[h]{lrrrccc}
      \hline \\[-5pt]
 Stream &N$_{init}$ &N$_{V_{R}}$&N$_{sel}$ &$\overline{U}$ $\pm \sigma_{u}$ & $\overline{V}$ $\pm \sigma_{v}$&$\overline{W}$ $\pm \sigma_{w}$\\
 & &&($\mid\kappa\mid$=3.)&($km\cdot s^{-1}$) &($km\cdot s^{-1}$) &($km\cdot s^{-1}$)\\
 \hline \\ [-5pt]
    ~1-1 &  14 &   7 &   3 &-14.9$\pm$ 0.5&-27.8$\pm$ 0.6& -1.7$\pm$ 1.9\\
    ~1-2~ Pleiades OCl&  40 &  17 &  12 & -4.7$\pm$ 2.0&-23.8$\pm$ 1.2& -8.9$\pm$ 2.6\\
    ~1-4 &  16 &   6 &   3 &-15.1$\pm$ 3.1&-24.1$\pm$ 2.9& -0.7$\pm$ 1.7\\
    ~1-6~Pleiades SCl (1)&  33 &  14 &   9 &-13.1$\pm$ 3.1&-21.9$\pm$ 3.3& -7.1$\pm$ 2.5\\
    ~1-7~Pleiades SCl (2)&  46 &  30 &  20 &-11.1$\pm$ 1.7&-21.9$\pm$ 3.0& -6.0$\pm$ 1.8\\
   1-10 &  22 &  10 &   3 &-31.0$\pm$ 5.2&-19.0$\pm$ 1.5& -6.0$\pm$ 2.7\\
   1-11 &  18 &   6 &   3 &-19.2$\pm$ 3.4&-19.0$\pm$ 2.4& -7.1$\pm$ 3.5\\
   1-12 &  22 &  11 &   3 & -9.9$\pm$ 0.4&-20.3$\pm$ 1.6& -0.5$\pm$ 1.8\\
   1-13 &  22 &  11 &   3 &-16.1$\pm$ 2.0&-20.9$\pm$ 0.1&  1.9$\pm$ 2.9\\
   1-15 &  32 &  17 &   9 &-18.8$\pm$ 4.2&-15.7$\pm$ 3.4& -1.8$\pm$ 2.3\\
   1-17~Cen-Lup &  61 &  17 &   5 & -9.4$\pm$ 3.3&-18.2$\pm$ 2.7& -5.9$\pm$ 1.1\\
   1-19~Hyades SCl (1)&  24 &  12 &  10 &-29.0$\pm$ 5.5&-12.5$\pm$ 6.0& -0.0$\pm$ 4.4\\
   1-20~IC 2391 SCl (1)&  33 &  15 &   8 &-20.1$\pm$ 2.9&-12.8$\pm$ 3.1& -5.0$\pm$ 1.8\\
   1-21 &   8 &   6 &   3 &-19.9$\pm$ 1.1&-14.8$\pm$ 3.2&-15.5$\pm$ 0.5\\
   1-22 &  28 &  12 &   3 &-27.2$\pm$ 2.4&-17.1$\pm$ 2.8& -8.4$\pm$ 3.2\\
   1-23~Hyades SCl (2)&  39 &  22 &   8 &-33.2$\pm$ 2.3&-13.1$\pm$ 3.5& -6.6$\pm$ 2.5\\
   1-24 &  32 &  12 &   3 &-13.8$\pm$ 2.7&-15.5$\pm$ 5.9&  1.1$\pm$ 3.4\\
   1-25~IC 2391 SCl (2)&  41 &  21 &   8 &-21.0$\pm$ 4.5& -10.0$\pm$ 3.7& -6.1$\pm$ 2.9\\
   1-26 &  23 &  11 &   6 & -2.4$\pm$ 1.6&-11.7$\pm$ 1.8& -4.5$\pm$ 3.2\\
   1-27 &  12 &   7 &   3 &-30.5$\pm$ 1.0&-10.8$\pm$ 0.6& -9.1$\pm$ 2.1\\
   1-29~Cen-Crux &  45 &  17 &   6 &-13.1$\pm$ 3.2& -9.4$\pm$ 3.7& -7.7$\pm$ 4.5\\
   1-31 &  53 &  21 &  12 &-19.5$\pm$ 3.4& -5.8$\pm$ 2.3&-10.5$\pm$ 3.0\\
   1-32 &  17 &   7 &   4 &-25.8$\pm$ 2.0& -9.9$\pm$ 3.2& -0.2$\pm$ 3.0\\
   1-34 &  24 &  11 &   7 & -8.0$\pm$ 1.0& -3.8$\pm$ 1.3&-13.2$\pm$ 2.2\\
   1-37 &  14 &   9 &   3 &  3.1$\pm$ 5.4& -1.1$\pm$ 1.8&-15.2$\pm$ 3.4\\
   1-39 &  13 &   6 &   5 & 22.1$\pm$ 1.5& -1.1$\pm$ 1.9& -0.5$\pm$ 2.2\\
   1-41~New SCl (2)&  39 &  18 &  11 &  4.1$\pm$ 4.2&  0.0$\pm$ 2.4& -7.7$\pm$ 1.8\\
   1-45 &  14 &   5 &   3 &  8.5$\pm$ 1.4&  0.9$\pm$ 1.3& -2.4$\pm$ 2.5\\
   1-46 &  27 &  13 &   4 &  1.4$\pm$ 4.3&  1.4$\pm$ 2.0& -3.4$\pm$ 2.0\\
   1-49 &  16 &   6 &   3 & 15.6$\pm$ 1.3&  2.1$\pm$ 4.4& -9.1$\pm$ 1.6\\
   1-51~Sirius SCl (2)&  16 &   7 &   4 &  8.9$\pm$ 3.1&  4.8$\pm$ 1.6&-12.9$\pm$ 0.6\\
   1-54 &  18 &  11 &   5 &  6.5$\pm$ 1.2&  5.9$\pm$ 1.8&-10.6$\pm$ 2.1\\
   1-55~New SCl (3)&  18 &   9 &   5 &  2.9$\pm$ 4.4&  3.9$\pm$ 1.8& -4.9$\pm$ 2.6\\
   1-56~Sirius SCl (1)&  13 &   8 &   4 & 12.0$\pm$ 2.7&  5.9$\pm$ 1.3& -9.2$\pm$ 1.1\\
   1-58 &  20 &  11 &   6 &  3.7$\pm$ 1.9&  7.2$\pm$ 3.5&-10.0$\pm$ 2.7\\
   1-59~New SCl (5)&  24 &   9 &   4 &  6.2$\pm$ 0.6&  6.0$\pm$ 1.2& -0.3$\pm$ 1.9\\
   1-61~New SCl (4)&  15 &   5 &   4 &  7.2$\pm$ 1.8&  9.9$\pm$ 2.2&-10.5$\pm$ 2.7\\
   1-62 &  17 &   9 &   5 & -3.9$\pm$ 7.6& 10.5$\pm$ 2.2& -7.2$\pm$ 1.1\\
   1-59~New SCl (5)&  24 &   9 &   4 &  6.2$\pm$ 0.6&  6.0$\pm$ 1.2& -0.3$\pm$ 1.9\\
   1-61~New SCl (4)&  15 &   5 &   4 &  7.2$\pm$ 1.8&  9.9$\pm$ 2.2&-10.5$\pm$ 2.7\\
   1-62 &  17 &   9 &   5 & -3.9$\pm$ 7.6& 10.5$\pm$ 2.2& -7.2$\pm$ 1.1\\
\hline \\[-5pt]
Total1& 1262&556 & && &  \\
      \hline \\[-5pt]
Total2& & &220&\multicolumn{3}{l}{($V_{R}$ confirmed stream members)} \\
Total3&521.2 & & & \multicolumn{3}{l}{(Total stream members)}\\
       \hline \\[-5pt]
Total4&{\bf 17.9$\%$}&&&\multicolumn{3}{l}{(Fraction of stars in streams)}\\
      \hline \\[-5pt]
\end{tabular}
\end{center}
\end{table*}
\begin{table*}
 \vspace{-0.1cm}
  \caption{\em Cross-identification data for known kinematical groups: Open clusters (OCl) and Superclusters (SCl)}
  \label{tab:cross}
    \leavevmode
    \footnotesize  
   \begin{center}
    \begin{tabular}[h]{lrrrcl}
      \hline \\[-5pt]
   &      $\overline{U}$  &    $\overline{V}$   &   $\overline{W}$   & Age &  Source\\[+5pt]
        &  \multicolumn{3}{c}{($km\cdot s^{-1}$)}& (yr) &\\[+5pt]
      \hline \\
OCl  &&&&&\\
      \hline \\
Hyades          &        -44.4     &      -17.0     &      -5.0      &  8$\pm$2$\cdot 10^{8}$ &   \cite{Pal77}\\
Pleiades           &     -5.8       &     -24.0       &    -12.4     &  5 $\pm$1$\cdot 10^{7}$ &   \cite{Pal77}\\
IC2391             &     -18.3     &      -13.5     &      -5.9     &  3 $\pm$1$\cdot 10^{7}$   &   \cite{Pal77}\\
U.Ma               &     +14.5      &     +2.5       &     -8.5      &  3 $\cdot 10^{8}$  &  \cite{Eg73}\\
Coma Ber.        &        -1.8      &      -8.2       &     -0.7      &  4 $\pm$1$\cdot 10^{8}$  &   \cite{Pal77}\\
IC2602           &       -0.7       &     -25.7      &     -1.4      & 1 $\pm$0.5$\cdot 10^{7}$   &   \cite{Pal77}\\
Praesepe         &       -37.1     &      -23.5     &      -7.0     & 7 $\pm$2$\cdot 10^{8}$   &    \cite{Pal77}\\
Alpha Per.    &        -10.8    &       -20.5     &      -0.7      & 4 $\pm$0.5$\cdot 10^{7}$   &   \cite{Pal77}\\
\hline\\
SCl& & & & &\\
\hline\\
Hyades          &        -40.0      &     -17.0     &      -2 .0     & 3- 6- 8$\cdot 10^{8}$  & \cite{Eg92b}\\
Hyades         &        -44.4      &     -17.8     &      -1.5  & 13.3$\cdot 10^{8}$  &    \cite{Chen97}\\
& & & & &\\
Pleiades        &        -11.6     &      -20.7    &       -10.4& 5$\cdot 10^{7}$ $\&$ 2$\cdot 10^{8}$& \cite{Eg92a}\\
Pleiades      &        -10.0   &      -19.0    &       -8.1&  0.7$\cdot 10^{8}$&   \cite{Chen97} \\
& & & & &\\
IC 2391        &          -22.4     &      -17.5     &      -9.4& 8$\cdot 10^{7}$ $\&$ 2.5$\cdot 10^{8}$ & \cite{Eg91}\\
IC 2391        &          -20.8     &      -15.9     &      -8.3&  &    \cite{Eg92d} \\
IC 2391    &          -15.9     &      -13.1     &      -4.5&  2.8$\cdot 10^{8}$&\cite{Chen97}\\
& & & & &\\
Sirius         &         +15.0     &        +1.0        &      -11.0 &  6.3$\cdot 10^{8}$ $\&$ $10^{9}$& \cite{Eg92c}\\
Sirius     &         +13.0   &        +3.2       &      -7.5  &  6.5$\cdot 10^{8}$& \cite{Chen97}\\
& & & & &\\
NGC 1901     &           -26.4      &     -10.4     &      -1.5 &  8$\cdot 10^{8}$ $\&$ 1.2$\cdot 10^{9}$& \cite{Eg96}\\
\hline \\[-5pt]
   \end{tabular}
   \end{center}
   \end{table*}
%
%
\section{Conclusions}
\label{sec:conclu} 
A systematic multi-scale analysis of both the space and velocity distributions of a thin disc young star sample 
has been performed. Over-densities in space distributions are 
mainly well known open clusters (Hyades OCl, Coma OCl and Ursa Major OCl) and associations of stars 
(Centaurus-Crux and Centaurus-Lupus). Evidence is given for three nearby loose clusters or associations 
not detected yet (Bootes and Pegasus 1 and 2). Evaporation of relatively massive stars (1.8 M$\odot$) out 
of the Hyades open cluster is also mapped in Paper II and form an asymmetric pattern. The sample is well 
mixed in position space since no more than 7$\%$ of the stars are in concentrated clumps with coherent 
tangential velocities.\\
The 3D velocity field reconstructed from a statistical convergent point method exhibits a strong 
structuring at typical scales of $\overline{\sigma}_{stream}\sim$ 6.3, 3.8 and 2.4 $km\cdot s^{-1}$. 
At large scale (scale 3) the majority of structures are Eggen's {\em superclusters} (Pleiades SCl, 
Hyades SCl and Sirius SCl) with the whole Centaurus association embedded in a larger kinematic structure. 
A new {\em supercluster}-like structure is found with a mean velocity between the Sun and Sirius SCl velocities. 
These large scale velocity structures are all characterized by a large age range which reflects 
the overall sample age distribution. Moreover, few old streams of $\sim$ 2 Gyr are also extracted at 
this scale with high U components towards the Galactic center. Taking into account the fraction of 
spurious members, evaluated with an observed radial velocity data set, into all these large velocity 
dispersion structures we show that they represent 63$\%$ of the sample. This percentage drops to 
46$\%$ if we remove the velocity background created by a smooth velocity ellipsoid in each structure. 
Smaller scales ($\overline{\sigma}_{stream}\sim$ 3.8 and 2.4 $km\cdot s^{-1}$) reveal that 
{\em superclusters} are always substructured by 2 or more streams which generally exhibits a 
coherent age distribution. The older the stream is the more difficult the age segregation between 
close velocity clumps is. At scale 2 and 1, background stars are negligible and percentages of stars 
in streams, after evaluating the fraction of spurious members, are 38$\%$ and 18$\%$ respectively.\\
%
%
\begin{acknowledgements}
This work was facilitated by the use of the Vizier service and the Simbad database developed at CDS.
E.C thanks A. Bijaoui and J.L. Starck for constructive discussions on wavelet analysis, and C. Pichon 
for interesting discussions and help on some technical points. E.C is very grateful to R. Asiain 
for providing the two codes of age calculation from Str\"omgren photometry and thanks 
J. Torra and F. Figueras for valuable discussions.
\end{acknowledgements}

\end{document}